\documentclass[prd, 10pt,aps,nofootinbib, floatfix, notitlepage,twocolumn]{revtex4-1}

\usepackage{amsmath,amsfonts,amssymb}
\usepackage{mathrsfs}
\usepackage{graphicx}
\usepackage[english]{babel} 
\usepackage{url} 
\usepackage{color}
\usepackage{bbold}
\usepackage{adjustbox}
 \usepackage[utf8]{inputenc} 
 \usepackage{float}
\usepackage[colorlinks=true,citecolor=blue,urlcolor=blue]{hyperref}

\newcommand{\be}{\begin{equation}}
\newcommand{\ee}{\end{equation}}
\newcommand{\bes}{\begin{equation*}}
\newcommand{\ees}{\end{equation*}}

\makeatletter
\def\l@subsubsection#1#2{}
\makeatother

\begin{document}

\title{Oscillating scalar fields and the Hubble tension: a resolution with novel signatures}
\author{Tristan L.~Smith$^1$}
\author{Vivian Poulin$^2$}
\author{Mustafa A.~Amin$^3$}
\affiliation{$^1$Department of Physics and Astronomy, Swarthmore College, 500 College Ave., Swarthmore, PA 19081, USA}
\affiliation{$^2$Laboratoire Univers \& Particules de Montpellier (LUPM), CNRS \& Universit\'e de Montpellier (UMR-5299),Place Eug\`ene Bataillon, F-34095 Montpellier Cedex 05, France}
\affiliation{$^3$Department of Physics \& Astronomy, Rice University, Houston, TX 77005, USA}

\date{\today}
\begin{abstract}
We present a detailed investigation of a sub-dominant oscillating scalar field (`early dark energy', EDE) in the context of resolving the Hubble tension. Consistent with earlier work, but without relying on fluid approximations, we find that a scalar field frozen due to Hubble friction until ${\rm log}_{10}(z_c)\sim3.5$, reaching $\rho_{\rm EDE}(z_c)/\rho_{\rm tot}\sim10$\%, and diluting faster than matter afterwards can bring cosmic microwave background (CMB), baryonic acoustic oscillations, supernovae luminosity distances, and the late-time estimate of the Hubble constant from the SH0ES collaboration into agreement. A scalar field potential which scales as $V(\phi) \propto \phi^{2n}$ with $2\lesssim n\lesssim 3.4$ around the minimum is preferred at the 68\% confidence level, and the {\em Planck} polarization places additional constraints on the dynamics of perturbations in the scalar field. In particular, the data prefers a potential which flattens at large field displacements. An MCMC analysis of mock data shows that the next-generation CMB observations (i.e., CMB-S4) can unambiguously detect the presence of the EDE at very high significance. This projected sensitivity to the EDE dynamics is mainly driven by improved measurements of the $E$-mode polarization. 

We also explore new observational signatures of EDE scalar field dynamics: (i) We find that depending on the strength of the tensor-to-scalar ratio, the presence of the EDE might imply the existence of isocurvature perturbations in the CMB. (ii) We show that a strikingly rapid, scale-dependent growth of EDE field perturbations can result from parametric resonance driven by the anharmonic oscillating field for $n\approx 2$. This instability and ensuing potentially nonlinear, spatially inhomogenoues, dynamics may provide unique signatures of this scenario.
\end{abstract}

\maketitle

\tableofcontents

\section{Introduction}

The standard cosmological model which includes a cosmological constant, $\Lambda$, cold dark matter (CDM), along with baryons, photons, and neutrinos (known as the $\Lambda$CDM model), is incredibly powerful at describing cosmological observables up to a very high degree of accuracy. This is especially true for our observations of the cosmic microwave background (CMB), the baryon acoustic oscillations (BAO) and the luminosity distances to Type Ia supernovae (SNe Ia). However, it remains a parametric model and the nature of its dominant components - dark matter and dark energy - still needs to be understood. 

In recent years, several tensions between probes of the early and late universe have emerged, possibly leading to a new understanding of these mysterious components. At the heart of this work is the  the long-standing `Hubble tension' \cite{Freedman:2017yms}. This is a statistically significant disagreement between the value of the current expansion rate (i.e., the Hubble constant) {\em measured} by the classical distance ladder (CDL) and that {\em inferred} from measurements of the CMB or the primordial element abundances established during big bang nucleosynthesis (BBN). In particular, the SH0ES team, using Cepheid-calibrated SNe Ia, has determined $H_0 = 74.03\pm1.42$ km/s/Mpc \cite{Riess:2019cxk}, while the $\Lambda$CDM cosmology deduced from {\em Planck} CMB data and BAO+Dark Energy Survey+BBN data predict $H_0=67.4\pm0.6$ km/s/Mpc \cite{Aghanim:2018eyx} and $H_0=67.4^{+1.1}_{-1.2}$ km/s/Mpc \cite{Abbott:2017smn}, respectively.

Additional, low-redshift, methods to determine the Hubble constant also point toward a value that is in disagreement with the value inferred from high redshift observations. One example is the measured strong-lens time delays, which yield $73.3\pm1.8$ km/s/Mpc \cite{Wong:2019kwg} within a flat $\Lambda$CDM cosmology. In combination with the classical distance ladder determination of $H_0$ this leads to a discrepancy with the CMB-inferred value that has now reached the $5.3\sigma$ level. A review of the various estimates of $H_0$ can be found in Ref.~\cite{Verde:2019ivm} and a combination of all late-time determinations gives $H_0 = 73.3 \pm 1.0$ km/s/Mpc. 

Attempts to resolve the Hubble tension modify either late-time ($z\lesssim 1$) or early-time ($z\gtrsim 1100$, pre-recombination) physics (see Ref.~\cite{Knox:2019rjx} for a review).  However, direct probes of the expansion rate at late-times from SNe Ia and BAO measurements place severe limitations on late-time resolutions \cite{Bernal:2016gxb,Poulin:2018zxs,Aylor:2018drw}. On the other hand, early-time resolutions affect the physics that determine the fluctuations in the CMB. At first glance, given the precision measurements of the CMB from {\it Planck}, this might appear to be even more constraining than the late-time probes of the expansion rate. Surprisingly, there are a few early-time resolutions which do not spoil the fit to current CMB temperature measurements (e.g., Refs.~\cite{Lin:2018nxe,Poulin:2018cxd,Kreisch:2019yzn,Lin:2019qug}). A model which can also provide a consistent fit to {\it Planck} CMB polarization measurements involves an anomalous increase in the expansion rate around matter/radiation equality due to some new component with perturbations that evolve as though they have a sound speed less than unity \cite{Lin:2018nxe,Poulin:2018cxd}. 

In this paper we explore the detailed phenomenology of one of these successful models, first proposed in Refs.~\cite{Karwal:2016vyq,Poulin:2018cxd}, which makes use of an oscillating scalar field playing the role of `early dark energy' (EDE).
Following previous work \cite{Poulin:2018dzj,Poulin:2018cxd}, we consider fields whose oscillations are {\em anharmonic} such that, once dynamical, they redshift faster than matter \cite{Turner:1983he}. The presence of this scalar field can increase the Hubble parameter for a limited amount of time. This, in turn, leads to a decrease in the acoustic sound horizon and the diffusion damping scale. Perturbations in the field have significant pressure support and therefore provide an additional non-collapsing source for the gravitational potentials, leading to distinct signatures in the CMB non-degenerate with those of $\Lambda$CDM parameters. 

Of particular interest, and in contrast to most past literature on this topic \cite{Poulin:2018dzj,Poulin:2018cxd}, we do not make any approximations and directly solve the linearized scalar field equations (as is also done in Ref.~\cite{Agrawal:2019lmo} for pure power-law potentials). We confirm that a frozen scalar field with up to $f_{\rm EDE}\equiv \rho_{\rm EDE}/\rho_{\rm tot}\sim10\%$ at a critical redshift $z_c\sim3500$ and diluting faster than matter afterwards can resolve the Hubble tension. The field becomes dynamical after the Hubble parameter drops below some critical value (determined by the effective mass of the field) and oscillates around its local minimum of its potential. Moreover, we show that solving for the full dynamics has striking consequences. 

We assume that the field initially is (almost) perfectly homogeneous and isotropic. This implies that whatever process established this scalar field had to have occurred well before the end of inflation. Such fields generically exhibit both `adiabatic' and `isocurvature' initial conditions. The adiabatic initial conditions arise due to the scalar field `falling' into the (adiabatic) gravitational potentials established during inflation. The isocurvature initial conditions arise due to fluctuations in the scalar field as a spectator during inflation. We show that for the potentials considered here, at large initial field displacements (favored by the data), the isocurvature initial conditions can be large, such that {\it Planck} data then place an upper limit on the amplitude of the isocurvature primordial power spectrum (which is identical to a limit on the tensor-to-scalar ratio). 

We also show that sub-dominant scalar fields following potentials $V\propto \phi^{2n}$ with $n\simeq 2$ around their minima experience significant `self-resonance' \cite{Lozanov:2017hjm}, where oscillations of the homogeneous field lead to resonant growth of perturbations in the scalar field. Such rapid growth can lead to a breakdown of perturbation theory (in the field), giving rise to spatially inhomogeneous dynamics. The analysis we present here is solely within the linear regime so that once the field becomes non-linear our analysis is no longer accurate. However, the presence of non-linear and highly inhomogeneous scalar field dynamics may provide unique observational signatures of this scenario which we plan to explore further in future work.

There have been criticisms of the SH0ES collaboration Cepheid calibration which, if valid, could bring the low and high redshift values into closer agreement \cite{Rigault:2014kaa,Rigault:2018ffm}; but subsequent analyses with larger SNe Ia samples have shown the reductions to be insignificant \cite{Jones:2018vbn,Rose:2019ncv}. Additionally, the recent measurement of $H_0$ from SNe Ia calibrated using the tip of the red giant branch method by the Chicago Carnegie Hubble Project (CCHP) sits right in between the early and late universe determination of the Hubble rate, with $H_0$=69.8 +/- 0.8 (stat) +/- 1.7 (sys) km/s/Mpc. However a recent re-analysis of the CCHP result quotes a value of $H_0 = 72.4 \pm 1.9$ \cite{Yuan:2019npk}. We also note that an inverse distance ladder combination of strong-lens time delays and (relatively) high-redshift supernovae yield $H_0 = 73-74$ km/s/Mpc \cite{Taubenberger:2019qna,Collett:2019hrr}. Future estimates of the Hubble constant using `gravitational wave sirens' may play a crucial role in determining the significance of the Hubble tension \cite{Schutz:1986gp,Holz:2005df,Abbott:2017xzu,Mortlock:2018azx}. 

Even without a clean, local, determination of $H_0$, any attempt to resolve the current Hubble tension leads to specific signatures in a variety of cosmological data. Detecting these signatures will therefore be essential to pin down the nature of the resolution to the Hubble tension. Here, we show that next-generation CMB experiments will be able to detect the presence of the EDE required to solve the Hubble tension at very high statistical significance, independently of SH0ES data, while {\em Planck} cannot. 

The results presented here are unexpected and novel since they demonstrate that current \textit{Planck} CMB measurements allow for a non-trivial amount ($\sim 10\%$) of the total energy density to consist of a cosmological scalar field around the time of matter/radiation equality. In this way, the use of the SH0ES prior on $H_0$ uncovers a set of degeneracies that were previously unrecognized. 

This paper is organized as follows. In Sec.~\ref{sec:cosmology}, we start by reviewing the cosmological evolution of a scalar field.   We then present the details of our MCMC analysis with current data in Sec.~\ref{sec:results}, and we show that a next generation CMB experiment can detect the proposed EDE at high statistical significance. In Sec.~\ref{sec:additional_obs}, we discuss two new signatures of an EDE. We show that an EDE naturally exhibits isocurvature modes that could spoil the success of the solution depending on the value of the scalar-to-tensor ratio $r$.  Furthermore, we show how the anharmonicity of the potential can lead to resonant growth of perturbations, and discuss the possibility of highly inhomogeneous, nonlinear dynamics of the scalar field. We conclude in Sec.~\ref{sec:conclusions}. We provide additional details of our numerical implementation, verification of our numerical code, discussion of parametric resonance in the EDE, and a detailed exploration of the $n=2$ model (i.e., massless scalar field) in the Appendix. 

\section{Cosmology of an oscillating scalar field}
\label{sec:cosmology}

We first reivew the background and linear dynamics of a cosmological scalar field and discuss our choice of potential.  

\subsection{Background dynamics}

The energy density and pressure of the scalar field affects the dynamics of other species through Einstein's equation. At the homogeneous and isotropic level, i.e., for the case of a Friedmann-Lema\^itre-Roberston-Walker metric, the expansion rate of the universe can be simply written as
\begin{equation}
    H=H_0 E(a) = H_0 \sqrt{\Omega_m(a) + \Omega_r(a) + \Omega_\Lambda + \Omega_\phi(a)},
\end{equation}
where $\Omega_X \equiv \rho_X/\rho_{\rm crit}$ and $\rho_{\rm crit} = 3H_0^2 M_P^2$, where $M_P \equiv (8\pi G)^{-1/2}$ is the reduced Planck mass. 
The energy-density and pressure of the scalar field at the homogeneous level is 
\begin{eqnarray}
    \rho_\phi &=& \frac{1}{2} \dot{\phi}^2 + V_n(\phi),\\
     P_\phi &=& \frac{1}{2} \dot{\phi}^2 - V_n(\phi),
\end{eqnarray}
where the dot indicates a derivative with respect to cosmic time. 
We consider a potential of the form
\begin{equation}\label{eq:potential}
    V_n(\phi) = m^2 f^2[1-\cos (\phi/f)]^n.
\end{equation} 
This functional form is inspired by ultra-light axions, fields that arise generically in string theory \cite{Arvanitaki:2009fg,Marsh:2015xka}.   The $n=1$ case is the well-established axion potential and the generalization to higher powers of $n$ has very interesting phenomenological consequences that we will develop, and may be generated by higher-order instanton corrections \cite{Kappl:2015esy}. We also note that potentials with power law minima and flattened ``wings" have been proposed and used in the context of inflationary physics as well as dark energy (see for example, Refs.~\cite{Dong:2010in,Kallosh:2013hoa,Carrasco:2015pla}).

Finally, to close the system of equations, one needs to solve the homogeneous Klein-Gordon (KG) equation of motion
\begin{equation}
   \ddot{\phi} + 3 H \dot{\phi} + V_{n,\phi}= 0,\label{eq:Vn}
\end{equation}
where the dot denotes a derivative with respect to cosmic time and $V_{n,\phi} \equiv dV_n/d\phi$. 

As already discussed in literature (e.g., Refs.~\cite{Griest:2002cu,Marsh:2010wq,Poulin:2018dzj}), the background dynamics of a cosmological scalar field can be described in the following way: at early times, Hubble friction dominates, such that the field is frozen at its initial value and its energy density is sub-dominant. It is only after the Hubble parameter drops below a critical value (which is related to the mass of the scalar field in the standard case), that the field starts evolving towards the minimum of the potential. In the case we study here, the fields then oscillates at the bottom of its potential, leading to a dilution of its energy density with an equation of state which depends on $n$ \cite{Turner:1983he}.
We modified the Einstein-Boltzmann code {\sf CLASS} \cite{Lesgourgues:2011re,Blas:2011rf} and implemented the potential given by Eq.~(\ref{eq:potential}). Details on the implementation, in particular regarding the numerical optimization, are given in Appendix~\ref{app:numerical}.

It is useful to define a re-normalized field variable, $\Theta \equiv \phi/f$, so that $-\pi \leq \Theta \leq \pi$. The KG equation can then be written 
\begin{equation}
    \ddot \Theta + 3 H \dot \Theta + \frac{1}{f^2} V_{n,\phi} = 0.
\end{equation}
Since the field always starts in slow-roll the background dynamics are specified by three parameters: $m$, $f$, and $\Theta_i$ (the initial field value in units of $f$), where without loss of generality we restrict $0\leq \Theta_i \leq \pi$.
\begin{figure*}
    \includegraphics[scale=0.45]{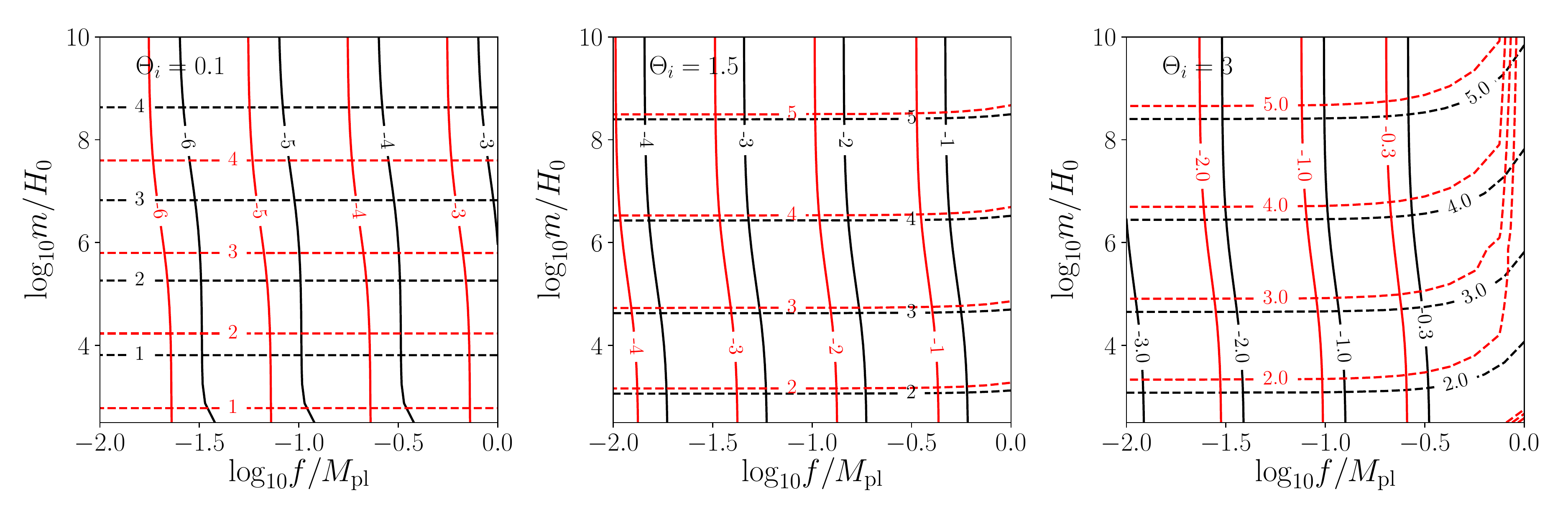}
    \caption{Contours of constant ${\rm log}_{10}f_{\rm EDE}(z_c)$ (vertical/solid) and ${\rm log}_{10}z_c$ (horizontal/dashed) as a function of the axion mass, $m$, and decay constant, $f$. The red lines show the contours for $n=2$ and the black for $n=3$. Since $H_0 = 100 h\ {\rm km/s/Mpc} = 2.13h \times 10^{-33}\ {\rm eV}$ the mass parameter of the potential that helps to resolve the Hubble tension ranges between $10^{-28}\ {\rm eV} \lesssim m \lesssim 10^{-26} \ {\rm eV}$ and $0.01 \lesssim f/M_{\rm pl}\lesssim 1$. \label{fig:contours}}
\end{figure*}

The observable consequences of the scalar field can be characterized by the maximum fraction of the total energy density in this field, $f_{\rm EDE}(z_c)$, and the redshift at which the energy density reaches this maximum, $z_c$. As shown in Fig.~\ref{fig:contours}, for any $\Theta_i$ we can always find a value of $m$ and $f$ which generates any given $\{f_{\rm EDE}(z_c),z_c\}$. There we can see that $m$ largely controls the value of $z_c$, while $f$ controls that of $f_{\rm EDE}(z_c)$. 

We can derive approximate equations to relate $m$ to $z_c$ and $f$ to $f_{\rm EDE}(z_c)$. Previous work on the dynamics of axions, which follow from the potential considered here with $n=1$, showed that in this case the field becomes dynamical around $m \simeq 3 H(z_c)$ \cite{Marsh:2010wq}. This approximate relation extends to more general potentials with $m \rightarrow |V_{n,\phi\phi}|$ so that 
\begin{equation}
    m^2 n \bigg|\left(1-\cos \Theta_i\right)^{n-1}\left(n-1+n\cos \Theta_i\right)\bigg| \simeq 9 H^2(z_c) \label{eq:meq},
\end{equation}
showing that for a fixed $\Theta_i$ a value of $m$ determines $z_c$.
Since the field only starts to become dynamical at $z_c$, the fraction of the total energy density in the field at $z_c$ is approximately given by 
\begin{equation}
    f_{\rm EDE}(z_c) \simeq \frac{V_n(\Theta_i)}{\rho_{\rm tot}(z_c)} = \frac{m^2f^2}{\rho_{\rm tot}(z_c)} (1-\cos\Theta_i)^n. 
\end{equation}
Eq.~(\ref{eq:meq}) shows $m^2 \propto \rho_{\rm tot}(z_c)$ which implies that $f_{\rm EDE}(z_c)$ is determined by $f$, $n$, and $\Theta_i$.
Additionally, the rate at which the field dilutes, i.e., the equation of state once the field oscillates, is simply set by $n$ through $w_\phi\equiv (n-1)/(n+1)$ \cite{Turner:1983he}.  

\begin{figure}
    \includegraphics[scale=0.56]{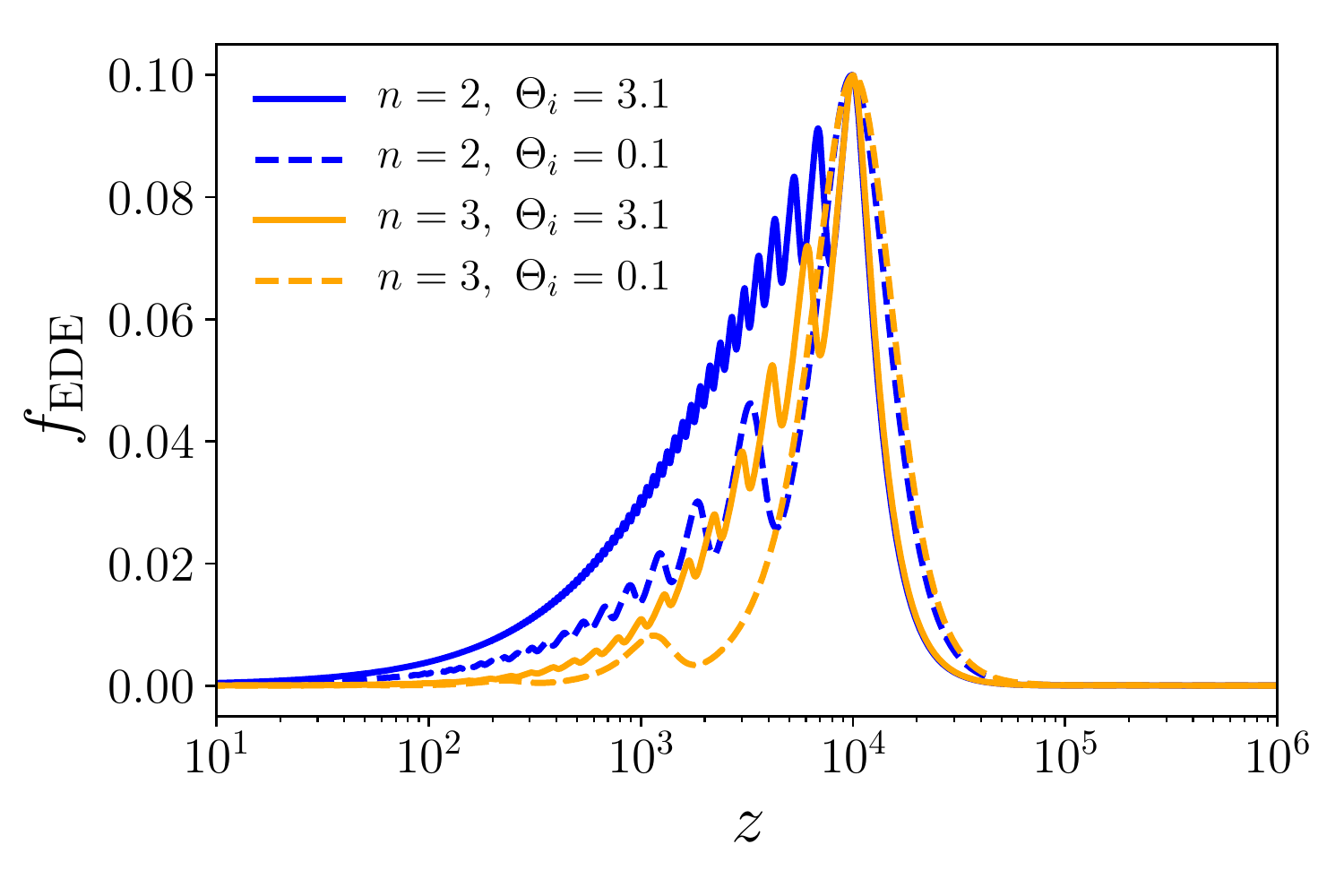}
    \caption{The evolution of the fraction of the total energy density in the EDE as a function of redshift for $z_c = 10^4$ and $f_{\rm EDE}(z_c) = 0.1$. Note that as the initial field displacement becomes larger the asymmetry of $f_{\rm EDE}(z)$ and oscillation frequency of the background field increases.}
    \label{fig:background_evo}
\end{figure}

The role of $\Theta_i$ is a little more subtle. As first discussed in Ref.~\cite{Poulin:2018dzj}, once we have fixed $n$, $z_c$ and $f_{\rm EDE}({z_c})$, the value of $\Theta_i$ controls the oscillation frequency of the background field and in turn, the effective sound speed of the perturbations. The change in the background oscillation frequency is clearly visible in Figure \ref{fig:background_evo}, where we plot the evolution of $f_{\rm EDE}$ with $z$ for various $n$ and $\Theta_i$, in a model where $f_{\rm EDE}(z_c=10^4)=0.1$. Note also that, at the background level, $\Theta_i$ has a suble impact on the redshift-asymmetry of the energy injection.

Finally we note that if the potential becomes too steep around its minimum then it is possible for the field to reach an attractor solution in which it will never oscillate. As discussed in Refs.~\cite{Ratra:1987rm,Liddle:1998xm} if $n>5$ during radiation domination or $n>3$ during matter domination there exists a power-law attractor for $\phi \propto t^{-\alpha}$ where $\alpha=2/(2n-2)$. Given that the resolution to the Hubble tension using a canonical scalar field requires oscillations (to make the effective sound speed smaller than one \cite{Lin:2019qug}), we expect $n>5$ to be disfavored by the data. As we discuss in Sec.~\ref{sec:results}, this is indeed what we find. 

\subsection{Linear Perturbations}
\label{sec:perturb}

Most previous work on the cosmological implications of scalar fields used an approximate set of fluid equations to evolve the scalar field perturbations \cite{Poulin:2018dzj,Poulin:2018cxd}. Once the field starts to oscillate we can average over the oscillations of the background field to produce a set of approximate `cycle-averaged' fluid equations with an effective sound speed in the field's local rest-frame, $c_s^2\equiv \langle \delta P_\phi\rangle/\langle \delta \rho_\phi \rangle$, which is both scale and time-dependent \cite{Hu:1998kj}. Here we do not make this approximation and instead solve the exact (linearized) KG equation,
\begin{equation}
    \delta\phi''_k + 2 H \delta \phi_k' + \left[k^2 + a^2 V_{n,\phi\phi}\right] \delta \phi_k = - h'  \phi'/2,\label{eq:linKG}
\end{equation}
where the prime denotes derivatives with respect to conformal time, we have written the metric potential in synchronous gauge (see, e.g., Ref.~\cite{Ma:1995ey}), and we can see that the perturbations evolve as a driven damped harmonic oscillator. 

The effective angular frequency, $\omega_{\rm eff}= \sqrt{k^2 + a^2 V_{n,\phi\phi}}$, is time-dependent. This frequency may be (for a limited amount of time) imaginary when $V_{n,\phi \phi}<0$ (i.e., `tachyonic') which may lead to exponential growth. We find that this growth only occurs if the homogeneous (undriven) solution is excited, which corresponds to scalar field isocurvature perturbations. As we discuss in detail in Sec.~\ref{sec:additional_obs}, isocurvature perturbations are generic but unimportant as long as the tensor-to-scalar ratio, $r \lesssim 5 \times 10^{-3}$. Since we do not incorporate isocurvature perturbations when constraining the EDE parameters, for the following Section we implicitly take $r \lesssim 5 \times 10^{-3}$. 

The time-dependence in $\omega_{\rm eff}$ (even without expansion) occurs when the potential is anharmonic (i.e., when $n>1$) -- arising from the oscillations of the background field. This can lead to the phenomenon of self-resonance, where the oscillating background field pumps energy into its perturbations in a scale-dependent manner. This transfer of energy can lead to an exponential growth of perturbations for $n\approx 2$ leading to the formation of non-linear scalar field perturbations. Since we are only solving linear equations, our analysis in the following Section is restricted to $n>2$, though includes linear resonant effects when they are present. We explore the $n\simeq 2$ case in more detail in Sec.~\ref{sec:res}.

\begin{table*}[!t]
  \begin{tabular}{|l|c|c|c|}
     \hline\hline Parameter &~~$\Lambda$CDM~~&~~~$n=3$~~~&~~~ $n$ free~~~\\ \hline \hline
    $H_0$ &$68.37~(68.21)\pm0.54$  & $71.49~(72.19)\pm1.20$ &$71.45~(72.81)_{1.40}^{1.10}$\\
    $100~\omega_b$ & $2.242~(2.253)\pm0.015$& $2.260~(2.253)\pm0.025$ & $2.261~(2.251)\pm0.024$\\
    $\omega_{\rm cdm}$& $0.1175~(0.1177)\pm0.0012$& $0.1295~(0.1306)_{-0.0043}^{+0.0039}$ &  $0.1290~(0.1320)_{-0.0045}^{+0.0041}$\\
    $10^{9}A_s$& $2.187~(2.216)\pm0.052$& $2.193~(2.215)\pm0.054$  & $2.196~(2.191)\pm0.055$ \\
    $n_s$&$0.9696~(0.9686)\pm0.0043$& $0.9863~(0.9889)\pm0.0078$ & $0.9853~(0.9860)_{-0.0079}^{+0.0073}$\\
    $\tau_{\rm reio}$& $0.078~(0.085)\pm0.013$& $0.069~(0.072)\pm0.014$ & $0.070~(0.068)\pm0.014$\\
    ${\rm Log}_{10}(z_c)$& $-$& $3.568~(3.562)_{-0.140}^{+0.056}$ & $3.558~(3.531)_{-0.110}^{+0.053}$\\
    $f_{\rm EDE}(z_c)$ & $-$& $0.107~(0.122)_{-0.030}^{+0.035}$ & $0.103~(0.132)\pm0.035$\\
    $\Theta_i$ &$-$ & $2.64~(2.83)_{-0.04}^{+0.36}$&$2.49~(2.72)_{-0.01}^{+0.52}$ \\
    $n$ &$-$ & $3$ (fixed) & $3.16~(2.60)_{-1.16}^{+0.18}$ \\
    \hline
    $100~\theta_s$ & $1.04202~(1.04215)\pm 0.0003$& $1.04138~(1.04152)\pm+0.00039$& $1.04139~(1.04106)_{-0.00036}^{+0.00041}$\\
    $r_s(z_{\rm rec})$& $145.15~(145.3)\pm0.27$& $139.1~(138.5)\pm1.9$ & $139.3~(137.7)_{-1.8}^{+2.1}$ \\
    $S_8$ & $0.820~(0.830)\pm0.012$& $0.842~(0.843)\pm0.014$ & $0.840~(0.832)\pm0.015$\\
    \hline
  \end{tabular}
  \caption{The mean (best-fit) $\pm1\sigma$ error of the cosmological parameters reconstructed from our combined analysis including high-$\ell$ (i.e., $\ell \geq 30$) polarization data in each model.}
  \label{table:param_values}
\end{table*}

\section{Implications for the Hubble tension}

In this Section we explore the resolution of the Hubble tension provided by the EDE using a variety of cosmological observations. The results presented here confirm the conclusions reached in Ref.~\cite{Poulin:2018cxd} where an approximate, `cycle-averaged', form of the scalar field evolution equations was used. Here we use full homogeneous and linear scalar field dynamics along with i) promoting the exponent of the potential to a free parameter, and show explicitly that the best-fit exponent is close to $n=3$, as previous results hinted at \cite{Poulin:2018cxd}; ii) for the $n=3$ case we compare the use of low-$\ell$ TEB ($\ell <30$) and high-$\ell$ TT ($\ell \geq 30$) data to the full {\em Planck} temperature and polarization measurements and show that the high-$\ell$ polarization data prefers a large initial scalar field displacement; iii) we compare the use of a pure power-law potential \cite{Agrawal:2019lmo} to the full cosine (i.e., the small $\Theta_i$ limit) and explain why the pure power-laws are disfavored by the data; iv) we perform a forecast for CMB-S4 in order to demonstrate that a CMB-only detection of the EDE cosmology is possible in the near future.

\label{sec:results}
\subsection{Analysis method}

We run a Markov-chain Monte Carlo (MCMC) using the public code {\sf MontePython-v3}\footnote{\url{https://github.com/brinckmann/montepython_public}} \citep{Audren:2012wb,Brinckmann:2018cvx}, interfaced with our modified version of {\sf CLASS}.
We perform the analysis with a Metropolis-Hasting algorithm, assuming flat priors on $\{\omega_b,\omega_{\rm cdm},\theta_s,A_s,n_s,\tau_{\rm reio},\log_{10}(z_c),f_{\rm EDE}(z_c) ,\Theta_i\}$ and allow $n$ free to vary or set $n=3$ (which is close to its best-fit value). As described in Appendix \ref{app:numerical}, we use a shooting method to map a choice of $\{\log_{10}(z_c), f_{\rm EDE}\}$ to the theory parameters $\{m,f\}$. 
We adopt the {\em Planck} collaboration convention and model free-streaming neutrinos as two massless species and one massive with $M_\nu=0.06$ eV \cite{Ade:2018sbj}. 
Unless specified otherwise, our data set includes {\em Planck} 2015 high-$\ell$ and low-$\ell$ TT,TE,EE and lensing likelihood \cite{Aghanim:2015xee}\footnote{As this work was close to completion, a new version of {\em Planck} likelihoods were released. We have checked that in a baseline $n=3$ run our results are unaffected.}; the latest SH0ES measurement of the present-day Hubble rate $H_0=74.03\pm1.42$ km/s/Mpc~\cite{Riess:2019cxk}; the isotropic BAO measurements from 6dFGS at $z = 0.106$~\cite{Beutler:2011hx} and from the MGS galaxy sample of SDSS at $z = 0.15$~\cite{Ross:2014qpa}; the anisotropic BAO and the growth function $f\sigma_8(z)$ measurements from the CMASS and LOWZ galaxy samples of BOSS DR12 at $z = 0.38$, $0.51$, and $0.61$~\cite{Alam:2016hwk}. Additionally, we use the Pantheon\footnote{\url{https://github.com/dscolnic/Pantheon}} supernovae dataset \cite{Scolnic:2017caz}, which includes measurements of the luminosity distances of 1048 SNe Ia in the redshift range $0.01 < z < 2.3$. As usual, we use a Choleski decomposition \citep{Lewis:2013hha} to deal with the numerous nuisance parameters associated with the likelihoods (not recalled here for brevity). We consider chains to be converged using the Gelman-Rubin \citep{Gelman:1992zz} criterion $R -1<0.1$. 

\subsection{Extracting the best-fit exponent}

\begin{figure*}
\centering
\includegraphics[scale=0.38]{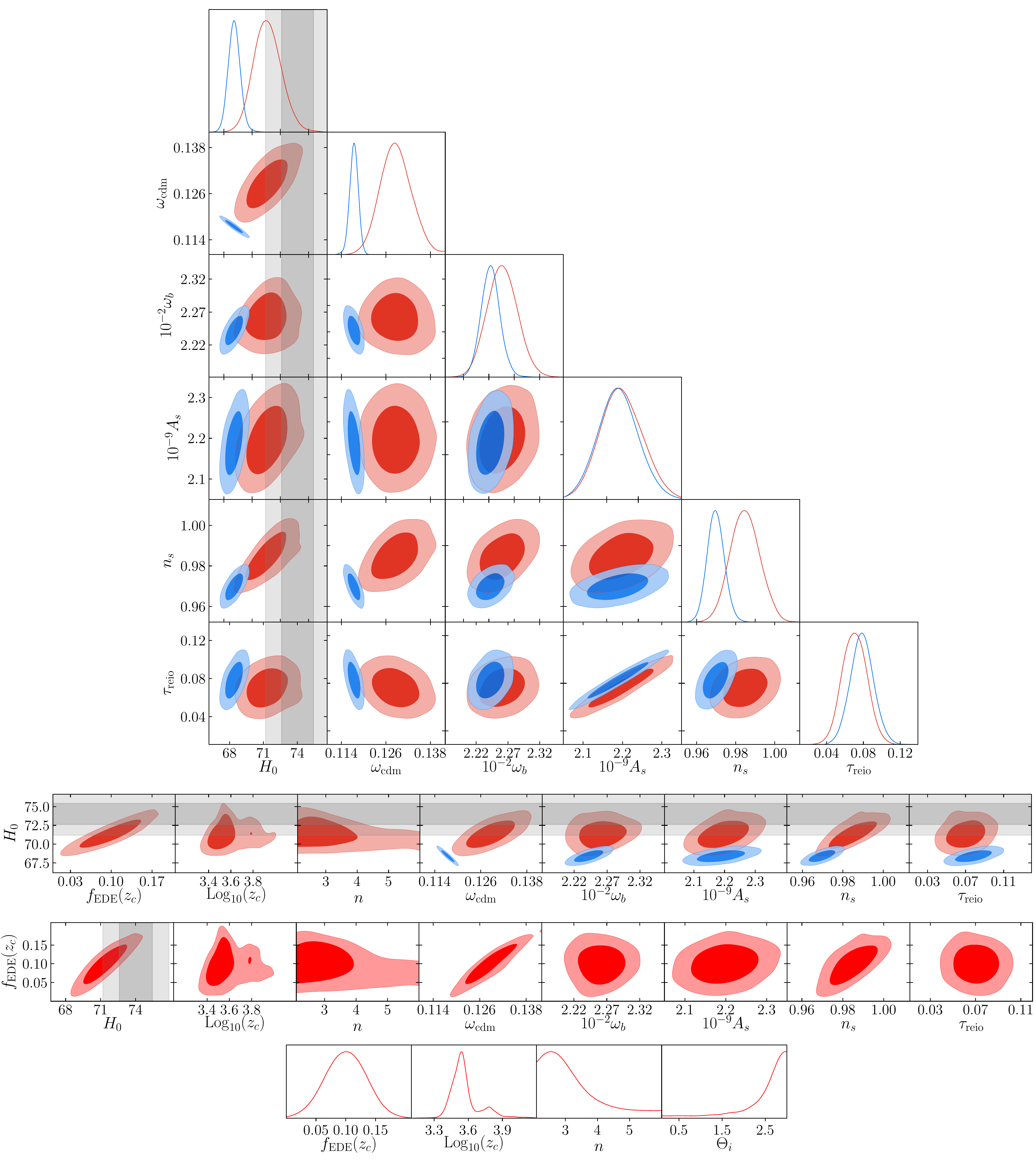}
\caption{Posterior distributions of the cosmological parameters reconstructed from a run to all data (including {\em Planck} high-$\ell$ polarization) in the $\Lambda$CDM (blue) and EDE (red) cosmology. From top to bottom we show: the $\Lambda$CDM parameters, 2D distributions of $H_0$ and $f_{\rm EDE}(z_c)$ vs a subset of parameters, the 1D posterior distribution of the EDE parameters. We show the SH0ES determination of $H_0$ in the gray bands.}\label{fig:nfree-vs-lcdm}
\end{figure*}

In the first analysis we perform we let the exponent $n$ of the potential vary freely with a flat prior, $2<n<6$. We leave out the region $n\in[1,2]$, for which the number of oscillations per Hubble time makes the computation time much longer and is not tractable in a MCMC analysis\footnote{Barring the effect of self-resonance discussed later, we anticipate that a fluid approximation following Ref.~\cite{Poulin:2018dzj} might be accurate in this regime and plan to address this part of parameter space in a future study.}.
We report the reconstructed parameters in Table~\ref{table:param_values} and the corresponding $\chi^2_{\rm min}$ in Table~\ref{table:chi2_preliminary}. We plot the reconstructed posterior distributions in $\Lambda$CDM and in the EDE cosmology in Fig.~\ref{fig:nfree-vs-lcdm}.

These constraints tell a very interesting story. First, they confirm the conclusions of Ref.~\cite{Poulin:2018cxd}: namely that an oscillating EDE scalar field which becomes dynamical around matter/radiation equality provides a good fit to both the CMB and the SH0ES determination of the Hubble constant. Because of the slight increase in the most recent best-fit SH0ES value of $H_0$ and the decreased uncertainty we now see evidence for the EDE at $>3 \sigma$ [$f_{\rm EDE}(z_c) \simeq 0.1 \pm 0.03$]. Additionally, our analysis yields a marginalized constraint of $n=3.16^{+0.18}_{-1.16}$ showing that a range of power-law indices can lead to dynamics which resolves the Hubble tension, but favors values of $n$ close to 3 as was found in Ref.~\cite{Poulin:2018cxd} for discrete values of $n$.
Second, it is striking that the $\Delta\chi^2_{\rm min}\!=\!-20.3$ when including the new value of SH0ES has increased {\em without} spoiling {\em Planck} data. In fact, as shown in the first two rows of Table \ref{table:chi2_preliminary}, we find that the fit to {\em Planck} data is improved with respect to that of $\Lambda$CDM {\em fit on Planck data only} by $-4$. This is far from statistically significant, but encouraging and would deserve more attention in future work in order to understand more precisely where the improvement comes from. 
 
As far as the $\chi^2_{\rm min}$ for individual likelihoods are concerned, both high-$\ell$ and small-$\ell$ data are slightly improved. The smallness of the improvement in the fit explains why {\em Planck} data alone do not allow to detect the EDE independently from SH0ES. This is related to an issue of sampling volume when $f_{\rm EDE}(z_c)>0$ as opposed to when $f_{\rm EDE}(z_c)=0$.  Indeed, with {\em Planck} data only,  when $f_{\rm EDE}(z_c) = 0$ any value in the $(z_c,\Theta_i)$ parameter space is identical to a $\Lambda$CDM model. On the contrary, when $f_{\rm EDE}(z_c)>0$  only a small region of the $(z_c,\Theta_i)$ parameter space provides a good fit to the {\em Planck} data. It seems plausible that the Metropolis-Hasting algorithm does not sufficiently explore such a small parameter volume and instead spends most of its time close to $\Lambda$CDM-like models when only {\em Planck} data are included. We therefore only present results that include the SH0ES likelihood. In Sec.~\ref{sec:future}, we show that this behavior also appears in mock data that includes an EDE signal.

The other contours in Fig.~\ref{fig:nfree-vs-lcdm} show shifts and degeneracies that are similar to previous analyses \cite{Poulin:2018cxd,Lin:2019qug}. In particular we can see that in the EDE scenario the presence of extra energy density around matter-radiation equality leads to an increase in the preferred value of the CDM physical energy density $\omega_{\rm CDM}$ and the scalar spectral index $n_s$. We can also see that the posterior for the EDE critical redshift, $z_c$, is slightly bimodal and correlated with $\Theta_i$. As shown in Fig.~\ref{fig:TT-vs-TTTEEE}, this bimodality is driven by the high-$\ell$ polarization data and is also present when we analyze synthetic data in Sec.~\ref{sec:future}. We plan to explore what properties of the polarization power spectra drives this curious feature of the posterior distribution in future work. 

\begin{table}[th]
\scalebox{0.9}{
  \begin{tabular}{|l|c|c|c|}
    \hline\hline
    Datasets &~~$\Lambda$CDM~~&~~~$n=3$~~~&~~~ $n$ free~~~\\ \hline \hline
    \textit{Planck} high-$\ell$ TT, TE, EE &  2446.66  &  2444&2445.53  \\
    \textit{Planck} low-$\ell$ TT, TE, EE &10496.65 & 10493.25&10493.65\\
    \textit{Planck} lensing& 10.37& 10.24&  9.14\\
    BAO-low $z$& 1.86& 2.53&2.77\\
    BAO-high $z$&1.84 &2.1 & 2.12 \\
    Pantheon &1027.04 &1027.11 &1026.96\\
    SH0ES & 16.80 & 1.68 & 0.73 \\
    \hline
    Total $\chi^2_\mathrm{min}$   & 14001.23 & 13980.94&13980.90\\
    $\Delta \chi^2_\mathrm{min}$ & 0 & -20.29 & -20.33\\ 
    \hline
  \end{tabular}
  }
  \caption{The best-fit $\chi^2$ per experiment for the standard $\Lambda$CDM model and the EDE cosmologies, with high-$\ell$ polarization data. The BAO-low $z$ and high $z$ datasets correspond to $z\sim 0.1-0.15 $ and $z \sim 0.4-0.6$, respectively.  For comparison, using the same {\sf CLASS} precision parameters and {\sf MontePython}, a $\Lambda$CDM fit to {\it Planck} data only yields $\chi^2_{{\rm high}-\ell}\simeq2446.2$, $\chi^2_{{\rm low}-\ell}\simeq10495.9$ and $\chi^2_{{\rm lensing}}\simeq9.4$ with $R-1<0.008$.}
  \label{table:chi2_preliminary}
\end{table}

\begin{table}
\scalebox{0.9}{
  \begin{tabular}{|l|c|c|}
    \hline\hline Parameter &~~$\Lambda$CDM~~&~~~$n=3$~~~\\ \hline \hline
    $H_0$ &$68.99~(68.87)\pm0.69$  & $71.82~(72.43)\pm1.2$ \\
    $100~\omega_b$ & $2.248~(2.245)\pm0.020$&  $2.248~(2.225)\pm0.041$\\
    $\omega_{\rm cdm}$& $0.1162~(0.1165)\pm0.0015$&$0.1304~(0.1328)\pm0.0061$ \\
    $10^{9}A_s$& $3.095~(3.097)_{-0.028}^{+0.025}$&$2.187~(2.174)\pm0.062$  \\
    $n_s$&$0.9733~(0.9723)\pm0.0052$&$0.9861~(0.9936)\pm0.0095$  \\
    $\tau_{\rm reio}$& $0.085~(0.085)\pm0.015$&$0.066~(0.063)\pm0.017$\\
    ${\rm Log}_{10}(z_c)$& $-$&$3.50~(3.62)_{-0.09}^{+0.15}$ \\
    $f_{\rm EDE}(z_c)$ & $-$&$0.108~(0.138)_{-0.044}^{+0.036}$ \\
    $\Theta_i$ &$-$ &$~(2.81)$  \\
    \hline
    $100~\theta_s$ & $1.04230~(1.04231)\pm 0.00042$&$1.04138~(1.04121)_\pm0.00054$\\
    $r_s(z_{\rm rec})$& $145.43~(145.39)\pm0.36$&$138.74~(137.47)\pm2.5$  \\
    $S_8$ & $0.811~(0.813)\pm0.014$&$0.842~(0.843)\pm0.019$  \\
    \hline
  \end{tabular}
  }
  \caption{The mean (best-fit) $\pm1\sigma$ error of the cosmological parameters reconstructed from our combined analysis without high-$\ell$ polarization data in each model.}
  \label{table:param_TT}
\end{table}
\begin{table}[th]
\scalebox{0.9}{
\begin{tabular}{|l|c|c|}
    \hline\hline
    Datasets &~~$\Lambda$CDM~~&~~~$n=3$~~~\\ \hline \hline
    \textit{Planck} high-$\ell$ TT & 770.03   & 770.12  \\
    \textit{Planck} low-$\ell$ TT, TE, EE & 10495.74&10492.43\\
    \textit{Planck} lensing& 9.27& 9.60\\
    BAO-low $z$& 2.7&2.19 \\
    BAO-high $z$&  2&2 \\
    Pantheon & 1027.13&1027.01\\
    SH0ES &13.22 &1.26 \\
    \hline
    Total $\chi^2_\mathrm{min}$   & 12320.09&12304.61\\
    $\Delta \chi^2_\mathrm{min}$ &0 &-15.48\\ 
    \hline
  \end{tabular}}
  \caption{The best-fit $\chi^2$ per experiment for the standard $\Lambda$CDM model and the EDE cosmologies, without high-$\ell$ polarization data. For comparison, using the same {\sf CLASS} precision parameters and {\sf MontePython}, a $\Lambda$CDM fit to {\it Planck} data only yields $\chi^2_{{\rm high}-\ell}\simeq2446.2$, $\chi^2_{{\rm low}-\ell}\simeq10495.9$ and $\chi^2_{{\rm lensing}}\simeq9.4$ with $R-1<0.008$.}
  \label{table:chi2_TT}
\end{table}

\subsection{A deeper analysis of the $n=3$ case}
\label{sec:n3}

We now turn to studying in more depth the case of the best-fit exponent, which is roughly $n=3$.

\subsubsection{Temperature-vs-polarization data}
\label{sec:tempvspol}

Relative to several previous attempts at resolving the Hubble tension the EDE scenario presented here is not degraded when we add the small-scale {\em Planck} polarization measurements. Instead, the small-scale polarization measurements place a tight constraint on the initial field displacement, $\Theta_i$. Here we explore this in detail by focusing on the $n=3$ EDE model. 

We start by comparing {\em Planck} high-$\ell$ temperature + low-$\ell$ TEB data (which we denote by `TT') to the full  {\em Planck} dataset (which we denote by `TT,TT,EE'). We show the 2D posterior distributions of $f_{\rm EDE}(z_c)$ against $\{{\rm log}_{10}(z_c),\Theta_i,H_0,\omega_{\rm cdm}\}$ as they exhibit the most interesting degeneracies. We report the reconstructed parameters with TT data in Table~\ref{table:param_TT} and the corresponding $\chi^2_{\rm min}$ in Table~\ref{table:chi2_TT}. The results with TT,TE,EE data are reported in Tables~\ref{table:param_values} and \ref{table:chi2_preliminary}.  

\begin{figure*}[t!]
    \includegraphics[scale=0.65]{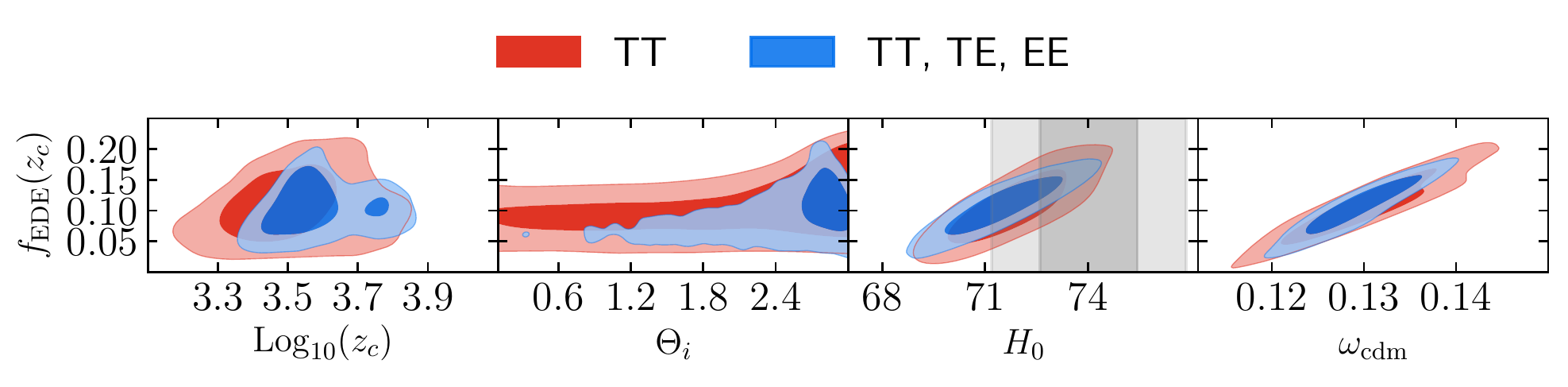}\\
    \caption{2D posterior distribution of a subset of parameters in the $n=3$ case. We compare the results with and without high-$\ell$ TT,TE,EE data.}
    \label{fig:TT-vs-TTTEEE}
\end{figure*}

\begin{figure}[t]
 \includegraphics[scale=0.55]{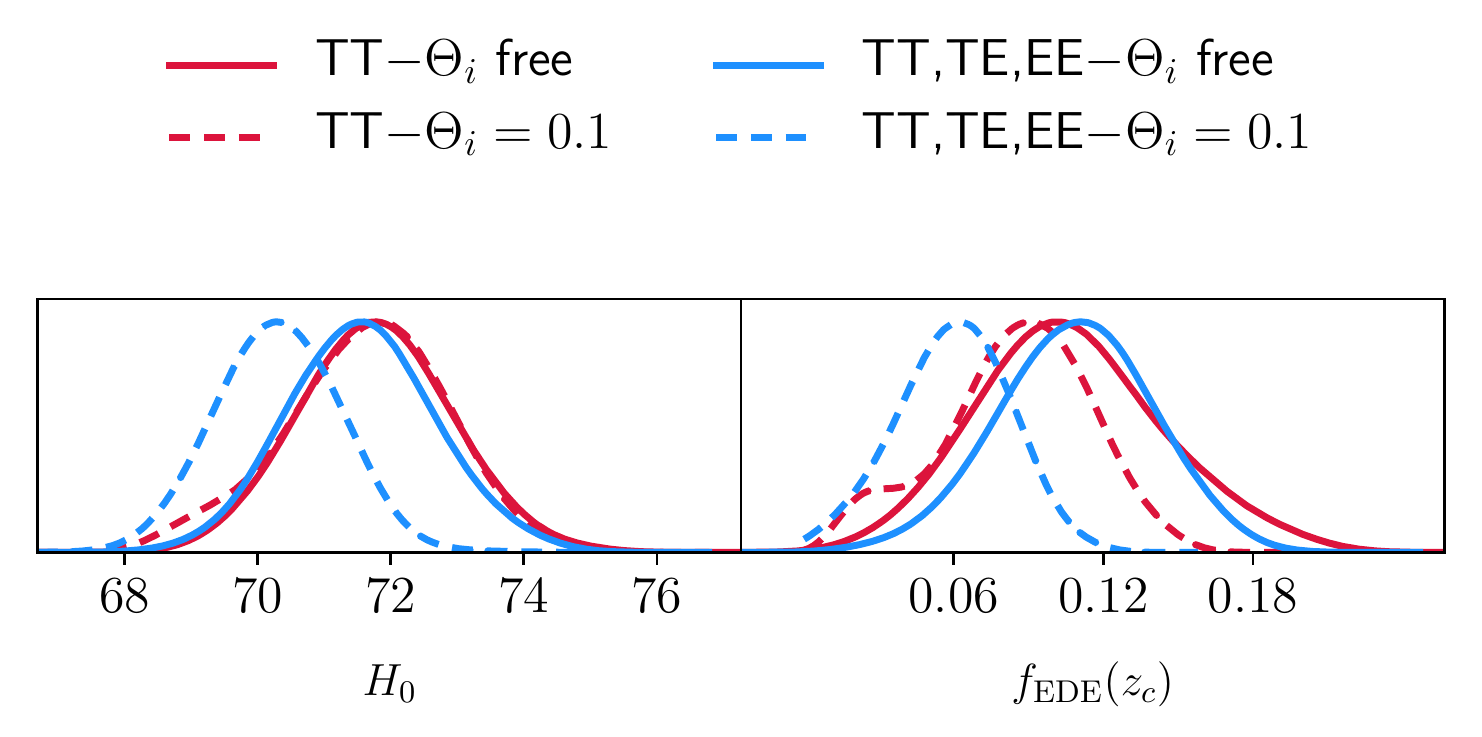}
    \caption{Reconstructed 1D posterior of $H_0$ and $f_{\rm EDE}(z_c)$. We compare the results with (blue) and without (red) high-$\ell$ TT,TE,EE data, as well as keeping $\Theta_i$ free (full lines) and enforcing $\Theta_i=0.1$, i.e., the power-law case (dashed lines).}
    \label{fig:small_TT}
\end{figure}

\begin{figure*}[t!]
    \includegraphics[scale=0.75]{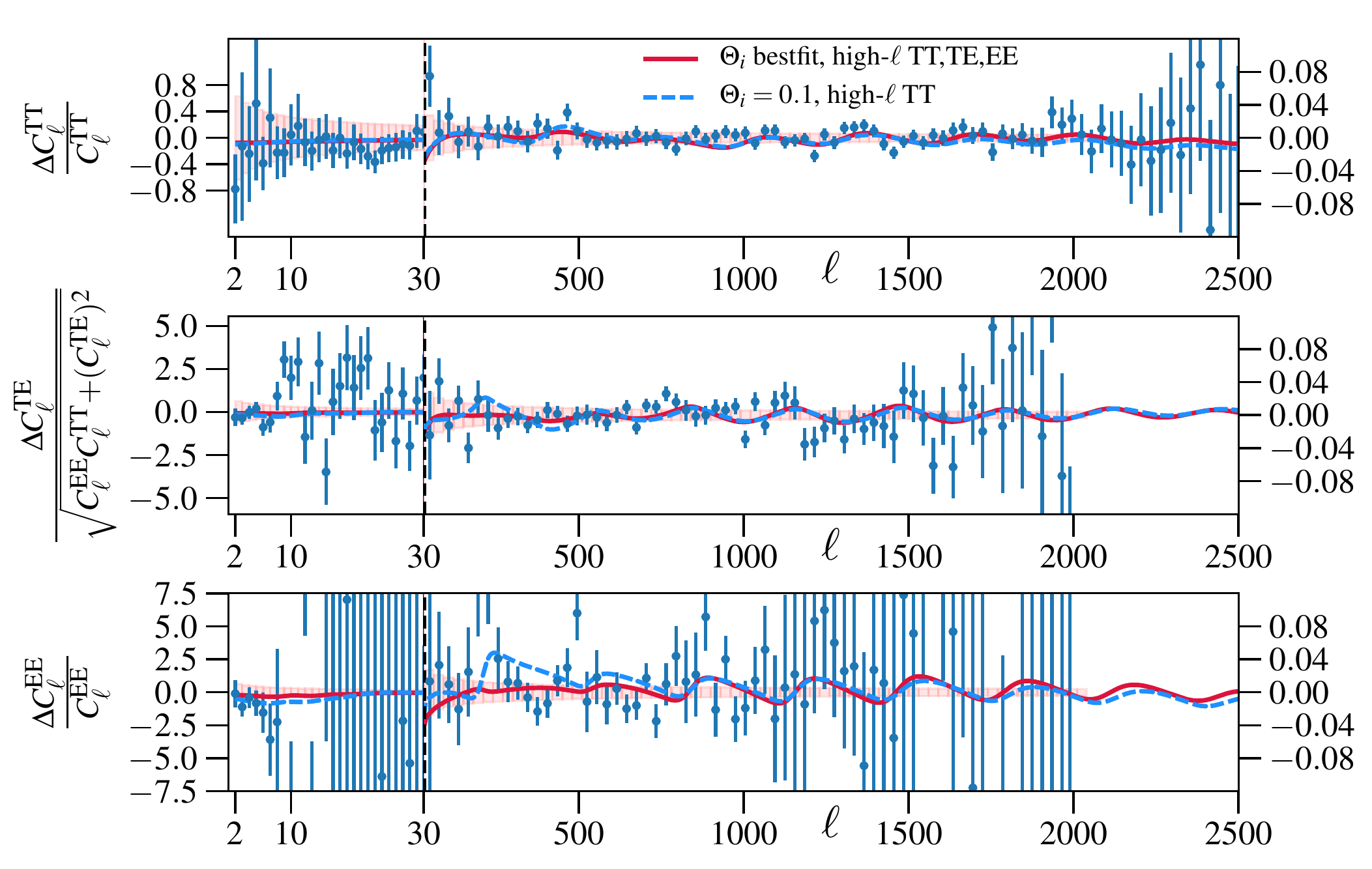}
    \caption{Power spectrum residuals between the best-fit $\Lambda$CDM and various best-fit EDE cosmologies with $n=3$. We compare the results without high-$\ell$ polarization data and enforcing $\Theta_i=0.1$ (blue dashed curves) to those obtained when including these data and letting $\Theta_i$ free to vary.}
    \label{fig:temp2} 
\end{figure*}

\begin{figure}[h!]
    \includegraphics[scale=0.85]{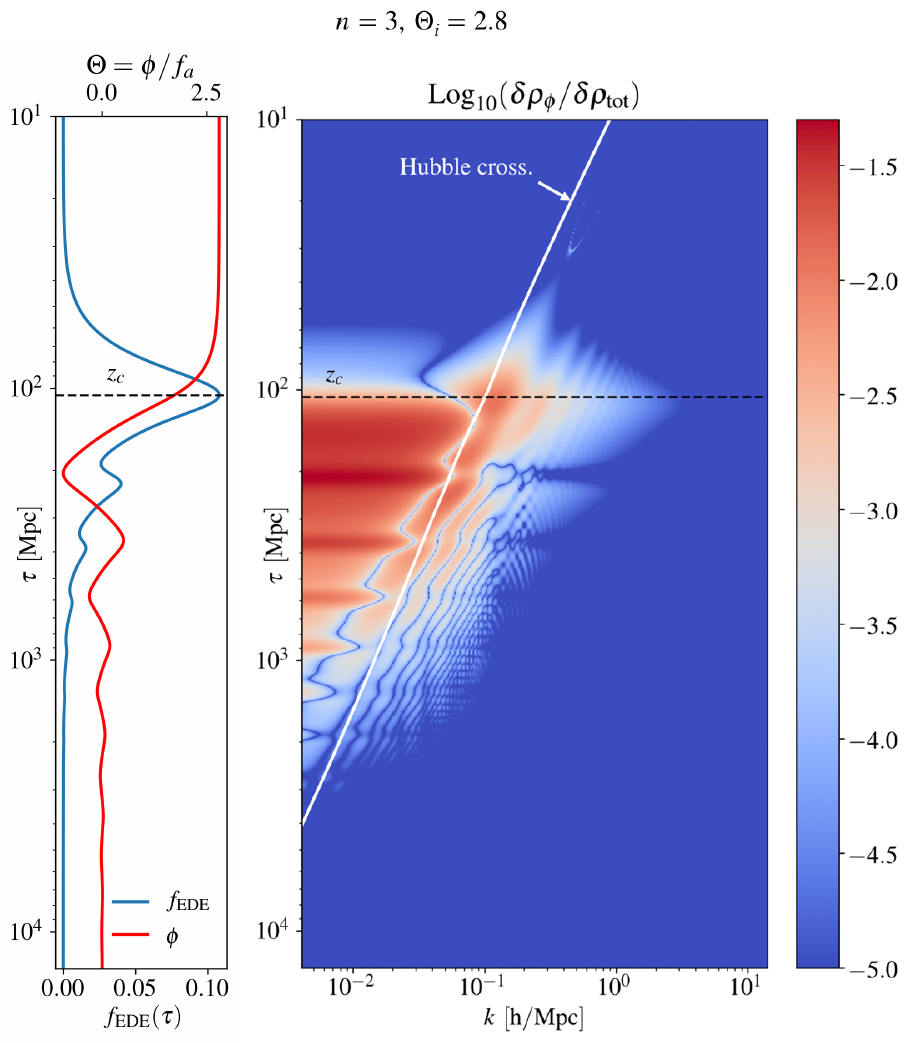}
    \caption{The fraction of the total energy density in the EDE (blue), the field evolution (solid-red), and the evolution of the field envelope (dashed-red),as a function of conformal time. \emph{Right}: The fraction of the total density perturbations in the EDE as a function of wavenumber and conformal time. Note that only a limited range of $\tau$ and sub-horizon $k$ (below the white line labeled `Hubble cross.') have a significant contribution from the EDE. This implies that the EDE effects in the CMB are localized in multipoles.}
    \label{fig:awesome}
\end{figure}

The addition of high-$\ell$ polarization data primarily places a constraint on the initial field displacement, $\Theta_i$, and does not lead to an increase in the Hubble tension-- see  Fig.~\ref{fig:TT-vs-TTTEEE}. It is interesting to see that polarization data forbids small values of $\Theta_i$, excluding the region $\Theta_i<1.8$ at 95\% C.L., and we shall now explore this in more detail (see also Ref.~\cite{Lin:2019qug}).

To explore how the addition of polarization data impacts the constraints to $\Theta_i$, we perform runs enforcing $\Theta_i = 0.1$ with and without high-$\ell$ polarization data. We compare the reconstructed 1D posterior of $H_0$ and $f_{\rm EDE}(z_c)$ to the ones obtained when letting $\Theta_i$ free to vary in a temperature-only analysis in  Fig.~\ref{fig:small_TT}. It is clear that, except for the case that includes polarization and enforces $\Theta_i=0.1$ (dashed blue lines), the allowed region of parameter space significantly overlap.  

The preference for large $\Theta_i$ when high-$\ell$ polarization data are included can be better understood by considering the residual between the best-fit $\Lambda$CDM model and the EDE models with $\Theta_i = 0.1$ fit to the TT data, as shown in the blue-dashed line in Fig.~\ref{fig:temp2}; in the solid-red line we show the best-fit EDE models using the full dataset where $\Theta_i = 2.72$. Given where these residuals differ the most, we can see that the large $\Theta_i$ preference comes from a pattern in the residuals of the TE and EE spectra in the multipole range $\ell\sim30-500$ that is disfavored by the data. This range of multipoles roughly corresponds to the modes that enter the horizon while the EDE contributes a significant fraction of the total density perturbation. As shown in Fig.~\ref{fig:awesome}, the EDE contributes a few percent of the total energy perturbation for $10^{-2} h {\rm/Mpc} \lesssim k \lesssim 10^{-1} h{\rm/Mpc}$, which, using the relationship between wavenumber and multipole ($k\tau_0 \simeq \ell$) corresponds to $100 \lesssim \ell \lesssim 1000$. 
 
Before further exploring the preference for large initial field value, let us mention that there has been some recent interest in potentials with a pure power-law \cite{Agrawal:2019lmo}
\begin{equation}
    V(\phi)=V_0\phi^{2n}\,.
\end{equation}
The dynamics of a power-law potential are specified by three parameters (as opposed to four for the potentials we consider): the power-law index $n$, the potential amplitude, $V_0$, and the initial field value $\phi_i$. Note that, when fixing $\Theta_i = 0.1$, the cosine potential we explore is well-approximated (to the sub-percent level) by a power-law in the small-angle approximation:
\begin{equation}
    V_n(\Theta) \simeq \frac{m^2 f^2}{2^n} \Theta^{2n}.
\end{equation}
In this case, we can map our parameters to that used in Ref.~\cite{Agrawal:2019lmo}, and one has $V_0 \equiv m^2 f^2/2^n$ and $\phi_i = f\Theta_i$. Our results in the small $\Theta_i$ limit are in excellent agreement with these of Ref.~\cite{Agrawal:2019lmo} (see also Fig.~\ref{fig:small_TT}). The dynamics of a power-law potential, in the small $\Theta_i$ limit of our potential, explains why that study could not fully recover the results of Ref.~\cite{Poulin:2018cxd}. In contrast to what was claimed in Ref.~\cite{Agrawal:2019lmo}, the difference in conclusions was {\em not} due to the use of an effective fluid approximation in Ref.~\cite{Poulin:2018cxd}, which as we have shown here (and noted in Ref.~\cite{Lin:2019qug}) is able to capture the main features (i.e., $z_c$, $f_{\rm EDE}(z_c)$, $\Theta_i$, and $n$) of the EDE scenario. 
 
 \subsubsection{The preference for a large initial field displacement}

The initial field value, $\Theta_i$, has two main effects on the EDE phenomenology. First, as is demonstrated in Fig.~\ref{fig:background_evo}, at fixed $z_c$ and $f_{\rm EDE}({z_c})$ the initial field value affects the asymmetry in the rise and fall of the fractional energy density contained within the EDE. In particular, smaller values of $m$ and $f$ required by a larger initial displacement yields a faster rise of the energy density towards the peak and a slower dilution along with more oscillations. 

The initial field value also affects the dynamics of perturbations in the EDE. The full dynamics are governed by the linearized KG equation which, in turn, depends on the time evolution of the background field. We can build an intuition for how that time evolution affects the EDE perturbations by using an approximate `cycle-averaged' set of fluid equations which depends on an effective sound speed \citep{Hu:2000ke,Hwang:2009js,Marsh:2010wq,Park:2012ru,Hlozek:2014lca,Marsh:2015xka,Noh:2017sdj,Poulin:2018dzj}
\begin{eqnarray}
 c_s^2 = \frac{2 a^2 (n-1) \varpi^2(a)+k^2}{2 a^2 (n+1) \varpi^2(a)+k^2},\label{eq:ceff2}
\end{eqnarray}
where $\varpi(a)$ is the angular frequency of the oscillating background field and is well-approximated by \cite{Johnson:2008se,Poulin:2018dzj}
\begin{eqnarray}
\varpi(a) &\simeq& m\frac{\sqrt{\pi} \Gamma(\frac{1+n}{2n})}{\Gamma\left(1+\frac{1}{2n}\right)}2^{-(1+n)/2}  \Theta^{n-1}_{\rm env}(a),\label{eq:omega}\\
&\simeq& 3 H(z_c)\frac{\sqrt{\pi} \Gamma(\frac{1+n}{2n})}{\Gamma \left(1+\frac{1}{2n}\right)}2^{-(1+n)/2} \frac{\Theta^{n-1}_{\rm env}(a)}{{\sqrt{|E_{n,\Theta \Theta}(\Theta_i)|}}},\nonumber
\end{eqnarray}
where the envelope of the background field ($\Theta_{\rm env} \equiv \phi_{\rm env}/f$) once it is oscillating is well-approximated by 
\begin{equation}
\phi_{\rm env}(a)=\phi_c\left(\frac{a_c}{a}\right)^{3/(n+1)}\label{eq:phi_env},
\end{equation}
where $\phi_c$ is the field value at $z_c$, and we have written the scalar field potential as $V_n(\phi) = m^2 f^2 E_n(\Theta = \phi/f)$. 

The effective sound speed introduces a new time-scale to the evolution of EDE perturbations. The linearized KG equation, Eq.~(\ref{eq:linKG}), shows that perturbations in the field will be driven at the frequency of the oscillation of the background field, $\varpi(a)$, and the effective sound-speed introduces a second frequency, $c_s k$. 

\begin{figure}
    \centering
    \includegraphics[scale=0.33]{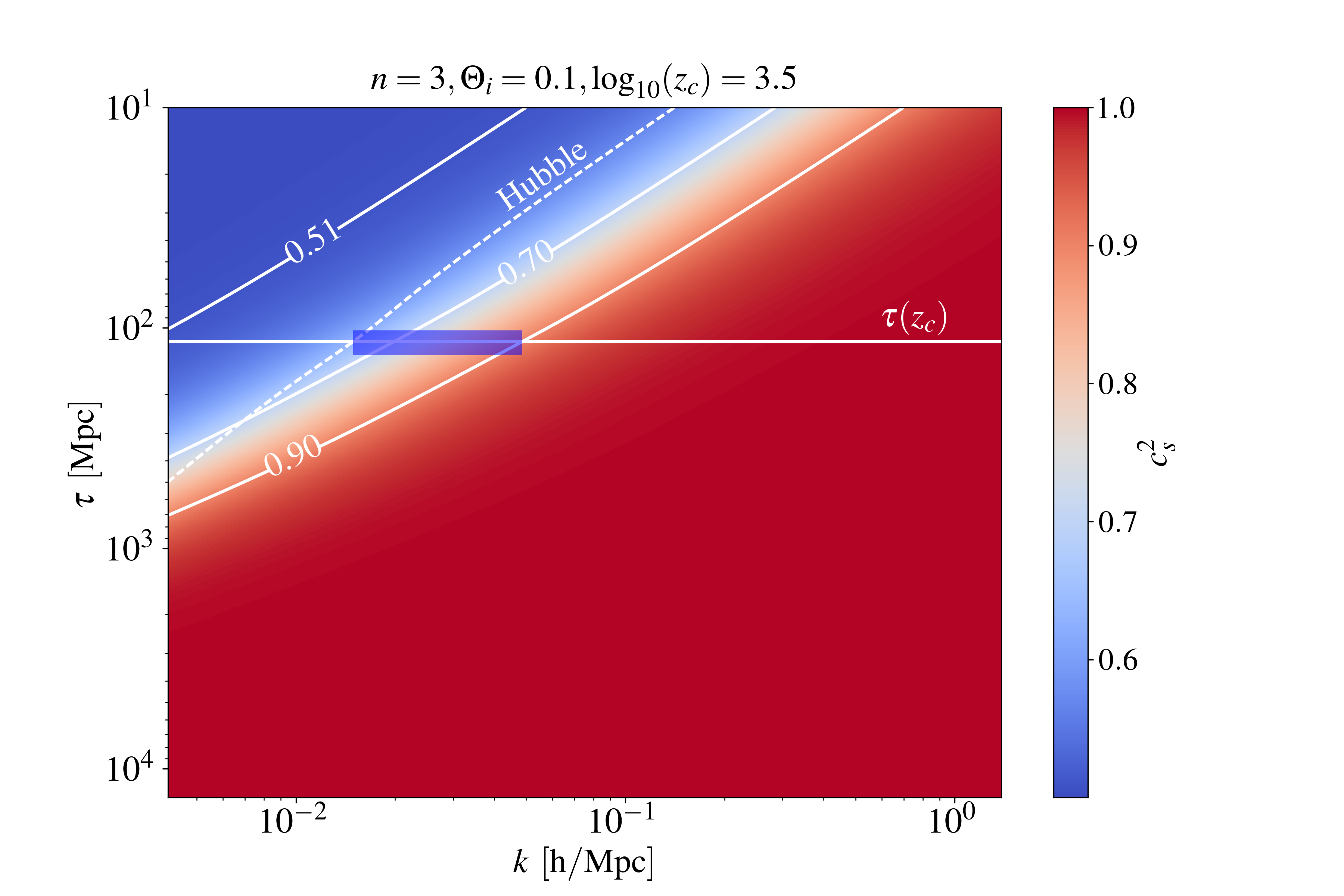}    \includegraphics[scale=0.33]{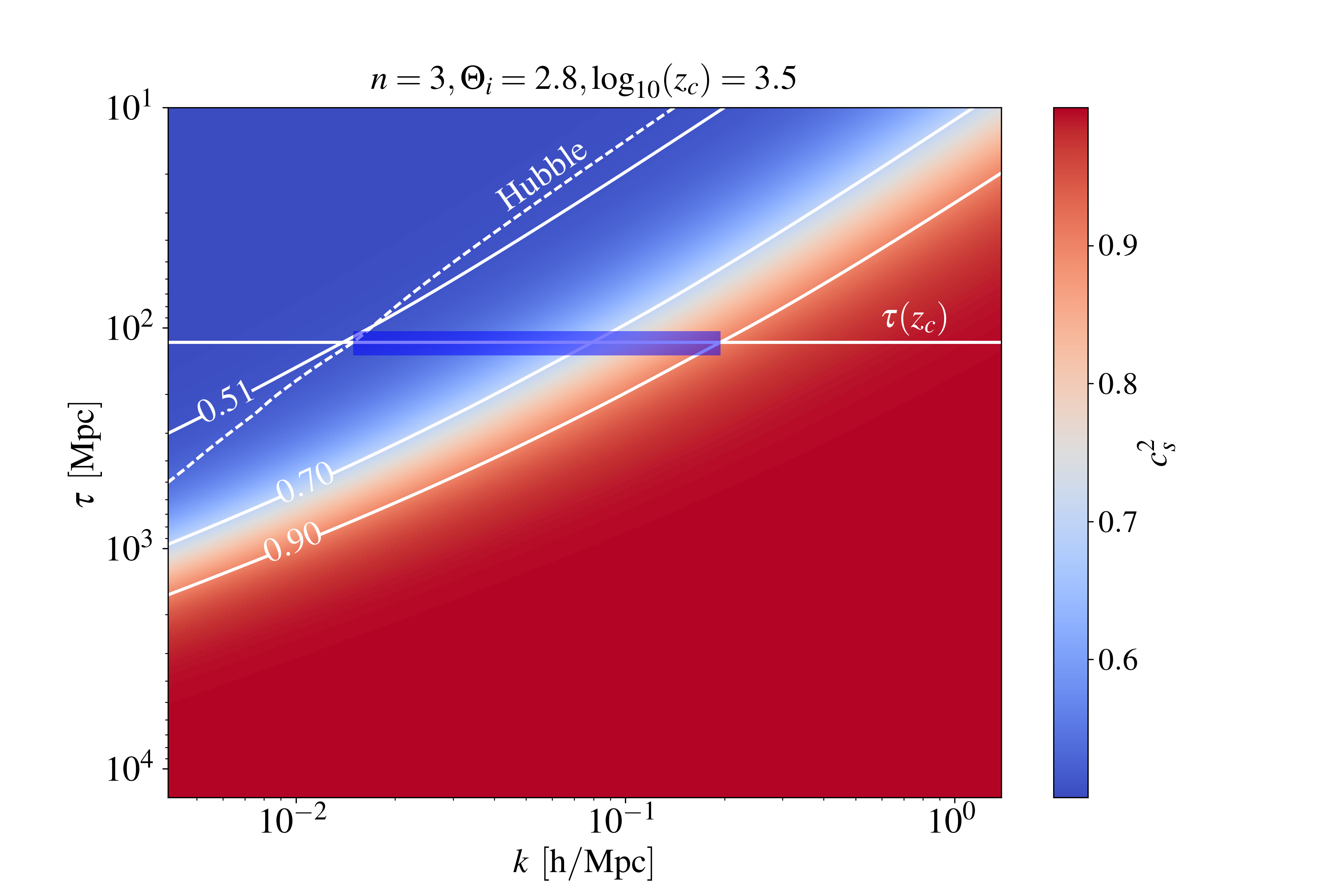}
    \caption{Effective sound speed from Eq.~(\ref{eq:ceff2}) for an EDE with $n=3$, ${\rm log}_{10}(z_c)=3.5$ and $\Theta_i=0.1$ (top panel) or  $\Theta_i=2.8$ (bottom panel). The blue shaded region show the range of $k$ within the horizon having $c_s^2<0.9$ around $z_c$. }
    \label{fig:cs2plot}
\end{figure}
\begin{figure}
    \centering
    \includegraphics[scale=0.4]{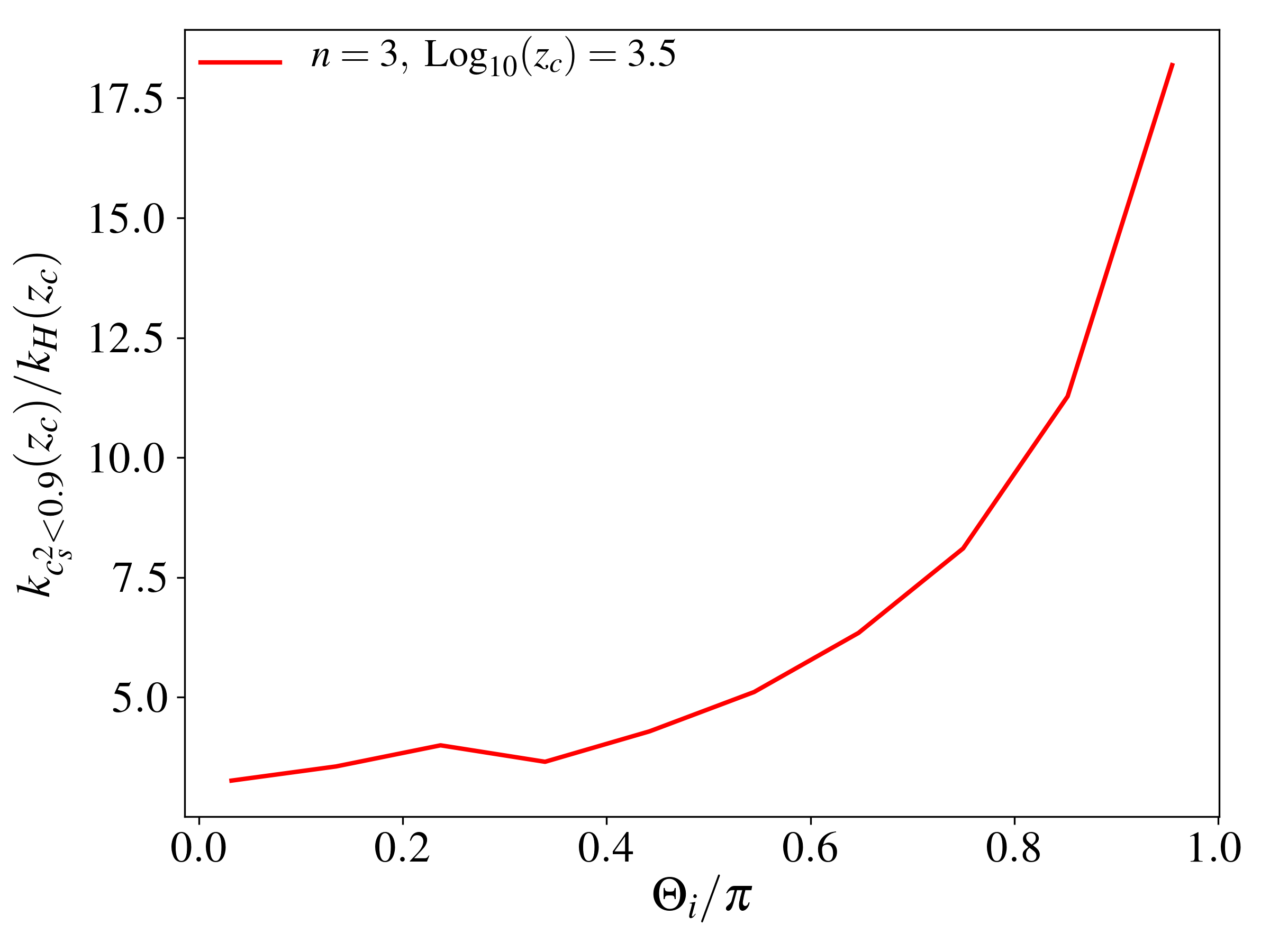}    
    \caption{The range of $k$ within the horizon having $c_s^2<0.9$ at $z_c$ as a function of $\Theta_i$. }
    \label{fig:cs2}
\end{figure}

As argued in Ref.~\cite{Lin:2019qug}, an `acoustic dark energy' with a constant effective sound-speed must have $c_s^2 \simeq 0.24 (n-1)/(n+1) + 0.6$ in order to resolve the Hubble tension. For example, with $n=3$ the best-fit (constant) sound-speed is $c_s^2 \simeq 0.72$. These results indicate that the data prefers an EDE which has modes inside of the horizon around $z_c$ with an effective sound-speed less than $\simeq 0.9$. As we show in Fig.~\ref{fig:cs2plot}, the range of modes that are inside of the horizon at $z_c$ and have $c_s^2<0.9$ is a strong function of $\Theta_i$. It is straightforward to show that the ratio $2n\varpi(a_c)/H(a_c)$ determines the range of modes within the horizon which have $c_s^2 <0.9$-- the larger this ratio is (compared to unity) the larger range of dynamical wavenumbers with $c_s^2 <0.9$. We show this ratio in Fig.~\ref{fig:cs2}:  more sub-horizon modes have $c_s^2<0.9$ as $\Theta_i \rightarrow \pi$. 

This provides an explanation as to why pure power-law potentials fail to provide as good of a resolution to the Hubble tension. If the potential can be approximated by a power-law then we will always have $\Theta_i^{n-1}/\sqrt{|E_{n,\Theta \Theta}(\Theta_i)|} \simeq 1$. In this case the only way to control the range of wavenumbers which have $c_s^2<0.9$ is by changing the power-law index $n$. Eq.~(\ref{eq:omega}) shows that as the power-law index $n$ decreases the range of wavenumbers which have $c_s^2<0.9$ increases, possibly explaining why Ref.~\cite{Agrawal:2019lmo} finds a slightly improved resolution of the Hubble tension for $n\rightarrow 2$ (see their Fig.~5).

This discussion, along with the results of Ref.~\cite{Lin:2019qug}, indicates that the EDE fit to current CMB measurements is improved as more sub-horizon modes evolve with $c_s^2<0.9$. This can be achieved as long as $\Theta_i^{n-1}/\sqrt{|E_{n,\Theta \Theta}(\Theta_i)|} \gg 1$. In the case of the potentials considered here this, in turn, requires $\Theta_i/\pi \simeq 1$ and can also be achieved by any potential with a second derivative that goes towards zero faster than $\Theta^{2n-2}$. 

\subsection{Detecting Early Dark Energy in the CMB}
\label{sec:future}
In previous Sections we have seen that the preference for an EDE is strong when including the SH0ES measurement, but only mild (and not statistically significant) within {\em Planck} data alone. However, new experiments such as CMB-S4\footnote{We take it as a proxy for next-generation ground based experiments. Given its planned characteristics, very similar results up to factors of order unity would be obtained with the Simons Observatory \cite{Ade:2018sbj} when doing these forecasts.} have been proposed as a way to improve our measurements of CMB polarization at large multipoles. In this Section, we show that an EDE model that resolves the Hubble tension can be detected with a (future) CMB-only analysis. The independent detection of the EDE in future cosmological data is an essential consistency test of such models, and would help to establish the Hubble tension (and its resolution).  

To perform this analysis, we use the mock CMB-S4 likelihood as provided in {\sf MontePython-v3.1} and follow the fiducial prescription: we include multipoles $\ell$ from 30 to 3000, assume a sky coverage of 40\%, uncorrelated Gaussian error on each $a_{\ell m}$’s (which is known to break at low-l), as well as uncorrelated temperature and polarization noise and perfect foreground cleaning up to $\ell_{\rm max}$. Given that there is no information at low-$\ell$, we add a Gaussian prior on the optical depth $\tau_{\rm reio}=0.065\pm0.012$ based on recent {\em Planck} data.  We choose a fiducial model compatible with our reconstructed best-fit model: $\{\omega_b = 0.02227, \omega_{\rm cdm} = 0.1293, h = 0.72, n_s = 0.9848, 10^9A_s = 2.1654, \tau_{\rm reio} = 0.065, \Theta_i = 2.91, f_{\rm EDE}(z_c) = 0.115, {\rm log}_{10}(z_c) = 3.53\}$. We perform fits of both the EDE and the $\Lambda$CDM cosmology. The latter runs will help us determine how much bias is introduced on $\Lambda$CDM parameters, when the ``true'' cosmological model contains an EDE. In order to check whether we should expect that a {\rm Planck}-only analysis is unable to detect the EDE, we perform an MCMC on synthetic {\rm Planck} data with the same fiducial EDE model. We generate the {\rm Planck} mock dataset with the simulated likelihood {\sf fake\_planck\_realistic} available in {\sf MontePython-v3.1}. 

\begin{figure}[h!]
    \centering
    \includegraphics[scale=0.8]{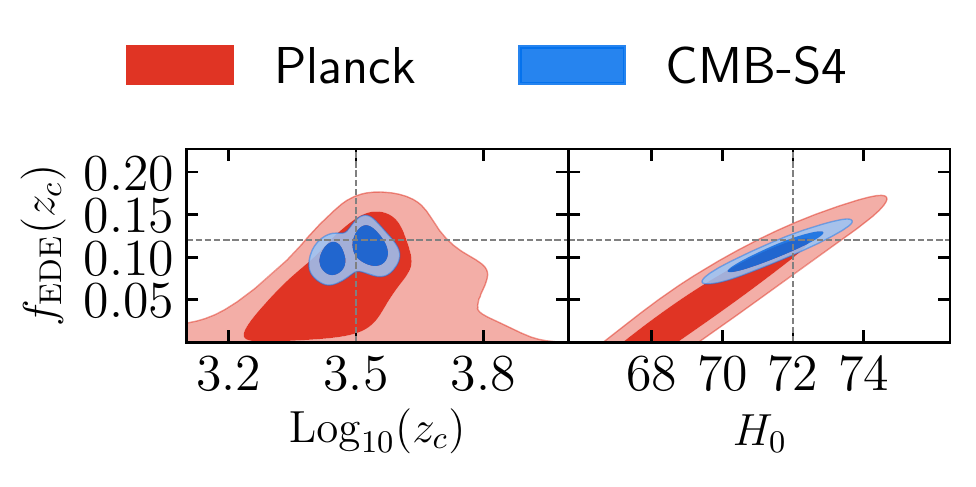}
    \caption{2D Posterior distributions of $\{{\rm log}_{10}(z_c),f_{\rm EDE}(z_c)\}$ and $\{H_0,f_{\rm EDE}(z_c)\}$ reconstructed from a fit to simulated {\em Planck} data and CMB-S4. The fiducial model has  $\{H_0\!=\! 72 ~{\rm km/s/Mpc}, f_{\rm EDE}(z_c)\!=\! 0.115, {\rm log}_{10}(z_c)\! =\! 3.53\}$.}
    \label{fig:planck-vs-cmbs4}
\end{figure}

Our reconstructed parameters are given in Tables~\ref{table:fake_planck} and \ref{table:fake_cmbs4}. In Fig.~\ref{fig:planck-vs-cmbs4}, we plot the 2D marginalized posterior distributions of $\{{\rm log}_{10}(z_c),f_{\rm EDE}(z_c)\}$ and $\{H_0,f_{\rm EDE}(z_c)\}$ reconstructed with simulated {\em Planck} or CMB-S4 data. From there and previous Tables one can read two very important pieces of information: i) CMB-S4 can un-ambiguously detect the presence of an oscillating EDE at more than $5\sigma$ (assuming Gaussian errors, we find a non-zero $f_{\rm EDE}(z_c)$ at $\sim10\sigma$); ii) {\em Planck} alone can only set an upper limit on the EDE fraction (we find $f_{\rm EDE}(z_c)<0.14$ at 95\% C.L.) and is compatible with the no-EDE hypothesis at 1$\sigma$.
Comparing with the $\Lambda$CDM reconstruction is also instructive. For simulated {\em Planck} data, we find a $\Delta\chi^2_{\rm min}=-7.8$ in favor of the EDE cosmology, which is in good agreement with what is found in real data (we recall from Table \ref{table:chi2_preliminary} that we found $\Delta\chi^2_{\rm min}=-4.85$ for real {\em Planck} data). Additionally the reconstructed $\Lambda$CDM parameters are all well within $1\sigma$ of what is obtained in the global fit of real data. This leads to bias in the reconstructed parameters that can be many $\sigma$ away from the injected ones. 

\begin{figure}[h!]
    \centering
    \includegraphics[scale=0.33]{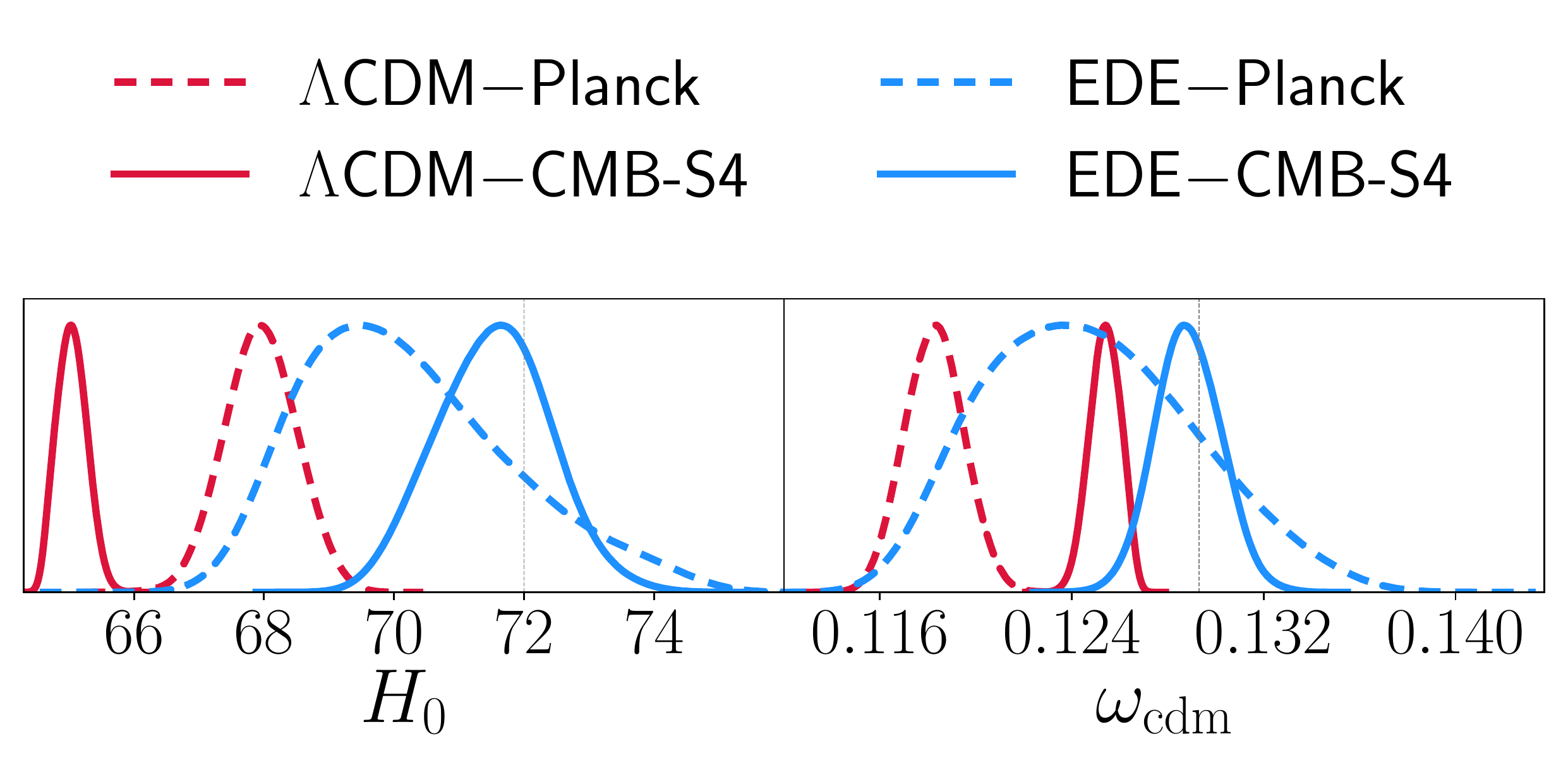}
    \caption{1D Posterior distributions of $H_0$ and $\omega_{\rm cdm}$ reconstructed from a fit to simulated {\em Planck} data (dashed lines) and CMB-S4 (full lines) in either the $\Lambda$CDM (blue) or EDE (red) cosmology. The fiducial model has  $\{H_0\!=\! 72 ~{\rm km/s/Mpc},\omega_{\rm cdm} = 0.1293\}$.}
    \label{fig:lcdm-vs-axiclass_fake}
\end{figure}

We report the biases on $\Lambda$CDM parameters in Tables.~\ref{table:fake_planck} and \ref{table:fake_cmbs4}. For instance, as shown in Fig.~\ref{fig:lcdm-vs-axiclass_fake}, with simulated {\em Planck} data the $\Lambda$CDM reconstructed $H_0=68\pm0.6$ km/s/Mpc is $6.7\sigma$ lower than the fiducial value of 72 km/s/Mpc. Similar shifts are seen for parameters strongly correlated with $f_{\rm EDE}$ such as $\omega_{\rm cdm}$. Naturally, with the much more precise CMB-S4 these biases increase tremendously as can be read off of Table~\ref{table:fake_cmbs4}. 
Reassuringly, for CMB-S4 we find such a large $\Delta\chi^2_{\rm min}=-496$ that any statistical test would strongly favor the EDE, as already discussed. Interestingly though, the reconstructed central value of $H_0$ in CMB-S4 with $\Lambda$CDM is much smaller than that deduced from {\em Planck}. Such a large shift from one experiment to another could be interpreted as a sign that $\Lambda$CDM is not the ``true'' model. We note that such a shift in the central value of $H_0$ already occured when going from WMAP9 ($70.0\pm2.2$ km/s/Mpc) to {\em Planck} ($67.37\pm0.54$ km/s/Mpc), and is attributed to pattern in the residuals at $\ell>1000$ not accessible with WMAP \cite{Aghanim:2016sns,Addison:2015wyg}.

\begin{table*}
  \begin{tabular}{|l|c|c|c|}
    \hline\hline Parameter &~~$\Lambda$CDM~~&~~~$n=3$~~~&~~~ $\Lambda$CDM bias ~~~\\ \hline \hline
    $H_0$/(km/s/Mpc) &$67.98~(67.95)\pm0.59$ &$70.17~(72.8)_{-2}^{+1.2}$ &$-6.81\sigma$\\
    $100~\omega_b$ & $2.226~(2.227)\pm0.015$& $2.237~(2.253)\pm0.023$&$-0.07\sigma$\\
    $\omega_{\rm cdm}$& $0.1183~(0.1182)\pm0.0013$&$0.1247~(0.1305)_{-0.0056}^{+0.0036}$ & $-8.46\sigma$ \\
    $10^{9}A_s$& $2.125~(2.124)\pm0.022$& $2.148~(2.174)\pm0.028$ & $-1.84\sigma$ \\
    $n_s$&$0.9672~(0.9674)\pm0.0038$&$0.9766~(0.9918)_{-0.011}^{+0.0068}$ & $-4.63\sigma$ \\
    $\tau_{\rm reio}$& $0.066~(0.065)\pm0.0055$& $0.0656~(0.0659)_{-0.0053}^{+0.0047}$&$0.02\sigma$\\
    ${\rm Log}_{10}(z_c)$& $-$&$3.51~(3.57)_{-0.1}^{+0.18}$ &$-$\\
    $f_{\rm EDE}(z_c)$ & $-$&$0.064~(0.129)_{-0.064}^{+0.018}$ &$-$\\
    $\Theta_i$ &$-$ & $2.22~(2.88)_{-0.11}^{+0.78}$& $-$\\
    \hline
    $\Delta\chi^2_{\rm min}$ & 0&$-7.8$&$-$\\
    \hline

  \end{tabular}
  \caption{The mean (best-fit) $\pm1\sigma$ error of the cosmological parameters reconstructed from a fit to simulated {\em Planck} data in $\Lambda$CDM and the EDE cosmology. In the $\Lambda$CDM case, we also give the shift in units of $\sigma$ between the reconstructed and fiducial parameters. The fiducial model has  $\{\omega_b = 0.02227, \omega_{\rm cdm} = 0.1293, h = 0.72, n_s = 0.9848, 10^9A_s = 2.1654, \tau_{\rm reio} = 0.065, \Theta_i = 2.91, f_{\rm EDE}(z_c) = 0.115, {\rm log}_{10}(z_c) = 3.53\}$.}
  \label{table:fake_planck}
\end{table*}

\begin{table*}
  \begin{tabular}{|l|c|c|c|}
    \hline\hline Parameter &~~$\Lambda$CDM~~&~~~$n=3$~~~&~~~ $\Lambda$CDM bias ~~~\\ \hline \hline
    $H_0$/(km/s/Mpc) &$65.03~(64.97)\pm0.26$ &$71.86~(71.86)\pm0.75$ &$-26.92\sigma$\\
    $100~\omega_b$ & $2.188~(2.187)\pm0.0034$& $2.227~(2.225)\pm0.005$&$-11.47\sigma$\\
    $\omega_{\rm cdm}$& $0.1254~(0.1256)\pm0.0007$&$0.1290~(0.1294)\pm0.0014$ & $-5.57\sigma$ \\
    $10^{9}A_s$& $3.041~(3.039)\pm0.01$& $2.163~(2.158)\pm0.026$ & $87.59\sigma$ \\
    $n_s$&$0.9643~(0.9643)\pm0.0022$&$0.9843~(0.9831)\pm0.004$ & $-9.32\sigma$ \\
    $\tau_{\rm reio}$& $0.052~(0.051)\pm0.006$& $0.065\pm0.007$&$-2.1\sigma$\\
    ${\rm Log}_{10}(z_c)$& $-$&$3.534~(3.526)\pm0.024$ &$-$\\
    $f_{\rm EDE}(z_c)$ & $-$&$0.112~(0.114)\pm0.013$ &$-$\\
    $\Theta_i$ &$-$ & $2.904~(2.914)_{-0.036}^{+0.046}$& $-$\\
    \hline
    $\Delta\chi^2_{\rm min}$ & 0&$-496$&$-$\\
    \hline

  \end{tabular}
  \caption{The mean (best-fit) $\pm1\sigma$ error of the cosmological parameters reconstructed from a fit to simulated CMB-S4 data in $\Lambda$CDM and the EDE cosmology. In the $\Lambda$CDM case, we also give the shift in units of $\sigma$ between the reconstructed and fiducial parameters. The fiducial model has  $\{\omega_b = 0.02227, \omega_{\rm cdm} = 0.1293, h = 0.72, n_s = 0.9848, 10^9A_s = 2.165, \tau_{\rm reio} = 0.065, \Theta_i = 2.91, f_{\rm EDE}(z_c) = 0.115, {\rm log}_{10}(z_c) = 3.53\}$.}
  \label{table:fake_cmbs4}
\end{table*}

\section{New signatures and observational consequences}

In this Section, we discuss two additional consequences of the existence of an EDE: i) isocurvature perturbations; ii) scale-dependent instabilities in scalar field perturbations, potentially leading to nonlinear dynamics in the EDE field.
\label{sec:additional_obs}
\subsection{Isocurvature perturbations}
\label{sec:iso}

A general solution to the linearized KG equation, Eq.~(\ref{eq:linKG}), can be divided into a sum of homogeneous and inhomogeneous terms, $\delta \phi = \delta \phi_H + \delta \phi_I$. The homogeneous term, where the initial gravitational potential perturbations are negligible compared to the field perturbation, is excited by isocurvature perturbations whereas the inhomogeneous term is excited by adiabatic perturbations-- we discuss adiabatic initial conditions in Appendix \ref{app:ad}.  

Generically, the field will have isocurvature initial conditions as a nearly massless spectator field during inflation. These perturbations will have primordial fluctuations, $\zeta_\phi(\vec k)$, which are uncorrelated with the adiabatic fluctuations, $\zeta_{\rm ad}(\vec k)$, and are drawn from a power spectrum \cite{Lyth:2001nq,Kobayashi:2013bna,Hlozek:2017zzf}
\begin{eqnarray}
    \langle \zeta_\phi(\vec k) \zeta_\phi^*(\vec k')\rangle &=& (2\pi)^3 P_{\phi}(k) \delta^{(3)}_D(\vec k - \vec k'),\\
    P_{\phi}(k)/P_\zeta(k) &=&  r \left(\frac{k}{k_0}\right)^{-r/8-(1-n_s)},
\end{eqnarray}
where $P_\zeta(k)$ is the standard (`adiabatic') primordial curvature perturbation power spectrum, $r$ is the tensor-to-scalar ratio, and we have used the fact that the effective mass of the scalar field is much less than the energy-scale of inflation. 

\begin{figure}[h!]
    \centering
    \includegraphics[scale=0.55]{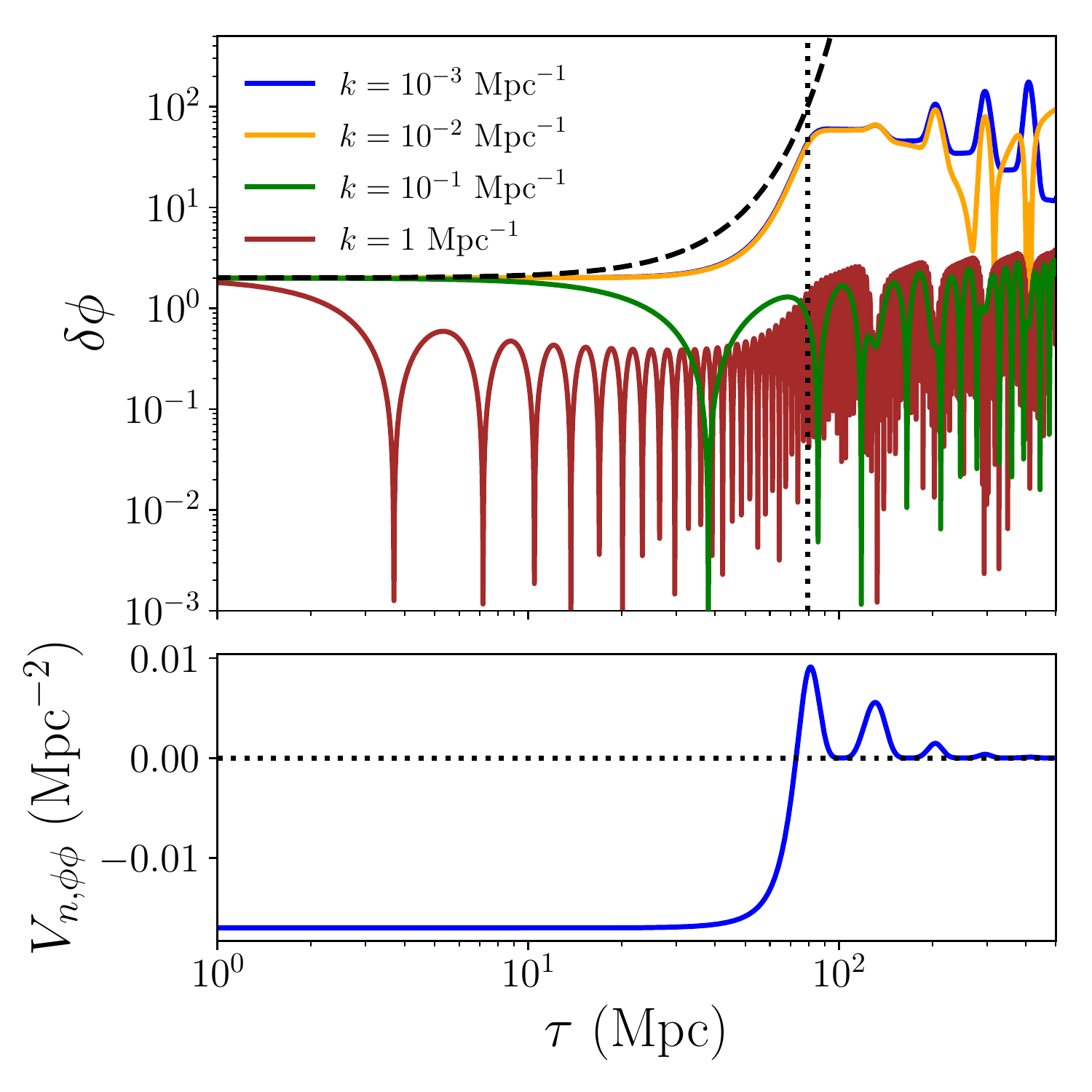}
    \caption{The evolution of the isocurvature (i.e., homogeneous) field perturbations in the case where $V_{n,\phi \phi}<0$, initially. In this case perturbations experience exponential growth for a limited amount of time. The dashed black curve shows the analytic solution in Eq.~(\ref{eq:homosol}) and the dotted-vertical curve shows the conformal time at which the background field starts to oscillate and $V_{n,\phi \phi} >0$; at this time the exponential growth stops.}
    \label{fig:iso_pert_comp}
\end{figure}

\begin{figure*}
     \includegraphics[scale=0.55]{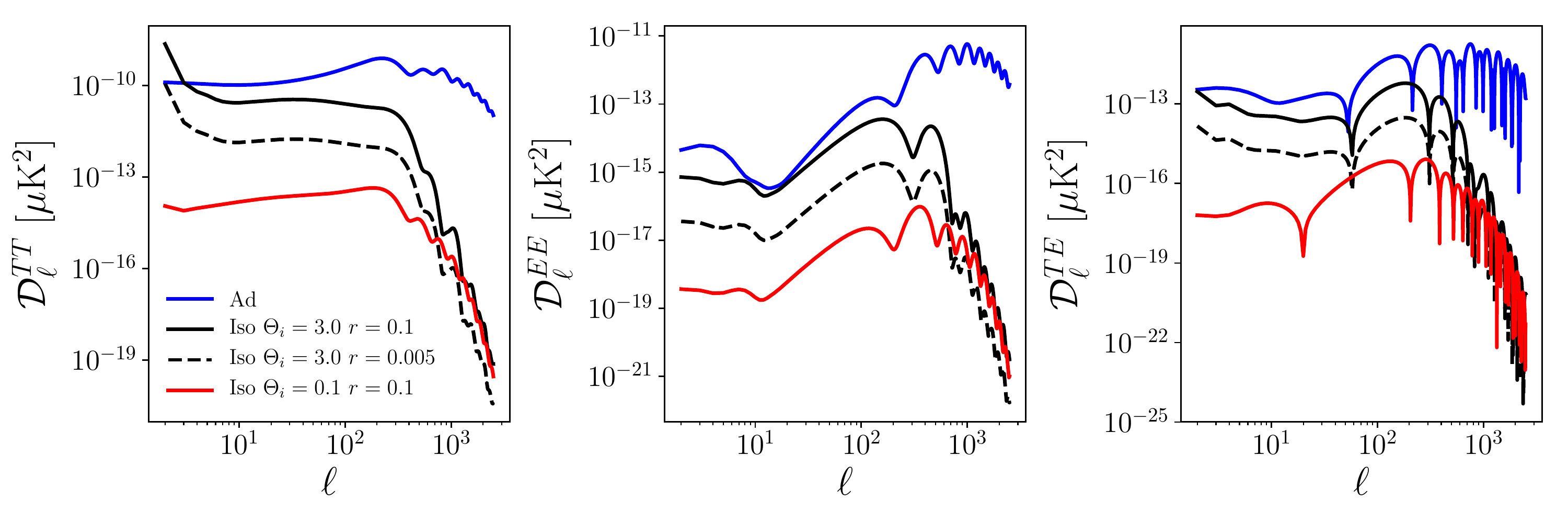}
    \caption{The standard adiabatic (blue) and EDE-isocurvature power spectra for $n=3$, $z_c = 10^{3.5}$, $f_{\rm EDE}(z_c) =0.1$. Since the EDE field is a spectator field during inflation, it naturally inherits both adiabatic and isocurvature initial conditions. As shown in this figure, the amplitude of isocurvature initial conditions are set by the tensor-to-scalar ratio, $r$, and are particularly sensitive to the initial field displacement as discussed in Sec.~\ref{sec:perturb}.}
    \label{fig:cls_comp}
\end{figure*}

To understand how the properties of the scalar field affect the isocurvature perturbations we solve for the superhorizon radiation dominated evolution of the field perturbations while the background field is undergoing slow-roll evolution. We can estimate this evolution by solving for the evolution of $\delta \phi$ with a vanishing driving term.  In this case it is straightforward to show that 
\begin{eqnarray}
    \delta \phi(a;\vec k) &=& \zeta_\phi(\vec k) e^{-i a^2 \sqrt{V_{n,\phi \phi}/(2 H_0 \sqrt{\Omega_{\rm rad}})}}\label{eq:homosol}
\\&\times& {\,}_1 F_1\left[\frac{3}{4}+\frac{ik^2}{4 H_0 \sqrt{V_{n,\phi \phi}\Omega_{\rm rad}}},\frac{3}{2},\frac{i a^2 \sqrt{V_{n,\phi \phi}}}{H_0}\right],\nonumber
\end{eqnarray}
where ${\,}_1 F_1$ is a hypergeometric function. In the case where $V_{n,\phi \phi}>0$ the exponential pre-factor produces oscillatory motion modulated by the hypergeometric function. On the other hand, when the initial field displacement is large we can have $V_{n,\phi \phi}<0$. In this case the perturbations in the field grow exponentially for $a>a_* = \sqrt{2 H_0}(\Omega_{\rm rad}/|V_{n,\phi \phi}|)^{1/4}$. 
If we approximate the critical scale-factor at which the background field becomes dynamical through $|V_{n,\phi \phi}| \simeq 9 H^2(z_c)$ we have that $a_c \simeq \sqrt{3 H_0}(\Omega_{\rm rad}/|V_{n,\phi \phi}|)^{1/4}$. Therefore we can see that in the case where $V_{n,\phi \phi}<0$, initially, linear perturbations experience a limited time of exponential growth until the background field becomes dynamical and falls to a value where $V_{n,\phi \phi}>0$; at this point the perturbations become stable. A similar statement can be made for the case where the field becomes dynamical during matter domination. This indicates that the amplitude of isocurvature perturbations will be highly dependent on the initial field value. 

We show the exponential growth of isocurvature field perturbations in Fig.~\ref{fig:iso_pert_comp} where we have used the isocurvature initial conditions presented in Ref.~\cite{Hlozek:2017zzf}. We choose a potential with $n=3$ and $\phi_i/f = 3.0$ so that initially $V_{n,\phi \phi}(\phi_i)/m^2 = -11.52$. The analytic solution in Eq.~(\ref{eq:homosol})-- shown as the dashed black curve-- indicates when these modes start to evolve exponentially. The vertical dotted curve shows when the background field starts to oscillate and, correspondingly, when $V_{n,\phi \phi} >0$; at this time the exponential growth in the field perturbation ends. 

Fig.~\ref{fig:cls_comp} shows the temperature and polarization power spectra (with $\mathcal{D}^{XY}_\ell \equiv \ell(\ell+1) C^{XY}_{\ell}/(2\pi)$) for the standard adiabatic perturbations and the scalar field isocurvature perturbations for a range of values of the initial field displacement, $\Theta_i$, and the tensor-to-scalar ratio $r$. We can see that when $\Theta_i/\pi \simeq 1$ the tachyonic instability is active and leads to an enhancement at large angular scales. In this case, in order to produce an effect within cosmic variance, the overall amplitude of the power isocurvature power-spectra must be at most $\simeq 10\%$ of the standard adiabatic power spectra on large angular scales; this occurs as long as $r \lesssim 5 \times 10^{-3}$. Since current observations of the CMB place an upper limit $r < 0.056$ at 95\% CL \cite{Akrami:2018odb}, a detection of $5 \times 10^{-3} \lesssim r<0.056$ could place significant constraints on the EDE scenario as a resolution to the Hubble tension. Given that we have yet to detect evidence of an inflationary gravitational wave background, in our analysis we have ignored the effects of the isocurvature mode, implicitly assuming that $r \lesssim 5 \times 10^{-3}$. 

\subsection{Self-resonance in anharmonic potentials}
\label{sec:res}

In this Section, we show that the anharmonicity of the oscillations of the background field lead to a scale-dependent, quasi-exponential, growth in perturbations due to self-resonance -- parametric resonance in the perturbations of a field driven by oscillations of the the field itself. In particular, there exists an instability leading to {\it significant} growth of perturbations for potentials which go as $V_n \propto \phi^{2n}$ with $n\simeq 2$ (near their minima). Similar resonant processes have been explored in previous work, e.g., Refs.~\cite{Johnson:2008se,Lozanov:2017hjm,Lozanov:2016hid}.  Here we focus on summarizing the main results of our analysis and direct the reader to Appendix \ref{app:parametric_resonance} for more details. 

\subsubsection{Parametric resonance preliminaries}

Parametric resonance occurs when the effective frequency of a harmonic oscillator varies at such a rate so as to pump energy into the oscillation. The phenomena is well-known by anyone who has been on a swing: as we pump our legs we change the moment of inertia of the pendulum and if we pump at the right rate we can increase the amplitude of the swing. The effective angular frequency of perturbations to the scalar field is given in Eq.~(\ref{eq:linKG}) as $\omega_{\rm eff}^2 \equiv k^2 + V_{n,\phi \phi}$ (ignoring expansion); if $V_n$ is anharmonic then $\omega_{\rm eff}^2$ will oscillate due to the oscillation of the amplitude of the background field, which will lead to an exponential growth of perturbations with certain wavenumbers $k$. 

In the context of a scalar field, there is another way of understanding the rapid growth of perturbations. The homogeneous oscillating field provides a time-dependent effective mass for its perturbations. As the effective mass changes (particularly when it passes through zero), we get enhanced particle production of certain momenta, that is, an increase in occupation number in certain $k$-modes. A previously occupied mode is further enhanced by Bose effects as the periodic changes in the effective mass repeats.

For the analysis of parametric resonance, we do not need to restrict ourselves to the regime where the potential is a power law. See for example, \cite{Lozanov:2017hjm} for treatment with the full shape of a flattened potential which cannot always be ignored (also see Appendix \ref{app:parametric_resonance}). However, restricting ourselves to power law potentials leads to more tractable and instructive expressions, as we present in this Section. Moreover, once the background field starts to oscillate, the amplitude of the oscillations quickly dilutes due to expansion such that the potential is well-approximated by a power law: $V_n(\phi) \simeq m^2 f^2/2^n(\phi/f)^{2n}$. 

In order to quantitatively understand the process of self-resonance in an oscillating scalar field, it is useful to start by ignoring both the expansion of the universe and metric perturbations, that is $a=1,h=0$ in Eq.~\eqref{eq:linKG}, which yields
\begin{equation}
    \delta\ddot{\phi}_k +\left[k^2 + V_{n,\phi \phi}(\phi)\right] \delta \phi_k = 0.
\end{equation}
Note that we have switched to cosmic time and $V_{n,\phi \phi}(\phi)$ will be periodic for an oscillatory background field $\phi$ for $n>1$.\footnote{For $n=1$,$V_{n,\phi \phi}(\phi)={\rm const.}$ which is trivially periodic, and Floquet's theorem still applies. But there are of course no instabilities.} In this case, Floquet's theorem guarantees that the solutions will have the form
\begin{equation}
\delta\phi_k(t)=e^{\mu_k t}P_+(k,t)+e^{-\mu_k t}P_-(k,t),
\end{equation}
where $P_\pm(k,t)$ are periodic functions of time with the same  period as $V_{n,\phi\phi}(\phi)$. Importantly, $\mu_k$ are the Floquet exponents; we have exponentially growing solutions when the real part of the Floquet exponent, $\Re[\mu_k]>0$. For a given potential $V(\phi)$, typically the Floquet exponent will depend on the amplitude of the oscillating field $\phi$ as well as the wavenumber $k$, and will form bands of instability where $\Re[\mu_k]>0$ in the $k-\phi$ plane (see Fig.~\ref{fig:phi4} and Fig.~\ref{fig:phi5} in Appendix \ref{sec:n2}). A simple algorithm for calculating the Floquet exponent can be found in, for example, Appendix A of Ref.~\cite{Amin:2011hu}, 
or a more general one in Sec.~3.2 of Ref.~\cite{Amin:2014eta} (also see references therein).

\begin{figure}[t!]
    \centering
    \includegraphics[scale=0.8]{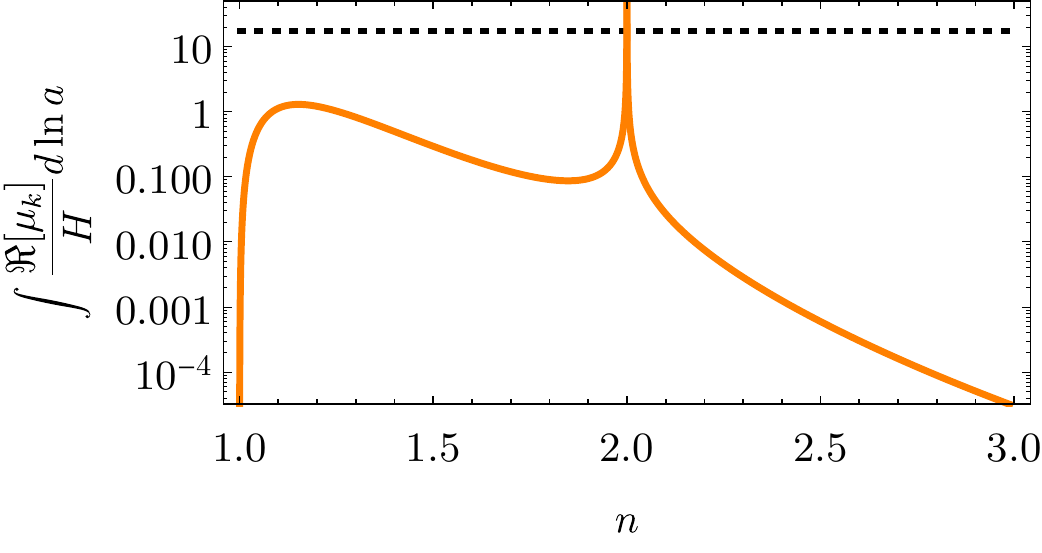}
    \caption{The shape of the integral of the growth ratio as a function of $n$ evaluated at $a=1$. The special nature of $n\approx 2$ is visible, with the dashed line indicating the value taken at $n=2$. The detailed shape near $n\approx 2$, as well as the magnitude of this ratio should be trusted only qualitatively. We assumed $a_c=a_{\rm eq}\approx 10^{-4}$ for the above plot.}
    \label{fig:growthratio}
\end{figure}

To include the effect of expansion (heuristically), we let $k\rightarrow k/a$ and $\phi\rightarrow \phi_{\rm env}\propto a^{-3/(1+n)}$. As a result, a typical co-moving mode now flows through the instability bands as the universe expands. See Fig.~\ref{fig:phi4} and Fig.~\ref{fig:phi5} in Appendix \ref{sec:n2} for examples. The following discussion should be interepreted within the assumption that the oscillatory timescale of the field is small compared to the expansion timescale of the universe. 

To get a sense of the behavior of a given mode, we need to compute the real part of its Floquet exponent integrated over time: $\int \Re[\mu_k]dt=\int H^{-1}\Re[\mu_k]d\ln a$. This integral is shown as a function of $n$ in Fig.~\ref{fig:growthratio}. To understand its relevance, note that heuristically, the evolution of the perturbations is given by 
\begin{equation}
\begin{aligned}
\label{eq:pertgrowth}
&k^{3/2}\delta\phi_k(a)\\
&\sim k^{3/2}\delta\phi_k(a_c)\left(\frac{a_c}{a}\right)^{\frac{3}{n+1}} \exp\left[{\int_{\Delta \ln a} \frac{\Re[\mu_k]}{H}d\ln b}\right]\,,
\end{aligned}
\end{equation}
where $\Delta \ln a(k)$ is the interval spent by the $k$ mode in the resonance band, and $a_c$ is the scalefactor when background oscillations of the field begin. The scaling with $a$ in front represents the approximate redshifting of the mode amplitudes without resonance. For there to be significant growth, the quantity appearing in the square brackets Eq.~\eqref{eq:pertgrowth} and shown in Fig.~\ref{fig:growthratio} should at the very minimum be larger than unity. The exponential has to overcome the usual decay of perturbation amplitudes in an expanding universe. Building on the work in \cite{Lozanov:2017hjm}, we derive useful analytic approximations in  Appendix \ref{app:parametric_resonance} for $\int H^{-1}\Re[\mu_k]d\ln a$  in a universe with matter/radiation. These same analytic expressions were used to obtain Fig.~\ref{fig:growthratio}.

For cases where there is significant growth, then at some point
\begin{equation}
\label{eq:anlwhen}
k^{3/2}\delta\phi_k(a_{\rm nl})\sim \phi_{\rm env}(a_{\rm nl})\qquad{\rm for}\qquad a_{\rm nl}<1\,,
\end{equation}
where $\phi_{\rm env}$ is the envelope of the homogeneous oscillating field, Eq.~(\ref{eq:phi_env}). When this approximate equality is reached, linear perturbation theory breaks down. One can expect mode-mode coupling and significant backreaction on the homogeneous field leading to spatially inhomogeneous dynamics which cannot be captured by linear perturbation theory. See Ref.~\cite{Lozanov:2017hjm} for lattice simulations of related models, but in the context of the early universe.  

Our analysis also allows us to roughly characterize the scales and redshifts at which non-linearity in the field appears. Of particular interest for the discussion here we find that the resonant wavenumber is approximately given by 
\begin{equation}
\frac{k_{\rm res}}{a}\approx m \left[\frac{\phi_{\rm env}(a)}{\sqrt{2}f}\right]^{n-1}\frac{2.54}{\sqrt{2}}.
\label{eq:kres_gen}
\end{equation}

From Fig.~\ref{fig:growthratio}, it should be evident that the $n\approx 2$ case is different. From Eq.~\eqref{eq:kres_gen} with $\phi_{\rm env}\propto a^{-3/(n+1)}$, the co-moving wavenumber that is resonant, $k_{\rm res}$, does not change with time for $n=2$. It reflects the special nature of $n=2$ case: if a co-moving mode is inside the narrow resonance band, it never leaves.  In contrast, for other $n$, a given $k$ mode can flow in and out of resonance bands. We again refer the interested reader to Appendix \ref{app:parametric_resonance}.

\subsubsection{A {\sf CLASS} comparison}

Using our modified version of {\sf CLASS}, which includes the effects from self-resonance in the $\phi$ field as well as gravitational effects from other components, we can check our analytic estimates for the resonant wavenumbers as well as growth-rate of perturbations. First, we have confirmed that for $n\gtrsim 2$ (but not too close to $n=2$), the perturbations remain linear at the resonant wavenumber, and never become comparable to the homogeneous field amplitude. Hence, a linear analysis is adequate. We did not check $n\lesssim 2$ since the number of oscillations over the Hubble time gets very large.

\begin{figure}[h!]
    \centering
    \includegraphics[scale=0.55]{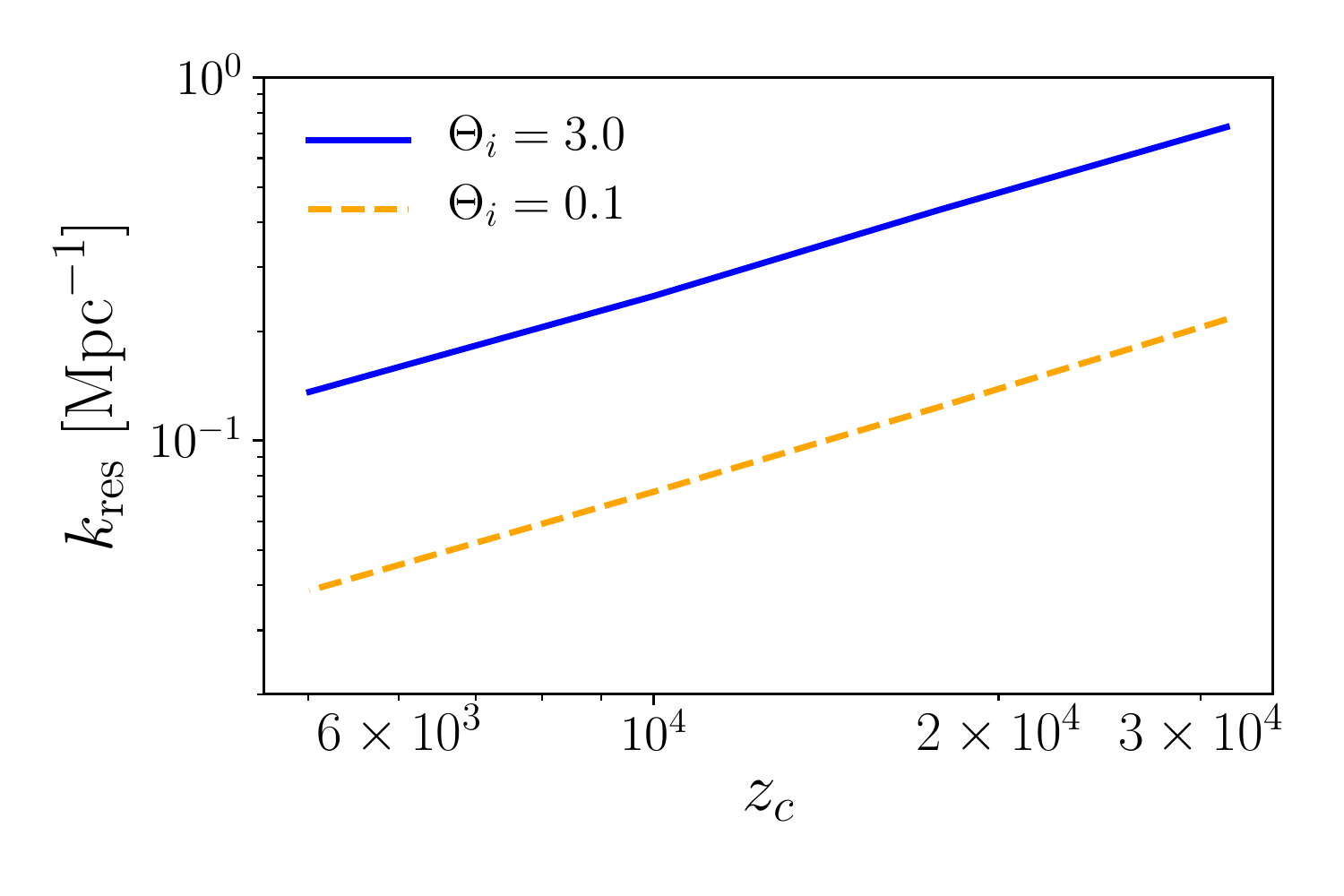}
    \caption{The resonant wavenumber as a function of $z_c$ for $n=2$ from evolving perturbations using ${\sf CLASS}$. These are in excellent agreement (better than $\sim 1\%$) with the analytic expectation provided in Eq.~\eqref{eq:kres_gen} for $n=2$.}
    \label{fig:res_k_n2}
\end{figure}

\begin{figure}[t!]
    \centering
    \includegraphics[scale=0.4]{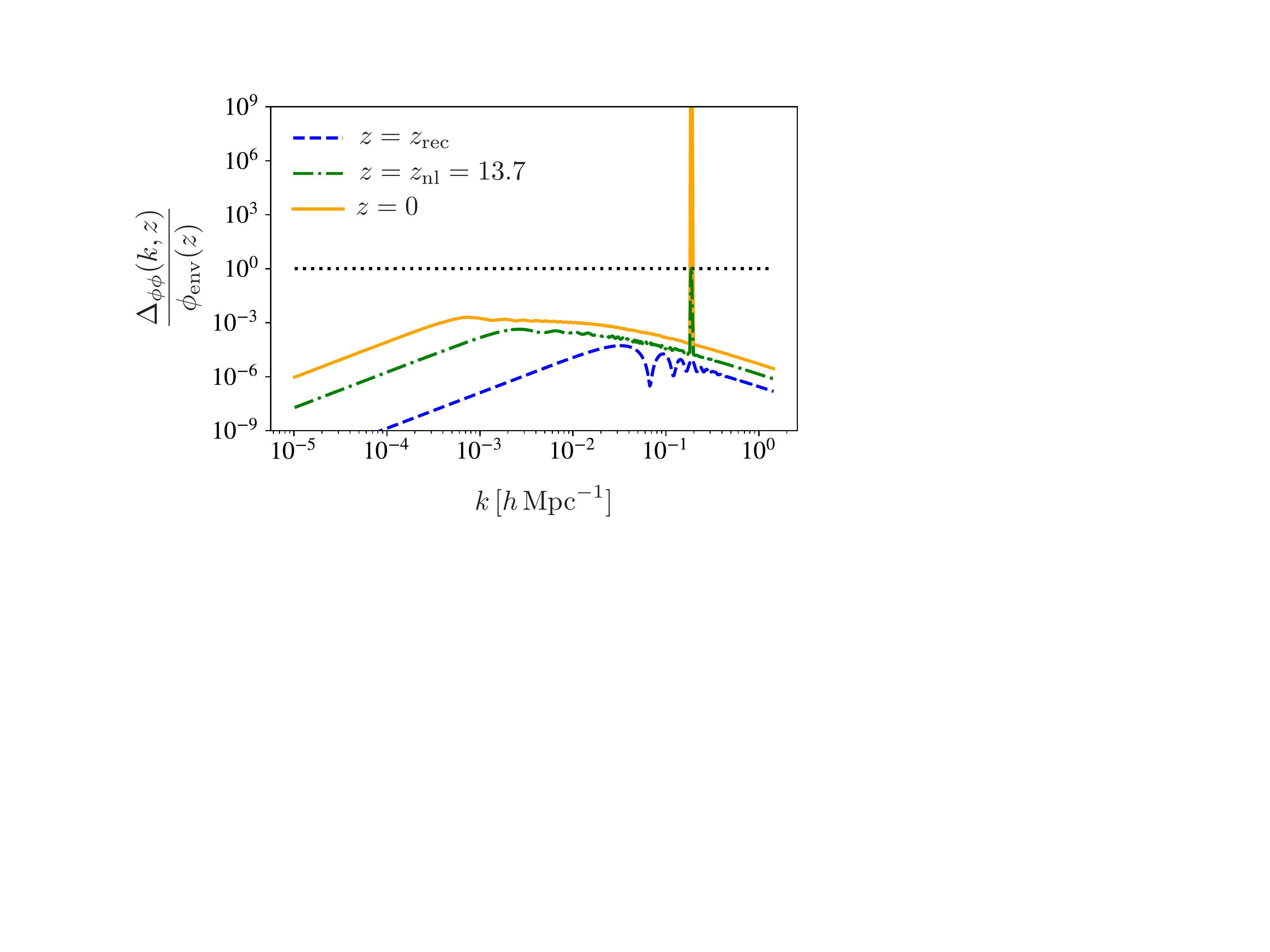}
     \includegraphics[scale=0.45]{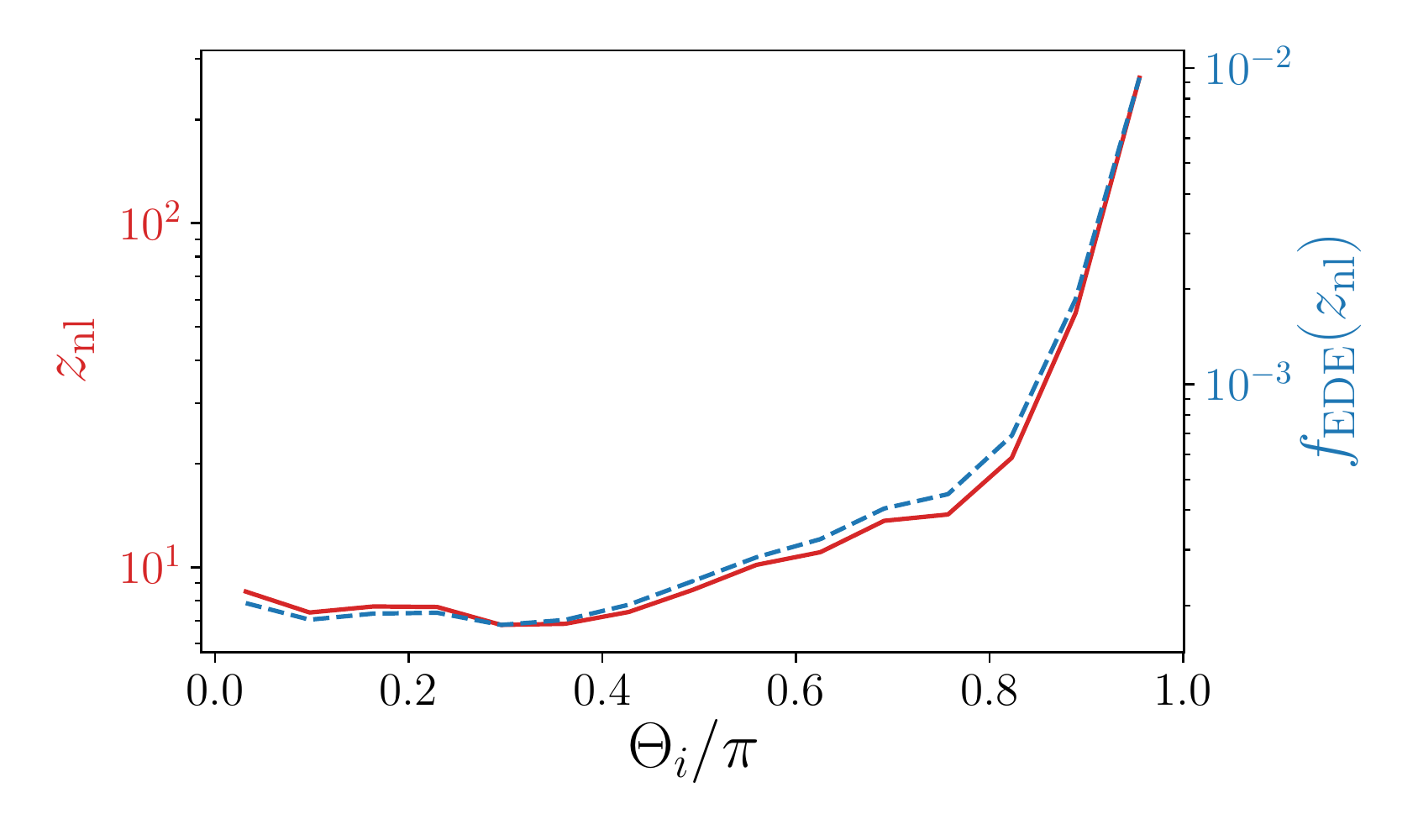}
    \caption{{\it Top}: The dimensionless power spectrum of the field for $n=2$, $\Theta_i = 2.4$, $z_c = 10^{4}$ and $f_{\rm EDE}(z_c) = 0.1$ obtained using ${\sf CLASS}$. The resonant wavenumber becomes non-linear only at late times when the fractional energy density in the field is approximately $10^{-3}$. {\it Bottom}: The redshift and fraction of the energy density when the field perturbations become non-linear for $n=2$, $z_c = 10^{4}$, and $f_{\rm EDE}(z_c) = 0.1$. Note that for $n=2$, during matter domination $f_{\rm EDE}(z)= \rho_{\rm EDE}/\rho_m \propto (1+z)$, so that in this case $f_{\rm EDE}(z_{\rm nl})$ follows a similar curve as $z_{\rm nl}$. }
    \label{fig:dimen_power}
\end{figure}

Let us focus further on the $n=2$ case. Using our numerical results from {\sf CLASS}, we have confirmed that Eq.~(\ref{eq:kres_gen}) is accurate to better than 1\% for $n=2$. We show the resonant wavenumber as a function of $z_c=1/a_c-1$ (and for a fixed $f_{\rm EDE}(z_c) = 0.1$) in Fig.~\ref{fig:res_k_n2}. As an important technical aside, we note that the resolution requirements in $k$ space to capture the resonant modes can be quite stringent. For $n\approx 2$, $\Delta k_{\rm res}\approx (\sqrt{3/2}-3^{1/3}) a_c(\phi_c/f)m=\mathcal{O}[10^{-4}]m$. (see Appendix \ref{sec:n2})

Similarly, we can compute the evolution of the power spectrum of $\phi$ perturbations using {\sf CLASS}. To do this we compute the scalar field dimensionless power spectrum, normalized by the envelope of the background field, $\Delta^2_{\phi \phi}/\phi_{\rm env}^2 \equiv k^3 P_{\phi \phi}(k)/(2\pi^2)/\phi_{\rm env}^2$. When this quantity becomes order unity, the field dynamics become nonlinear. In the top-panel of Fig.~\ref{fig:dimen_power} we show the dimensionless power spectrum at three different redshifts for $n=2$ and in the bottom-panel we show the redshift and EDE fraction at which the dimensionless power spectrum is equal to unity as a function of $\Theta_i$ and $z_c$. 

Given that the EDE density contrast when the field becomes non-linear is of order unity, its contribution to the gravitational potential (though the Poisson equation) is approximately equal to its fraction of the total energy density at this time. Fig.~\ref{fig:dimen_power} shows that we can have as much as a percent of the total energy density contained within the EDE field when the field perturbations become non-linear. Given that the fractional perturbation in the energy density of the other constituents of the universe on these scales are of order $\sim 10^{-3}-10^{-4}$, this implies that the resonance may leave an observable imprint on the CMB. It may also have an impact on other late-time probes of large-scale structure and gravitational radiation. 

However, in order to make progress with our current linear code for $n=2$, in Appendix \ref{sec:n2} we make use of a switch that simply ignores the EDE contribution to the perturbed Einstein's equation all together once the energy density fraction drops below $10^{-3}$.  Clearly where these novel non-linear scalar field dynamics may have an observable impact on current and future probes, a more careful analysis is warranted. These nonlinear aspects will be taken up in future work.

\section{Discussion and Conclusions}
\label{sec:conclusions}

In this paper we have studied the ability for an extension of the standard cosmological model (that we have called `early dark energy'-- EDE--) to address the so-called Hubble tension between the measurement of $H_0$ using a variety of low-redshift probes of the expansion rate (Cepheid-calibrated Type 1a supernovae, time-delays of strongly lensed quasars, megamasers, and galaxy surface brightness \cite{Verde:2019ivm})
and its inference from CMB data within the $\Lambda$CDM model. This tension now reaches the $4-6\sigma$ level and a resolution, physical or systematic, is not easy to come by \cite{Verde:2019ivm}. 

Specifically, we have investigated the cosmological evolution of a scalar field with a potential $V_n(\phi)=m^2f^2[1-\cos(\phi/f)]^n$ and its impact on the CMB and other cosmological observations. In addition to the stadard six $\Lambda$CDM parameters, this model is specified by four model-parameters: the mass, $m$, `decay constant, $f$, initial field value, $\phi_i$, and index $n$. These four model-parameters can be mapped on a set of `observed'-parameters: the redshift at which the field contributes the largest fractional energy density, $z_c$, the fractional density at that redshift, $f_{\rm EDE}(z_c)$, the effective sound-speed of the perturbations, $c_s^2$, and the effective equation of state, $w_\phi$. The background dynamics of the field can be described succinctly: the field is frozen until $\simeq z_c$ where it reaches a peak fractional contribution of $f_{\rm EDE}(z_c)$ and then dilutes with an equation of state $w_\phi = (n-1)/(n+1)$. The initial field value, $\phi_i$, controls the dynamics of the perturbations through its effects on the effective sound speed. 
Using exact (linearized) dynamics, we find that with {\it Planck} temperature and polarization, {\it Planck} estimates of the lensing potential, a variety of high and low $z$ BAO measurements, the Pantheon supernova dataset, and the SH0ES estimate of the Hubble constant the presence of this scalar field is indicated at $\simeq 3.5 \sigma$. If we fix $n=3$ then we have $\log_{10}(z_c)=3.5^{+0.051}_{-0.11}$, $f_{\rm EDE}(z_c)=0.107^{+0.036}_{-0.029}$, $\Theta_i\equiv\phi_i/f = 2.6^{+0.36}_{-0.04}$ can resolve the Hubble tension.  We have identified that a range of $n=3.16^{+0.18}_{-1.1}$ are favored by the data with $n<5$ at 95\% C.~L.  These constraints, when translated into the model parameters for $n=3$, give $f=0.18\pm 0.06\ M_{\rm pl}$ and $m = 3.4_{-3.0}^{+2.3}\times 10^{-27}\ {\rm eV}$. We stress that, as shown in Table \ref{table:chi2_preliminary}, while the EDE model brings both early and late estimates of $H_0$ into agreement, it {\it does not degrade} the overall fit to the {\it Planck} CMB measurements. 
We note that the changes in $H_0$, $\omega_m$, $n_s$, and $A_s$ leave signatures in the matter power spectrum that can potentially be probed by surveys such as KiDS. These effects can be summarized through the parameter $S_8\equiv\sigma_8(\Omega_m/0.3)^{0.5}$, which is shifted by about 1$\sigma$ upwards from its $\Lambda$CDM value. This slightly increases the so-called ``$S_8$ tension'' (e.g.~\cite{Raveri:2018wln}). For example, the tension with the most recent KiDS cosmic-shear measurement \cite{Hildebrandt:2018yau} increases from 2.3$\sigma$ to 2.5$\sigma$. Note that the Dark Energy Survey finds a larger value of $S_8$ \cite{Abbott:2017wau} which reduces the tension with our best-fit EDE model to $\sim 2 \sigma$. Finally we note that the updated {\em Planck} analysis find a smaller value of $S_8$ which will further reduce this tension. 

It is interesting to note how the small-scale polarization measurements affect constraints to the EDE scenario. We find that CMB temperature power spectrum and large-scale polarization is fairly insensitive to the initial field displacement. Only when one includes the small-scale polarization measurements does the initial field displacement become constrained to take on relatively large values (see Sec.~\ref{sec:tempvspol}). We identified that this preference is due to the fact that at high initial field values, the potential we study flattens. This in turn affects the effective sound-speed of the scalar field around the time it becomes dynamical, making it less than 1 for a broader range of scales \cite{Lin:2019qug}.

The presence of an EDE parameter, $\Theta_i$, that is uncorrelated with any LCDM parameter and yet is well-constrained by CMB polarization data is exactly what we expect to see if we are seeing the effects of new physics. We anticipate that near-future small-scale measurements of the CMB polarization with ACTPol and SPTPol will also have the sensitivity to shed additional light on the EDE scenario. Since the EDE scenario posits a change in the expansion rate over a limited amount of time its effects are relatively localized in scale, leading to changes in the CMB power spectrum for $50 \lesssim \ell \lesssim 1000$ (see Fig.~\ref{fig:pert_comp}). This localization may provide an explanation for the way in which cosmological parameters exhibit a shift when extracted from {\it Planck} data for $\ell < 1000$ and $\ell > 1000$ \cite{Addison:2015wyg,Aghanim:2016sns}. 

The fact that the CMB $\chi^2$ is nearly unchanged whether we fit it with $\Lambda$CDM or an EDE cosmology that resolves the Hubble tension (with $f_{\rm EDE}(z_c) >0$ at more than 3$\sigma$-- see Table \ref{table:chi2_preliminary}) clearly indicates that there is a significant degeneracy between $\Lambda$CDM and the EDE cosmology in {\em Planck} data. However, with the addition of SH0ES data, the $\chi^2$-degeneracy is broken and the sampler is forced to live in the region with (relatively) high $f_{\rm EDE}(z_c)$, uncovering this degeneracy. It is reassuring that this behavior is also seen with synthetic {\em Planck} data that contains an EDE signal.

While {\em Planck} data alone do not allow a detection of the EDE, we have shown that future CMB experiments such as CMB-S4 will be able to identify the presence of the EDE at high significance on its own. Additionally, we find that if synthetic $\Lambda$CDM+EDE data is analyzed in the context of $\Lambda$CDM the CMB-inferred value of $H_0$ is biased low and that this bias increases as the noise and angular resolution of the CMB observations decrease. It is interesting to note that this mimics what we find when we compare the $H_0$ analyze WMAP and {\em Planck} data. 

We have discussed two other aspects of the EDE scenario which provide additional predictions.  First, the presence of a spectator scalar field during inflation leads to a spectrum of isocurvature perturbations whose amplitude is controlled by the tensor-to-scalar ratio, $r$, and  the initial field displacement $\Theta_i$. A future measurement of $r$ might therefore set interesting constraints on the scenario proposed here. 

Finally, we have shown that perturbations in the scalar field grow rapidly due to self-resonance for a limited range of wavenumbers. Using a Floquet analysis, we have shown that  $n\simeq 2$ can lead to modes becoming non-linear sometime before today; we confirmed this analysis with {\sf CLASS}. The same analysis indicates that we can safely explore the oscillating EDE scenario at the linear perturbations level for $n\not\approx 2$.\footnote{As long as there is no significant perturbation growth in the ``wings" of the potential} Our analysis should apply to a wider range of scalar field potentials with power law minima and which are flattened at large field displacements \cite{Lozanov:2017hjm,Dong:2010in,Kallosh:2013hoa,Carrasco:2015pla}.

When nonlinear, spatially inhomogeneous dynamics occur, they can provide new signatures of EDE. The sharp scale-dependence of the resonant modes, and ensuing nonlinear dynamics could be searched for in future observations based on their gravitaional effects. For a concrete example of such nonlinear dynamics, see \cite{Lozanov:2017hjm,Lozanov:2016hid,Khlebnikov:1996mc}, where  numerical simulations that consider the full nonlinear dynamics of an energetically dominant field on a lattice (not directly in the context of EDE) were carried out. See the footnote\footnote{In \cite{Lozanov:2017hjm,Lozanov:2016hid}, it was shown  that when the field becomes nonlinear, the equation of state for the scalar field becomes $w_\phi\approx 1/3$, even when $n\ne 2$, as long as $n\not\approx 1$. Note that this differs from the usual $w_\phi=(n-1)/(n+1)$ result for homogeneous field. For $n\approx 2$, $w_\phi\approx 1/3$ is obtained with or without the nonlinear dynamics as expected.  If the shape of the potential and parameters are chosen so that resonance/growth of perturbations mainly takes place due to the flattened ``wings" of the potential ({\it not} the power law bottom), short-lived, spatially-localized, nonlinear structures were shown to form for $n\ne 1$ (``transients" \cite{Lozanov:2017hjm}). For $n=1$, oscillons -- which are long-lived can form \cite{Amin:2011hj}. However, $n=1$ would not provide a successful EDE. Also, see Ref.~\cite{Amin:2011hu} in this context.} below for more details. In general, the rapid nonlinear dynamics in the types of models considered here also lead to the generation of a stochastic background gravitational \cite{Lozanov:2019ylm,Khlebnikov:1997di}, which could provide another additional observational signature/constraint for these models. While the fact that the scalar field is a subdominant source of energy density can hinder some of the above dynamics, and reduce their observational impact, it provides an exciting new avenue to pursue. We will analyze these phenomena in upcoming work.
 
We are living a very exciting moment in cosmology. The tension between late and early determinations of the current rate of expansion, $H_0$, has opened up the possibility that we are seeing hints of new physical processes.  There are only a handful of beyond-$\Lambda$CDM models which can `explain' this discrepancy while providing a good statistical fit to all datasets, of which the EDE scenario is one.

This scenario may fit into a broader picture where the early inflationary epoch, a short EDE period around matter/radiation equality, and the current epoch of accelerated expansion are connected. One possibility is  that there exists a collection of cosmological scalar fields whose parameters (masses and decay constants) are pulled from some distribution, similar to the `axiverse' scenario \cite{Svrcek:2006yi,Arvanitaki:2009fg,Cicoli:2012sz,Stott:2017hvl}.
Variations of such scenarios have been proposed as a possible resolution of the so-called `coincidence problem' \cite{Griest:2002cu,Kamionkowski:2014zda}.
Moreover, the fact that the field reaches its maximum right around matter-radiation equality might provide clue to understanding the nature of the EDE.  As we have shown, the EDE scenario makes unique predictions which are accessible to near-future CMB experiments. 

Future experimental efforts to detect these new signatures will therefore be essential to verify whether an EDE was present in the early universe and have the potential to shed new light on the dark universe.


\begin{acknowledgments}
We thank Marco Raveri for helpful discussions on many aspects of this research and Graeme Addison, Francis-Yan Cyr-Racine, Daniel Grin, and Adam Riess for useful comments on the draft. We thank Thejs Brinckmann for providing help with the use of mock data in {\sc MontePython-v3}. We also thank K.~Lozanov for conversations regarding resonance in power law potentials, and help with Fig.~\ref{fig:rn}. We thank the organizers and participants of the workshop `Tensions Between the Early and the Late Universe' held at the Kavli Institute for Theoretical Physics on July 15-17 2019 where part of this work was presented and interesting comments helped us to improve it. This research used resources of the IN2P3/CNRS and the Dark Energy computing Center funded by the OCEVU Labex (ANR-11-LABX-0060) and the Excellence Initiative of Aix-Marseille University - A*MIDEX, part of the French “Investissements d’Avenir” programme. TLS acknowledges support in part from NASA 80NSSC18K0728 and from the Provost's office at Swarthmore College. TLS and VP thank Johns Hopkins University where part of this work has been completed.  MA is supported by a DOE grant DE-SC0018216. Part of this work by MA was carried out at the Aspen Center for Physics, which is supported by National Science Foundation grant PHY-1607611. MA thanks the Yukawa Institute for Theoretical Physics at Kyoto University; part of this work was carried out during the YITP-T-19-02 workshop on ``Resonant instabilities in cosmology".
\end{acknowledgments}

\begin{appendix}
\section{Numerical implementation}
\label{app:numerical}

To incorporate the dynamics of an oscillating scalar field into {\sf CLASS} we obtained approximate analytic expressions for various quantities. 

In order to search on the observable parameters $z_c$ and $f_{\rm EDE}(z_c)$ we must numerically solve for the corresponding model parameters $m$ and $f$ given some initial field displacement $\Theta_i = \phi_i/f$. We do that using a shooting method that requires an initial `first guess' for these parameters. We can determine an approximate first guess by solving for the field dynamics while it is in slow-roll and we find the following (approximate) equations:
\begin{widetext}
For $z_c>z_{\rm eq}$
\begin{eqnarray}
z_c &\simeq& C\left[\frac{20(1-F)\Theta_i \Omega_{r,0}(1-\cos\Theta_i)^{-n} \tan \Theta_i/2}{n\mu^2}\right]^{-1/4},\\
f_{\rm EDE}(z_c) &\simeq& \frac{4(1-F)\alpha^2 \Theta_i (1-\cos \Theta_i)^{-n}}{3n}\left[5(1-\cos F\Theta_i)^n +2(1-F) n \Theta_i(1-\cos \Theta_i)^n \cot\Theta_i/2\right]\tan\Theta_i/2;
\end{eqnarray}
for $z_c < z_{\rm eq}$
\begin{eqnarray}
z_c &\simeq&C\left[\frac{27(1-F) \Theta_i \Omega_{M,0} (1-\cos \Theta_i)^{-n} \tan \Theta_i/2}{2 n \mu^2}\right]^{-1/3},\\
f_{\rm EDE}(z_c) &\simeq& \frac{3(1-F) \alpha^2 \Theta_i (1-\cos \Theta_i)^{-n}}{2 n} \left[3(1-\cos F \Theta_i)^n +(1-F) n \Theta_i (1-\cos \Theta_i)^n \cot \Theta_i/2\right] \tan \Theta_i/2,
\end{eqnarray}
where $\mu \equiv m/H_0$, $\alpha \equiv f/M_{\rm pl}$, $C = 0.6$, and $F=0.8$. 
\end{widetext}
We have verified that these expressions are accurate enough to provide a first guess when shooting for the mass, $m$, and decay constant, $f$, given $z_c$ and $f_{\rm EDE}(z_c)$. 

Given $n$, $z_c$ and $f_{\rm EDE}(z_c)$, and $\Theta_i$ we can use the above equations to approximately solve for the corresponding model parameters $m$ and $f$ as a first guess. The shooting method then uses a Newton-Cotes rule to iteratively find more exact model parameters.  

The oscillations in the scalar field introduce a time-scale into the problem which is not present in the standard cosmological model. We therefore need to ensure that the time-steps used in the numerical solution are smaller than the oscillation period. We derive an approximate expression for the oscillation period following the steps outlined in Refs.~\cite{Johnson:2008se,Poulin:2018dzj} and find that the cosmic-time period is 
\begin{equation}
T_{\rm osc}(a) \simeq \frac{\Gamma[1+1/(2n)]}{m_a\Gamma[(1+n)/(2n)]}2^{2+(n-1)/2} \sqrt{\pi} \left[\frac{\phi_{\rm env}(a)}{f}\right]^{1-n},
\end{equation}
where $\phi_{\rm env}(a)$ is given in Eq.~(\ref{eq:phi_env}). 
To ensure that the time-step resolves these oscillations when computing the effects of the oscillating scalar field, we require that $\Delta t < T_{\rm osc}(a)/100$. 

\section{Adiabatic initial conditions}
\label{app:ad}

In this Section we derive and verify analytic expressions for the scalar field adiabatic initial conditions.  

The perturbations evolve according to the linearized Klein-Gordon (KG) equation,
\begin{equation}
    \delta\phi''_k + 2 H \delta \phi_k' + \left[k^2 + a^2 V_{,\phi\phi}\right] \delta \phi_k = - h'  \phi'/2,\label{eq:linKG2}
\end{equation}
where the prime denotes derivatives with respect to conformal time, we have written the metric potential in synchronous gauge (see, e.g., Ref.~\cite{Ma:1995ey}) and we can see that the perturbations evolve as driven damped harmonic oscillators. It is also possible to write these equations of motion in terms of two coupled first order differential equations. In this form, this second order equation of motion is equivalent to the conservation of the linearly perturbed scalar field stress-energy:
\begin{eqnarray}
\rho_\phi &=& \frac{1}{2} a^{-2} \phi'^2 + V,\\
p_\phi &=& \frac{1}{2} a^{-2} \phi'^2 - V,\\
\delta \rho_\phi &=& a^{-2}( \phi' \delta  \phi'+V_{,\phi}\delta \phi),\\ 
\delta p_\phi &=& \delta \rho_\phi - 2 V_{,\phi} \delta \phi,\\
(\rho_\phi + p_\phi) \theta_\phi  &=& k^2 a^2  \phi' \delta \phi,\\
p_\phi \sigma_\phi &=& 0,
\end{eqnarray}
where in the last line we have explicitly noted that the scalar field does not produce any anisotropic stress. From this it is straightforward to show that the conservation of the linearly perturbed scalar field stress energy follows that of a `generalized fluid' \cite{Hu:1998kj} with an effective sound-speed equal to unity:
\begin{eqnarray}
\delta^\prime_\phi &=& - (1+w_\phi)\left(\theta_\phi + \frac{1}{2} h'\right)- 6\mathcal{H} \delta_\phi \label{eq:deltaphi}\\ &-& 9(1-c_\phi^2)(1+w_\phi) \mathcal{H}^2 \frac{\theta_\phi}{k^2},\nonumber \\
\theta^\prime_\phi &=& 2 \mathcal{H} \theta_\phi +  \frac{\delta_\phi}{1+w_\phi}, \label{eq:deltatheta}
\end{eqnarray}
where $u_\phi \equiv (1+w_\phi) \theta_\phi$, the prime denotes a derivative with respect to conformal time, $\mathcal{H} \equiv a'/a$, $w_\phi \equiv p_\phi/\rho_\phi$ and $c_\phi^2$ is the scalar-field `adiabatic sound speed' given by 
\begin{equation}
    c_\phi^2 \equiv \frac{\dot{p}_\phi}{\dot{\rho}_\phi} = 1+\frac{2}{3} a^2 \frac{V_{,\phi}}{\mathcal{H}^2 \phi'}.
\end{equation}
Note that even though the conservation of scalar field stress-energy [Eqs~(\ref{eq:deltaphi}) and (\ref{eq:deltatheta})] is mathematically equivalent to the linearized KG equation [Eq.~(\ref{eq:linKG2})] it is not as useful when seeking numerical solutions with an oscillating scalar field. It is simple to see this: once the scalar field is oscillating its adiabatic sound speed becomes infinite every time the field velocity goes to zero. This formal infinity does not affect the full equations of motion because at the same time $\theta_\phi \propto \phi'$ also vanishes. However this behavior makes the fluid equations numerically unstable for an oscillating scalar field. On the other hand in the limit that the field is monotonically evolving (such as when it is in slow-roll) the fluid form of the equations of motion can be used. 

The RHS of Eq.~\ref{eq:linKG2} implies that the inhomogeneous solution will be sourced by the superhorizon gravitational potential, $h(\vec k) = \zeta_{\rm ad}(\vec k) k^2 \tau^2$, and the slow-roll field `velocity' $\phi ' \simeq - \frac{1}{5} H_0^2 V_{,\phi} \tau^3 \Omega_{\rm rad}$, where $\Omega_{\rm rad} h^2=4.15 \times 10^{-5}$ for photons (with a temperature of $\sim 2.7$ K today) plus three standard ultra-relativistic neutrinos. 
In this limit, it is easiest to solve for the evolution of the fluid variables, where the scalar field adiabatic sound speed is approximately given by $c_\phi^2 \simeq -7/3$ \cite{Hlozek:2014lca} and the equation of state of the background field evolves as 
\begin{equation}
    1+w_\phi \simeq \frac{H_0^2 V'^2 \tau^4 \Omega_{\rm rad}}{25 V}.
\end{equation}
We find that fluid variables evolve to leading order in $k\tau$ as
\begin{eqnarray}
\delta_\phi(\vec k,\tau)&\simeq& - \zeta_{\rm ad}(\vec k)  \frac{H_0^2 V_{,\phi}^2 \Omega_{\rm rad}}{1050 k^4 V}(k\tau)^6,\label{eq:SHdelta}\\
\theta_\phi(\vec k, \tau) &\simeq& -\zeta_{\rm ad}(\vec k) \frac{k }{42}(k\tau)^3,\label{eq:SHtheta}
\end{eqnarray}
where the potential and its derivative are evaluated at the initial field value $\phi_i$. 

\begin{figure}[h!]
    \centering
    \includegraphics[scale=0.55]{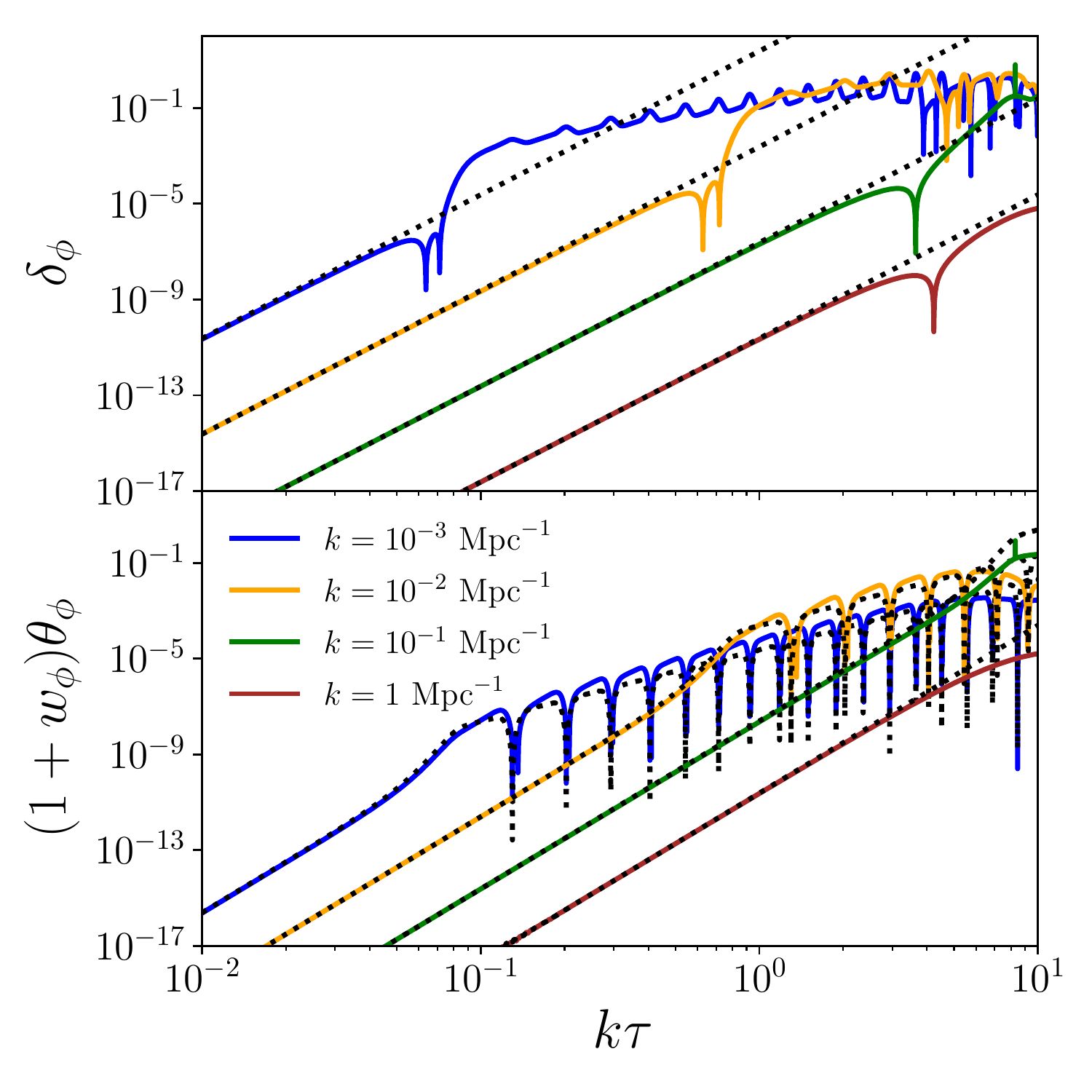}
    \caption{Analytic and numerical evolution of several adiabatic modes establishing the accuracy of the analytic set of initial conditions derived in the text.}
    \label{fig:pert_comp}
\end{figure}
We compare our super-horizon analytic adiabatic solutions in Eqs.~(\ref{eq:SHdelta}) and (\ref{eq:SHtheta}) to the output of our numerical code in Fig.~\ref{fig:pert_comp}. We can see that for small-scale modes (which enter the horizon before the background field begins to oscillate) these solutions are good approximations up until horizon entry ($k \tau \simeq 1$). For larger-scale modes the background field starts to oscillate before horizon entry and those oscillations provide a modulation of both the density and velocity perturbations. The initial conditions for adiabatic perturbations given in Eqns.~(\ref{eq:SHdelta}) and (\ref{eq:SHtheta}) also appear (in a less explicit form) in Ref.~\cite{Ballesteros:2010ks}. 

The agreement indicates that the code is solving the relevant equations correctly. Our analytic and numerical results show that there is no tachyonic instability for the inhomogeneous solution due to the presence of a driving term (and corresponding to adiabatic initial conditions). As discussed in Sec.~\ref{sec:iso}, the tachyonic instability may be present for the homogeneous solution (i.e., isocurvature initial conditions) while the background field is in a part of the potential where $V_{n, \phi \phi}<0$ (i.e. for a relatively large field displacement). 

\section{Parametric resonance}
\label{app:parametric_resonance}
We have three goals for this Appendix. First, for the $V_n(\phi)$ under consideration, we want to provide approximate analytic expressions for the growth rate of perturbations (captured by a scale-dependent integral of the Floquet exponent). We also wish to provide Floquet instability charts for two sample cases, $n=2.5$ and $n=2$, and discuss the special case with $n=2$ in more detail both analytically and from the point of view of observational constraints. 

\subsection{Analytic approximations, general $n$.}
A detailed instability analysis of parametric resonance in power law potentials $V_n\propto \phi^{2n}$ in an expanding universe was carried out in Ref.~\cite{Lozanov:2017hjm}\footnote{The calculation there also includes field displacements in the flattened part of the potential away  from the power law regime}. In that work, the Floquet exponents as a function of wavenumber and amplitude were provided for different $n$. We quote the main results necessary here without re-deriving them. 

\begin{figure}[t!]
    \centering
    \includegraphics[scale=0.7]{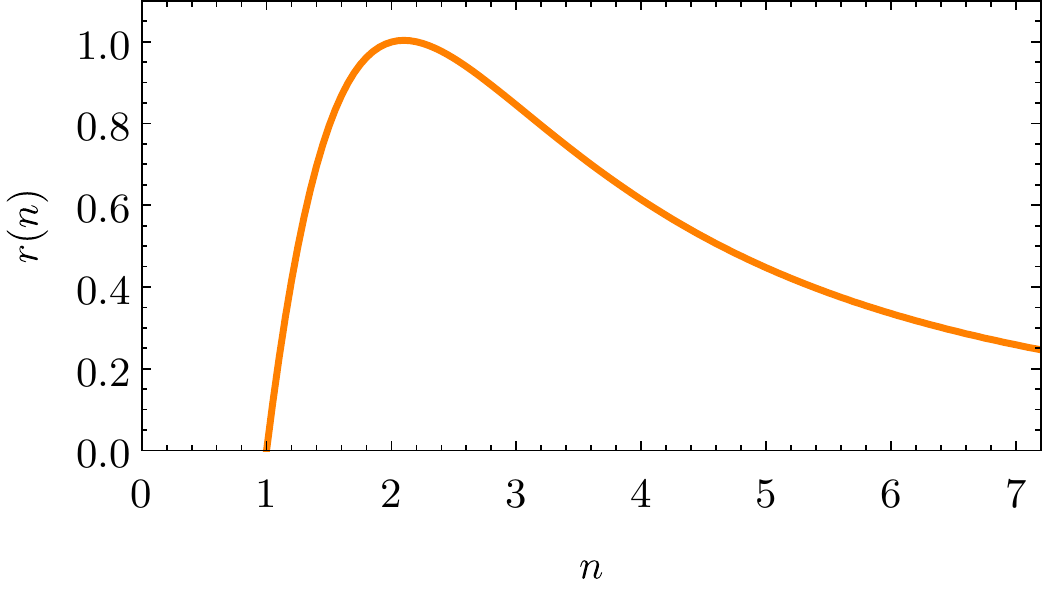}
    \caption{The essential features of the maximum Floquet exponent characterizing the growth rate of field perturbations for $V(\phi)\propto \phi^{2n}$, are captured by $r(n)$ shown above. For details, see the text and Fig. 4 of \cite{Lozanov:2017hjm}.}
    \label{fig:rn}
\end{figure}

From Fig.~3 of Ref.~\cite{Lozanov:2017hjm}, the maximal Floquet exponent for the first and most dominant, narrow instability band at small field oscillation amplitudes is given by\footnote{Note that $m_{\rm eff}$ is denoted by $m$ in \cite{Lozanov:2017hjm}. In the present paper $m$ is a constant, wheres in \cite{Lozanov:2017hjm} $m\rightarrow m_{\rm eff}$ was field dependent.}
\begin{equation}
\label{eq:mu}
\frac{\Re[\mu_k]_{\rm max}}{(m_{\rm eff}/ \sqrt{2n})}\approx 0.072 \times r(n)\,,\quad \textrm{with} \quad m_{\rm eff}^2\equiv V_{n,\phi}/\phi,
\end{equation} 
and $r(n)$ is such that $r(2)=1>r(n\ne 2)$. For detailed shape of $r(n)$ see  Fig. \ref{fig:rn} (reproduced from the top panel of Fig. 4 in \cite{Lozanov:2017hjm}). Similarly, again using Fig.~3 of Ref.~\cite{Lozanov:2017hjm}, the resonant wavenumber and the width of the resonant band is given by
\begin{equation}\
\begin{aligned}
\label{eq:kappa}
&\kappa \sqrt{2n}\approx 2.54\,,\qquad\textrm{and}\qquad\frac{\Delta \kappa}{\kappa}\approx 0.072\times r(n)\,,\\
&{\rm where}\qquad \kappa = \frac{k}{a\,m_{\rm eff}}\,.
\end{aligned}
\end{equation}
As mentioned in the main text, we reiterate that these results should be interpreted within the assumption that the expansion time-scale is slow compared to the oscillatory time scales in the equations. 

Translating these results to our parameters, we have
\begin{equation}
\begin{aligned}
\label{eq:kres_gen_app}
\frac{k_{\rm res}}{a}
&\approx m \left[\frac{\phi_{\rm env}(a)}{\sqrt{2}f}\right]^{n-1}\frac{2.54}{\sqrt{2}}\,,\\
\Re[\mu_k]_{\rm max}
&\approx  m\left[\frac{\phi_{\rm env}(a)}{\sqrt{2}f}\right]^{n-1} \frac{0.072}{\sqrt{2}}\times r(n)\,,\\
\end{aligned}
\end{equation}
where $\phi_{\rm env}(a)$
is the envelope of the background field after it has started to oscillate $a=a_c$ and is well-approximated by Eq.~(\ref{eq:phi_env}).
If $n>2$, then smaller co-moving wavenumbers get excited later and if $n<2$ the opposite is true (see Fig.~2 in \cite{Lozanov:2017hjm}). Note that for $n=2$, the above equations reduce to $k_{\rm res}\approx  1.27 m(\phi_c/f)a_c$, and $\Re[\mu_k]_{\rm max}\approx 0.036 m (\phi_c/f)(a_c/a)$, consistent with our analysis of the $n=2$ case presented in Appendix \ref{sec:n2}. 

We approximately identify the start of the oscillations when $V_{n,\phi \phi}(\phi_c)=9H^2(a_c)$, which yields
\begin{equation}
H(a_c)=\frac{m}{3}\sqrt{n(2n-1)}\left(\frac{\phi_c}{\sqrt{2}f}\right)^{n-1}\,.
\end{equation}
On the other hand, $H(a)=H_0\sqrt{\Omega_m}a^{-2}\sqrt{a+a_{\rm eq}}$ where we have ignored the energy density in the scalar field and late-time dark energy. Hence
\begin{equation}
\begin{aligned}
H(a)
&=\frac{m}{3}\sqrt{n(2n-1)}\left(\frac{\phi_c}{\sqrt{2}f}\right)^{n-1}\left(\frac{a_c}{a}\right)^{3/2}\sqrt{\frac{1+a_{\rm eq}/a}{1+a_{\rm eq}/a_c}}\,.
\end{aligned}
\end{equation}

The ratio relevant for the growth of perturbations
\begin{equation}
\label{eq:GrowthRatio}
\frac{\Re[\mu_k]_{\rm max}}{H}\approx \frac{3^3}{5^3}\sqrt{\frac{r^2(n)}{2n(2n-1)}}\left(\frac{a_c}{a}\right)^{\frac{3(n-3)}{2(n+1)}}\sqrt{\frac{1+a_{\rm eq}/a_c}{1+a_{\rm eq}/a}}\,.
\end{equation}
where we used Eq.~\eqref{eq:kres_gen_app} and Eq.~\eqref{eq:phi_env}. Repeating some of the analysis in Section~\ref{sec:res}, the evolution of the perturbations is given by 
\begin{equation}
\begin{aligned}
\label{eq:apertgrowth}
&k^{3/2}\delta\phi_k(a)\\
&\sim k^{3/2}\delta\phi_k(a_c)\left(\frac{a_c}{a}\right)^{\frac{3}{n+1}} \exp\left[{\int_{\Delta \ln a} \frac{\Re[\mu_k]}{H}d\ln b}\right]\,,
\end{aligned}
\end{equation}
where $\Delta \ln a(k)$ is the interval spent by the $k$ mode in the resonance band. Note that the exponent in square brackets is simply $\int \Re[\mu_k]dt$. The scaling with $a$ in front represents the approximate redshifting of the mode amplitudes without resonance.

\begin{figure}[t!]
    \centering
    \includegraphics[scale=0.8]{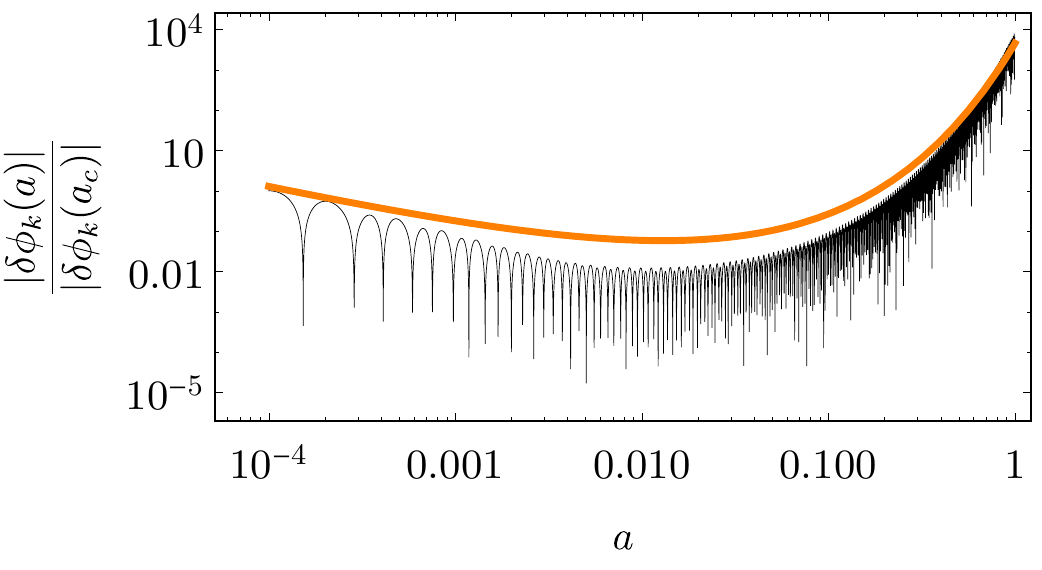}
    \caption{The evolution of the perturbation as a function of scalefactor for $n=2$, $k=k_{\rm res}\approx 1.27a_c(\phi_c/f)m$. The growth due to self-resonance is evident. The orange line is the analytic estimate using Eq. \eqref{eq:pertgrowth} and Eq.~\eqref{eq:growthn2}, the thin black line is obtained by numerical evolution. For the above plot we assume $a_{\rm eq}=a_c\approx 10^{-4}$ and $\phi_c/f\lesssim 1$.}
    \label{fig:modegrowthratio}
\end{figure}

For a given wavenumber, $k$, using the definition of $\kappa$ and the width of the instability band in Eq.~\ref{eq:kappa}, we can estimate the time spent in the instability band in terms of the fractional width of instability band as follows:\footnote{We caution that the following are approximate expressions, however, they are very useful to get a qualitative understanding.}
\begin{equation}
\begin{aligned}
\frac{d\ln \kappa}{d\ln a}\approx \frac{|4-2n|}{n+1}\longrightarrow d\ln a &\sim \frac{n+1}{|4-2n|}\frac{\Delta\kappa}{\kappa} \qquad n\not\approx 2\,,\\
& \approx\frac{n+1}{|4-2n|}0.072\times r(n) 
\end{aligned}
\end{equation}
Note that this expression gets a large contribution near $n=2$. While qualitatively this is fine, it not be trusted in detail too close to $n=2$. Integrating over  the interval spent in the band, we have 
\begin{equation}
\begin{aligned}
\label{eq:growthn!2}
\int_{\Delta \ln a} \frac{\Re[\mu_k]_{\rm max}}{H}d\ln b&\sim \frac{3^5}{5^6}\left(\frac{a_c}{a}\right)^{\frac{3(n-3)}{2(n+1)}}\sqrt{\frac{1+a_{\rm eq}/a_c}{1+a_{\rm eq}/a}}\\
&\times \sqrt{\frac{1}{2n(2n-1)}} \frac{n+1}{|4-2n|}r^2(n)\,,
\end{aligned}
\end{equation}
where, since $\Delta \kappa/\kappa \ll 1$, we did not need to integrate; we just replaced the integral over $\Delta \ln a(k)$ by a multiplication of the integrand with $d\ln a(k)$. 

The expression for $n=2$ is different, since if a $k$ mode is inside the resonance band it never leaves. As a result
\begin{equation}
\begin{aligned}
\label{eq:growthn2}
\int_{a_c}^a \frac{\Re[\mu_k]_{\rm max}}{H}d\ln b&\approx \frac{3^2\sqrt{3}}{5^3}\left(\frac{a}{a_c}\right)^{1/2}\sqrt{\frac{1+a_{\rm eq}/a}{1+a_{\rm eq}/a_c}}\\
&\times \left(1+\frac{a_{\rm eq}}{a_c}\right)\,.
\end{aligned}
\end{equation}
where we assumed $a\gg a_{\rm eq}\sim a_c$. A combination of the results in Eq.~\eqref{eq:growthn2} and Eq.~\eqref{eq:growthn!2} were used in Fig.~\ref{fig:growthratio} in the main text.

For numerical evolution, the resolution requirements in $k$ space to capture the resonant modes can be quite stringent. Using Eq.~\eqref{eq:kres_gen_app} and evaluating $k_{\rm res}$ at $a=a_c$ and $a=1$, we obtain that the resonant wavenumbers lie in an interval $\Delta k_{\rm res}\sim 2.54\times 2^{-n/2} a_c(\phi_c/f)^{n-1}[1-(a_c)^{2(n-2)/(n+1)}]m\sim\mathcal{O}[10^{-4}]m$ for $3>n\gtrsim 2$. Hence the $k$ bins should be at least significantly smaller than this value. For $n\approx 2$, $\Delta k_{\rm res}\approx (\sqrt{3/2}-3^{1/3}) a_c(\phi_c/f)m$ (also see the Floquet charts in Fig.~\ref{fig:phi4} and \ref{fig:phi5}).

\subsection{$n=2$ case and Floquet charts}
\label{sec:n2}

We have performed an analysis for the $n=2$ case for two reasons. First, the growth of perturbations due to parametric resonance discussed in Sec.~\ref{sec:res} is strongest in this case.  Second, this case is particularly compelling, given that the field evolves with a potential $V = \lambda \phi^4/4$ around its minimum, which has been well studied.

\begin{figure*}[t!]
    \centering
    \includegraphics[scale=0.5]{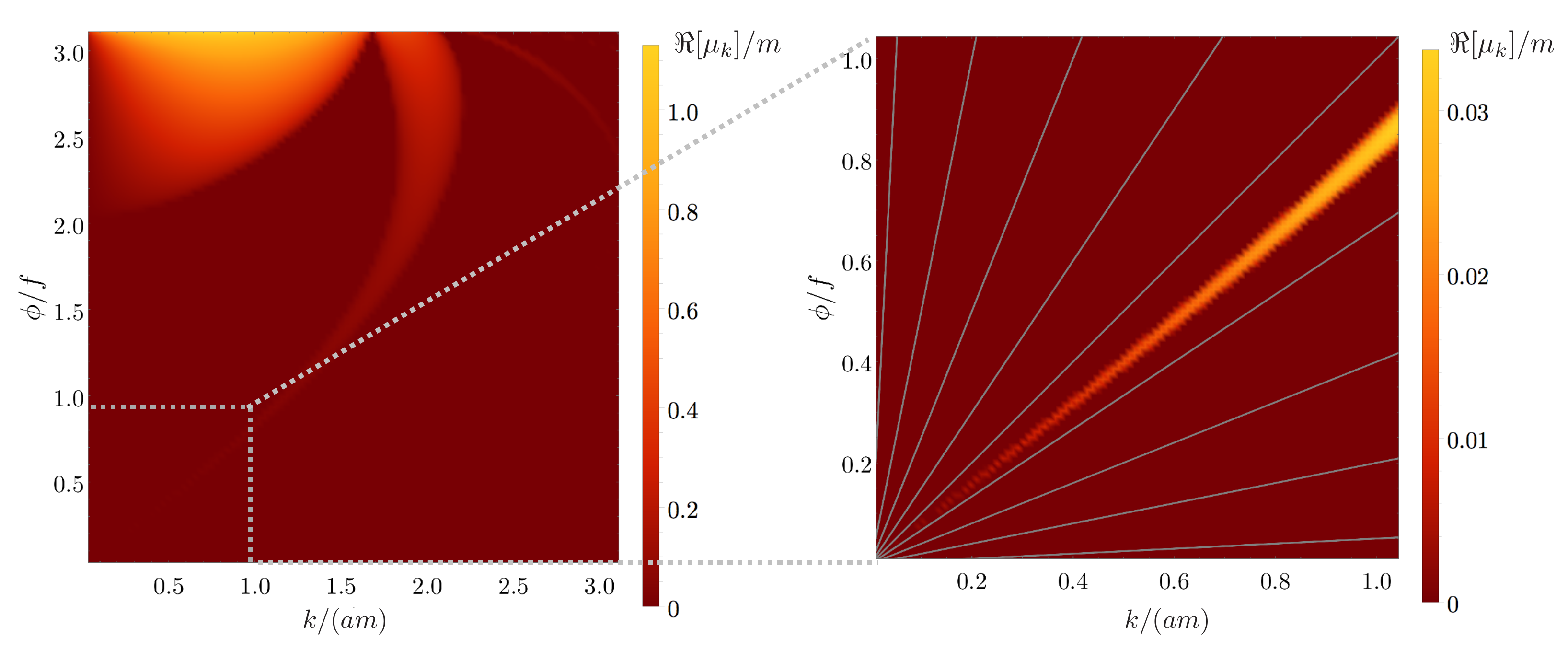}
    \caption{The Floquet chart for $V(\phi)=m^2f^2[1-\cos(\phi/f)]^2$. The left panel shows a broader range of field values and wavenumbers, including the large field amplitude instabity band $\phi/f\gtrsim 1$. The zoom in near the origin is the band structure for $\phi/f\ll 1$, that is for $V(\phi)=(m^2/4f^2)\phi^4$. Note the difference in scale for the Floquet exponent for the two panels. In the right panel we also show ``flow-lines" which indicate how any given co-moving wavenumebr passes through the resonance bands as field amplitude and physical wavenumeber redshift. For $n=2$, the field amplitude and wavenumeber redshift as $1/a$. In the small amplitude regime, once a mode is inside the resonance band, it stays inside, leading to a large amplification of the perturbations. Compare with the case where $n=2.5$ in Fig.~\ref{fig:phi5}.}
    \label{fig:phi4}
\end{figure*}

\begin{figure*}[t!]
    \centering
    \includegraphics[scale=0.5]{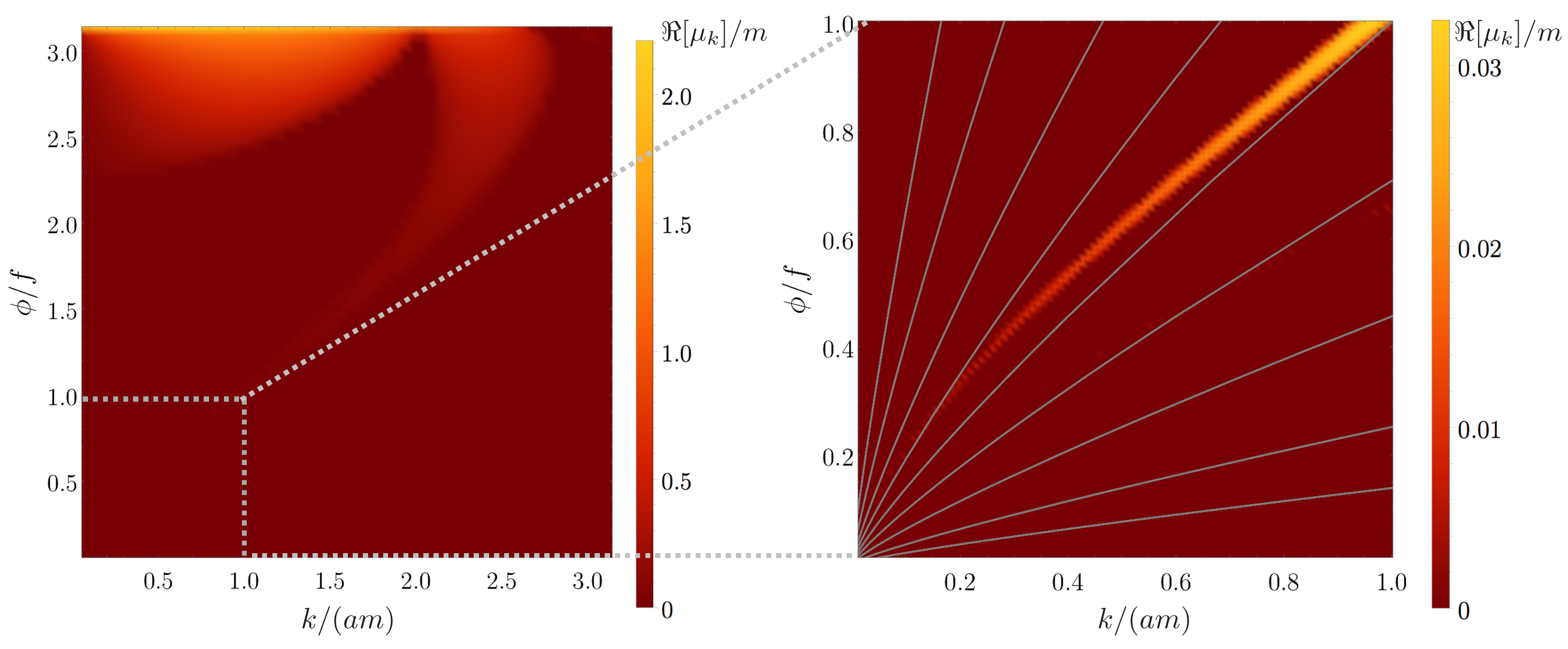}
    \caption{The Floquet chart for $V(\phi)=m^2f^2[1-\cos(\phi/f)]^n$ where $n=2.5$. Compare with the case with $n=2$ in Fig.~\ref{fig:phi4}.  In the right panel we also show ``flow-lines" which indicate how any given co-moving wavenumeber passes through the resonance bands as field amplitude and physical wavenumeber redshift. Unlike the $n=2$ case, the co-moving modes can flow in and out of resonance bands. Typically, the time spent in the resonance band is large at late times.}
    \label{fig:phi5}
\end{figure*}

We start by ignoring expansion and consider $V(\phi)=\lambda\phi^4/4$ where $\lambda=m^2/f^2$. For this potential, we have closed form solutions for the Floquet exponents \cite{Greene:1997fu}:
\begin{equation}
\begin{aligned}
\mu_k
&=\frac{2\sqrt{2}}{9K\left(\frac{1}{\sqrt{2}}\right)}k\sqrt{\left\{\left(\frac{k^2}{\lambda\phi_{\rm env}^2}\right)^{\!2}-\frac{9}{4}\right\}\left\{3-\left(\frac{k^2}{\lambda\phi_{\rm env}^2}\right)^2\right\}}\\
&\times\mathcal{J}\left(\frac{k^2}{\lambda\phi_{\rm env}^2}\right),
\end{aligned}
\end{equation}
with 
\begin{equation}
\mathcal{J}=\int_0^{\pi/2}du \frac{\sin^{2/3}u}{1+\frac{2}{3}\frac{k^2}{\lambda\phi_{\rm env}^2}\sin u+\left(\frac{4}{9}\frac{k^4}{\lambda^2\phi_{\rm env}^4}-1\right)\sin^2u},
\end{equation}
and where the envelope of the oscillating field, $\phi_{\rm env}$, is well-approximated by Eq.~(\ref{eq:phi_env}).
One can check that $\Re[\mu_k]>0$  for $3^{1/4}\sqrt{\lambda}\phi_{\rm env}<k<\sqrt{3/2}\sqrt{\lambda}\phi_{\rm env}$
and 
\begin{equation}
\Re[\mu_k]_{\rm max}\approx0.036\sqrt{\lambda}\phi_{\rm env}\ \ {\textrm{at}}\ \ k_{\rm res} \approx1.27\sqrt{\lambda}\phi_{\rm env}.\label{eq:resk}
\end{equation}
A Floquet diagram which shows $\Re[\mu_k]$ as a function of $k$ and $\phi$ is shown in the right panel of Fig.~\ref{fig:phi4}.

Let us now re-introduce the effect of expansion. In this regard, our $V(\phi)\propto \phi^4$ potential is quite special. In this case, the field $\phi$ redshifts as $\phi_{\rm env}\propto 1/a$, and as always, the physical momentum redshifts as $k/a$. Hence, if a given co-moving wavenumber is in the resonance band at some point, it remains in the resonance band for all times! Contrast this with the case for $n\ne 2$, where a given co-moving wavenumbers moves in and out of the resonance band (see Fig.~\ref{fig:phi4} and \ref{fig:phi5}).

The perturbations will approximately grow as
\begin{equation}
\begin{aligned}
    k^{3/2}\delta\phi_k(a)
   &\sim k^{3/2}\delta\phi_k(a_c)(a_c/a)e^{\int_{a_c}^a\frac{\Re[\mu_k]_{\rm max}}{H} d\ln b}\,,\\
    \end{aligned}
\end{equation}
To estimate the amount of resonant growth we consider the ratio of the maximum Floquet exponent to the Hubble rate (see the expression for general $n$ in Eq. \eqref{eq:GrowthRatio}).

\begin{equation}
\frac{\Re[\mu_k]_{\rm max}}{H}\approx \frac{3^3}{5^3}\sqrt{\frac{1}{12}}\left(\frac{a}{a_c}\right)^{1/2}\sqrt{\frac{1+a_{\rm eq}/a_c}{1+a_{\rm eq}/a}}\,.
\end{equation}
Integrating the above expression, we have
\begin{equation}
\begin{aligned}
\int_{a_c}^a \frac{\Re[\mu_k]_{\rm max}}{H}d\ln b&\approx \frac{3^2\sqrt{3}}{5^3}\left[\left(\frac{a}{a_c}\right)^{1/2}\sqrt{\frac{1+a_{\rm eq}/a}{1+a_{\rm eq}/a_c}}-1\right]\\
&\times \left(1+\frac{a_{\rm eq}}{a_c}\right)\,.
\end{aligned}
\end{equation}
which at late times is $\sim 10^{-1}(a/a_c)^{1/2}$ (assuming $a_c\sim a_{\rm eq}$). If $a_c\ll a_{\rm eq}$, significant growth is also possible during radiation domination.
As the growth continues, at some point the standard deviation of the perturbations, $k^{3/2}|\delta\phi_k|^2$, will become comparable to the field amplitude,$\phi_{\rm env}$, and linear perturbation theory breaks down. 

\subsection{Current constraints to $n=2$}

\begin{figure*}[t!]
    \centering
    \includegraphics[scale=0.7]{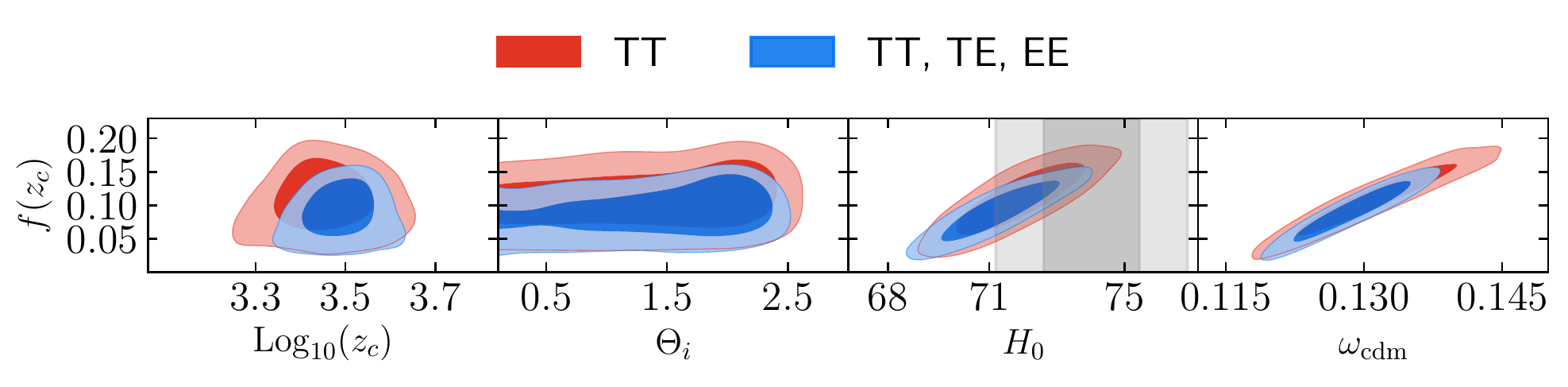}
    \caption{2D posterior distribution of a subset of parameters in the $n=2$ case. We compare the results with and without high-$\ell$ polarization data.}
    \label{fig:n2}
\end{figure*}

We perform the same analysis as in Sec.~\ref{sec:n3} and run a MCMC analysis with flat priors on $\{\omega_b,\omega_{\rm cdm},\theta_s,A_s,n_s,\tau_{\rm reio},f_{\rm EDE}(z_c) ,\log_{10}(z_c),\Theta_i\}$ and setting $n=2$. We include all previously mentioned datasets and compare the use of high-$\ell$ TT and TT,TE,EE data. Our results are reported in Table \ref{table:param_n2} together with the $\Delta\chi^2_{\rm min}$. We show the 2D posterior distributions of $f_{\rm EDE}(z_c)$ vs $\{{\rm Log}_{10}(z_c),\Theta_i,H_0\}$ in Fig.~\ref{fig:n2}. Barring the neglected effects of the non-linearities, our results show that the $n=2$ case can also resolve the Hubble tension. However the $|\Delta\chi^2_{\rm min}|$ is slighly smaller than in the $n=3$ case. This confirms the results of Ref.~\cite{Poulin:2018cxd}. We note one main difference between the $n=2$ and $n=3$ case: in the former case, large values of $\Theta_i$ are excluded. As we discussed in Sec.~\ref{sec:n3}, this is related to the evolution of perturbations in the EDE fluid and in particular the values of the effective sound speed. It is interesting to note that in the case of $n=2$ the preferred perturbation evolution is achieved for an initial field displacement which is only mid-way up the field's potential.  

\begin{table}[htb!]
\scalebox{0.85}{
  \begin{tabular}{|l|c|c|}
   \hline\hline Parameter &~~~$n=2$ (TT)~~~&~~~ $n=2$  (TT,TE,EE)~~~\\ \hline \hline
   $H_0$ &$72.40~(73.87)_{-1.40}^{+1.30}$ & $71.34~(71.63)_{-1.20}^{+1.10}$\\
    $100~\omega_b$ &  $2.219~(2.196)_{-0.039}^{+0.043}$&  $2.252~(2.237)\pm0.02$\\
    $\omega_{\rm cdm}$& $0.1327~(0.1397)\pm0.0061$ & $0.1288~(0.1269)_{-0.0041}^{+0.0044}$ \\
    $10^{9}A_s$&  $2.215~(2.243)\pm0.055$ & $2.215~(2.224)\pm0.013$\\
    $n_s$&$0.9825~(0.9846)\pm0.0076$ & $0.9794~(0.9774)_{-0.0061}^{+0.0064}$\\
    $\tau_{\rm reio}$& $0.072~(0.071)\pm0.015$& $0.075~(0.082)\pm0.013$\\
    $f_{\rm EDE}(z_c)$ & $0.12~(0.17)\pm0.04$ & $0.09~(0.09)_{-0.028}^{+0.032}$ \\
    ${\rm Log}_{10}(z_c)$ & $3.52~(3.51)_{-0.11}^{+0.08}$ &  $3.50~(3.52)\pm0.06$\\
    $\Theta_i$ & $1.80~(2.37)_{-1.80}^{+0.58}$& $1.53~(2.18)_{-0.37}^{+0.84}$\\
    \hline
   $100~\theta_s$ & $1.04117~(1.04063)_{-0.00057}^{+0.00053}$ & $1.04126~(1.04123)\pm0.00040$ \\
    $r_s(z_{\rm rec})$& $137.7~(134.7)_{-2.7}^{+2.4}$ & $139.4~(140.0)\pm2.0$  \\
    $S_8$ &$0.835~(0.843)\pm0.017$ & $0.834~(0.825)\pm0.015$  \\
    \hline
    $\Delta\chi^2_{\rm min}(\Lambda{\rm CDM})$ & -14.7  & -16.0 \\
    \hline
  \end{tabular}}
  \caption{The mean (best-fit) $\pm1\sigma$ error of the cosmological parameters reconstructed from our combined analysis in each model. We also report the $\Delta\chi^2_{\rm min}$ with respect to the best-fit $\Lambda$CDM model of the same combination of datasets. }
 \label{table:param_n2}
\end{table}


\end{appendix}

\let\oldaddcontentsline\addcontentsline
\renewcommand{\addcontentsline}[3]{}
\bibliography{biblio.bib}

\begin{thebibliography}{82}%
\makeatletter
\providecommand \@ifxundefined [1]{%
 \@ifx{#1\undefined}
}%
\providecommand \@ifnum [1]{%
 \ifnum #1\expandafter \@firstoftwo
 \else \expandafter \@secondoftwo
 \fi
}%
\providecommand \@ifx [1]{%
 \ifx #1\expandafter \@firstoftwo
 \else \expandafter \@secondoftwo
 \fi
}%
\providecommand \natexlab [1]{#1}%
\providecommand \enquote  [1]{``#1''}%
\providecommand \bibnamefont  [1]{#1}%
\providecommand \bibfnamefont [1]{#1}%
\providecommand \citenamefont [1]{#1}%
\providecommand \href@noop [0]{\@secondoftwo}%
\providecommand \href [0]{\begingroup \@sanitize@url \@href}%
\providecommand \@href[1]{\@@startlink{#1}\@@href}%
\providecommand \@@href[1]{\endgroup#1\@@endlink}%
\providecommand \@sanitize@url [0]{\catcode `\\12\catcode `\$12\catcode
  `\&12\catcode `\#12\catcode `\^12\catcode `\_12\catcode `\%12\relax}%
\providecommand \@@startlink[1]{}%
\providecommand \@@endlink[0]{}%
\providecommand \url  [0]{\begingroup\@sanitize@url \@url }%
\providecommand \@url [1]{\endgroup\@href {#1}{\urlprefix }}%
\providecommand \urlprefix  [0]{URL }%
\providecommand \Eprint [0]{\href }%
\providecommand \doibase [0]{http://dx.doi.org/}%
\providecommand \selectlanguage [0]{\@gobble}%
\providecommand \bibinfo  [0]{\@secondoftwo}%
\providecommand \bibfield  [0]{\@secondoftwo}%
\providecommand \translation [1]{[#1]}%
\providecommand \BibitemOpen [0]{}%
\providecommand \bibitemStop [0]{}%
\providecommand \bibitemNoStop [0]{.\EOS\space}%
\providecommand \EOS [0]{\spacefactor3000\relax}%
\providecommand \BibitemShut  [1]{\csname bibitem#1\endcsname}%
\let\auto@bib@innerbib\@empty
\bibitem [{\citenamefont {Freedman}(2017)}]{Freedman:2017yms}%
  \BibitemOpen
  \bibfield  {author} {\bibinfo {author} {\bibfnamefont {W.~L.}\ \bibnamefont
  {Freedman}},\ }\href {\doibase 10.1038/s41550-017-0121} {\bibfield  {journal}
  {\bibinfo  {journal} {Nat. Astron.}\ }\textbf {\bibinfo {volume} {1}},\
  \bibinfo {pages} {0121} (\bibinfo {year} {2017})},\ \Eprint
  {http://arxiv.org/abs/1706.02739} {arXiv:1706.02739 [astro-ph.CO]}
  \BibitemShut {NoStop}%
\bibitem [{\citenamefont {Riess}\ \emph {et~al.}(2019)\citenamefont {Riess},
  \citenamefont {Casertano}, \citenamefont {Yuan}, \citenamefont {Macri},\ and\
  \citenamefont {Scolnic}}]{Riess:2019cxk}%
  \BibitemOpen
  \bibfield  {author} {\bibinfo {author} {\bibfnamefont {A.~G.}\ \bibnamefont
  {Riess}}, \bibinfo {author} {\bibfnamefont {S.}~\bibnamefont {Casertano}},
  \bibinfo {author} {\bibfnamefont {W.}~\bibnamefont {Yuan}}, \bibinfo {author}
  {\bibfnamefont {L.~M.}\ \bibnamefont {Macri}}, \ and\ \bibinfo {author}
  {\bibfnamefont {D.}~\bibnamefont {Scolnic}},\ }\href@noop {} {\  (\bibinfo
  {year} {2019})},\ \Eprint {http://arxiv.org/abs/1903.07603} {arXiv:1903.07603
  [astro-ph.CO]} \BibitemShut {NoStop}%
\bibitem [{\citenamefont {Aghanim}\ \emph {et~al.}(2018)\citenamefont {Aghanim}
  \emph {et~al.}}]{Aghanim:2018eyx}%
  \BibitemOpen
  \bibfield  {author} {\bibinfo {author} {\bibfnamefont {N.}~\bibnamefont
  {Aghanim}} \emph {et~al.} (\bibinfo {collaboration} {Planck}),\ }\href@noop
  {} {\  (\bibinfo {year} {2018})},\ \Eprint {http://arxiv.org/abs/1807.06209}
  {arXiv:1807.06209 [astro-ph.CO]} \BibitemShut {NoStop}%
\bibitem [{\citenamefont {Abbott}\ \emph
  {et~al.}(2018{\natexlab{a}})\citenamefont {Abbott} \emph
  {et~al.}}]{Abbott:2017smn}%
  \BibitemOpen
  \bibfield  {author} {\bibinfo {author} {\bibfnamefont {T.~M.~C.}\
  \bibnamefont {Abbott}} \emph {et~al.} (\bibinfo {collaboration} {DES}),\
  }\href {\doibase 10.1093/mnras/sty1939} {\bibfield  {journal} {\bibinfo
  {journal} {Mon. Not. Roy. Astron. Soc.}\ }\textbf {\bibinfo {volume} {480}},\
  \bibinfo {pages} {3879} (\bibinfo {year} {2018}{\natexlab{a}})},\ \Eprint
  {http://arxiv.org/abs/1711.00403} {arXiv:1711.00403 [astro-ph.CO]}
  \BibitemShut {NoStop}%
\bibitem [{\citenamefont {Wong}\ \emph {et~al.}(2019)\citenamefont {Wong} \emph
  {et~al.}}]{Wong:2019kwg}%
  \BibitemOpen
  \bibfield  {author} {\bibinfo {author} {\bibfnamefont {K.~C.}\ \bibnamefont
  {Wong}} \emph {et~al.},\ }\href@noop {} {\  (\bibinfo {year} {2019})},\
  \Eprint {http://arxiv.org/abs/1907.04869} {arXiv:1907.04869 [astro-ph.CO]}
  \BibitemShut {NoStop}%
\bibitem [{\citenamefont {Verde}\ \emph {et~al.}(2019)\citenamefont {Verde},
  \citenamefont {Treu},\ and\ \citenamefont {Riess}}]{Verde:2019ivm}%
  \BibitemOpen
  \bibfield  {author} {\bibinfo {author} {\bibfnamefont {L.}~\bibnamefont
  {Verde}}, \bibinfo {author} {\bibfnamefont {T.}~\bibnamefont {Treu}}, \ and\
  \bibinfo {author} {\bibfnamefont {A.~G.}\ \bibnamefont {Riess}}\ }(\bibinfo
  {year} {2019})\ \Eprint {http://arxiv.org/abs/1907.10625} {arXiv:1907.10625
  [astro-ph.CO]} \BibitemShut {NoStop}%
\bibitem [{\citenamefont {Knox}\ and\ \citenamefont
  {Millea}(2019)}]{Knox:2019rjx}%
  \BibitemOpen
  \bibfield  {author} {\bibinfo {author} {\bibfnamefont {L.}~\bibnamefont
  {Knox}}\ and\ \bibinfo {author} {\bibfnamefont {M.}~\bibnamefont {Millea}},\
  }\href@noop {} {\  (\bibinfo {year} {2019})},\ \Eprint
  {http://arxiv.org/abs/1908.03663} {arXiv:1908.03663 [astro-ph.CO]}
  \BibitemShut {NoStop}%
\bibitem [{\citenamefont {Bernal}\ \emph {et~al.}(2016)\citenamefont {Bernal},
  \citenamefont {Verde},\ and\ \citenamefont {Riess}}]{Bernal:2016gxb}%
  \BibitemOpen
  \bibfield  {author} {\bibinfo {author} {\bibfnamefont {J.~L.}\ \bibnamefont
  {Bernal}}, \bibinfo {author} {\bibfnamefont {L.}~\bibnamefont {Verde}}, \
  and\ \bibinfo {author} {\bibfnamefont {A.~G.}\ \bibnamefont {Riess}},\ }\href
  {\doibase 10.1088/1475-7516/2016/10/019} {\bibfield  {journal} {\bibinfo
  {journal} {JCAP}\ }\textbf {\bibinfo {volume} {1610}},\ \bibinfo {pages}
  {019} (\bibinfo {year} {2016})},\ \Eprint {http://arxiv.org/abs/1607.05617}
  {arXiv:1607.05617 [astro-ph.CO]} \BibitemShut {NoStop}%
\bibitem [{\citenamefont {Poulin}\ \emph
  {et~al.}(2018{\natexlab{a}})\citenamefont {Poulin}, \citenamefont {Boddy},
  \citenamefont {Bird},\ and\ \citenamefont {Kamionkowski}}]{Poulin:2018zxs}%
  \BibitemOpen
  \bibfield  {author} {\bibinfo {author} {\bibfnamefont {V.}~\bibnamefont
  {Poulin}}, \bibinfo {author} {\bibfnamefont {K.~K.}\ \bibnamefont {Boddy}},
  \bibinfo {author} {\bibfnamefont {S.}~\bibnamefont {Bird}}, \ and\ \bibinfo
  {author} {\bibfnamefont {M.}~\bibnamefont {Kamionkowski}},\ }\href {\doibase
  10.1103/PhysRevD.97.123504} {\bibfield  {journal} {\bibinfo  {journal} {Phys.
  Rev.}\ }\textbf {\bibinfo {volume} {D97}},\ \bibinfo {pages} {123504}
  (\bibinfo {year} {2018}{\natexlab{a}})},\ \Eprint
  {http://arxiv.org/abs/1803.02474} {arXiv:1803.02474 [astro-ph.CO]}
  \BibitemShut {NoStop}%
\bibitem [{\citenamefont {Aylor}\ \emph {et~al.}(2018)\citenamefont {Aylor},
  \citenamefont {Joy}, \citenamefont {Knox}, \citenamefont {Millea},
  \citenamefont {Raghunathan},\ and\ \citenamefont {Wu}}]{Aylor:2018drw}%
  \BibitemOpen
  \bibfield  {author} {\bibinfo {author} {\bibfnamefont {K.}~\bibnamefont
  {Aylor}}, \bibinfo {author} {\bibfnamefont {M.}~\bibnamefont {Joy}}, \bibinfo
  {author} {\bibfnamefont {L.}~\bibnamefont {Knox}}, \bibinfo {author}
  {\bibfnamefont {M.}~\bibnamefont {Millea}}, \bibinfo {author} {\bibfnamefont
  {S.}~\bibnamefont {Raghunathan}}, \ and\ \bibinfo {author} {\bibfnamefont
  {W.~L.~K.}\ \bibnamefont {Wu}},\ }\href@noop {} {\  (\bibinfo {year}
  {2018})},\ \Eprint {http://arxiv.org/abs/1811.00537} {arXiv:1811.00537
  [astro-ph.CO]} \BibitemShut {NoStop}%
\bibitem [{\citenamefont {Lin}\ \emph {et~al.}(2018)\citenamefont {Lin},
  \citenamefont {Raveri},\ and\ \citenamefont {Hu}}]{Lin:2018nxe}%
  \BibitemOpen
  \bibfield  {author} {\bibinfo {author} {\bibfnamefont {M.-X.}\ \bibnamefont
  {Lin}}, \bibinfo {author} {\bibfnamefont {M.}~\bibnamefont {Raveri}}, \ and\
  \bibinfo {author} {\bibfnamefont {W.}~\bibnamefont {Hu}},\ }\href@noop {} {\
  (\bibinfo {year} {2018})},\ \Eprint {http://arxiv.org/abs/1810.02333}
  {arXiv:1810.02333 [astro-ph.CO]} \BibitemShut {NoStop}%
\bibitem [{\citenamefont {Poulin}\ \emph {et~al.}(2019)\citenamefont {Poulin},
  \citenamefont {Smith}, \citenamefont {Karwal},\ and\ \citenamefont
  {Kamionkowski}}]{Poulin:2018cxd}%
  \BibitemOpen
  \bibfield  {author} {\bibinfo {author} {\bibfnamefont {V.}~\bibnamefont
  {Poulin}}, \bibinfo {author} {\bibfnamefont {T.~L.}\ \bibnamefont {Smith}},
  \bibinfo {author} {\bibfnamefont {T.}~\bibnamefont {Karwal}}, \ and\ \bibinfo
  {author} {\bibfnamefont {M.}~\bibnamefont {Kamionkowski}},\ }\href {\doibase
  10.1103/PhysRevLett.122.221301} {\bibfield  {journal} {\bibinfo  {journal}
  {Phys. Rev. Lett.}\ }\textbf {\bibinfo {volume} {122}},\ \bibinfo {pages}
  {221301} (\bibinfo {year} {2019})},\ \Eprint
  {http://arxiv.org/abs/1811.04083} {arXiv:1811.04083 [astro-ph.CO]}
  \BibitemShut {NoStop}%
\bibitem [{\citenamefont {Kreisch}\ \emph {et~al.}(2019)\citenamefont
  {Kreisch}, \citenamefont {Cyr-Racine},\ and\ \citenamefont
  {Doré}}]{Kreisch:2019yzn}%
  \BibitemOpen
  \bibfield  {author} {\bibinfo {author} {\bibfnamefont {C.~D.}\ \bibnamefont
  {Kreisch}}, \bibinfo {author} {\bibfnamefont {F.-Y.}\ \bibnamefont
  {Cyr-Racine}}, \ and\ \bibinfo {author} {\bibfnamefont {O.}~\bibnamefont
  {Doré}},\ }\href@noop {} {\  (\bibinfo {year} {2019})},\ \Eprint
  {http://arxiv.org/abs/1902.00534} {arXiv:1902.00534 [astro-ph.CO]}
  \BibitemShut {NoStop}%
\bibitem [{\citenamefont {Lin}\ \emph {et~al.}(2019)\citenamefont {Lin},
  \citenamefont {Benevento}, \citenamefont {Hu},\ and\ \citenamefont
  {Raveri}}]{Lin:2019qug}%
  \BibitemOpen
  \bibfield  {author} {\bibinfo {author} {\bibfnamefont {M.-X.}\ \bibnamefont
  {Lin}}, \bibinfo {author} {\bibfnamefont {G.}~\bibnamefont {Benevento}},
  \bibinfo {author} {\bibfnamefont {W.}~\bibnamefont {Hu}}, \ and\ \bibinfo
  {author} {\bibfnamefont {M.}~\bibnamefont {Raveri}},\ }\href@noop {} {\
  (\bibinfo {year} {2019})},\ \Eprint {http://arxiv.org/abs/1905.12618}
  {arXiv:1905.12618 [astro-ph.CO]} \BibitemShut {NoStop}%
\bibitem [{\citenamefont {Karwal}\ and\ \citenamefont
  {Kamionkowski}(2016)}]{Karwal:2016vyq}%
  \BibitemOpen
  \bibfield  {author} {\bibinfo {author} {\bibfnamefont {T.}~\bibnamefont
  {Karwal}}\ and\ \bibinfo {author} {\bibfnamefont {M.}~\bibnamefont
  {Kamionkowski}},\ }\href {\doibase 10.1103/PhysRevD.94.103523} {\bibfield
  {journal} {\bibinfo  {journal} {Phys. Rev.}\ }\textbf {\bibinfo {volume}
  {D94}},\ \bibinfo {pages} {103523} (\bibinfo {year} {2016})},\ \Eprint
  {http://arxiv.org/abs/1608.01309} {arXiv:1608.01309 [astro-ph.CO]}
  \BibitemShut {NoStop}%
\bibitem [{\citenamefont {Poulin}\ \emph
  {et~al.}(2018{\natexlab{b}})\citenamefont {Poulin}, \citenamefont {Smith},
  \citenamefont {Grin}, \citenamefont {Karwal},\ and\ \citenamefont
  {Kamionkowski}}]{Poulin:2018dzj}%
  \BibitemOpen
  \bibfield  {author} {\bibinfo {author} {\bibfnamefont {V.}~\bibnamefont
  {Poulin}}, \bibinfo {author} {\bibfnamefont {T.~L.}\ \bibnamefont {Smith}},
  \bibinfo {author} {\bibfnamefont {D.}~\bibnamefont {Grin}}, \bibinfo {author}
  {\bibfnamefont {T.}~\bibnamefont {Karwal}}, \ and\ \bibinfo {author}
  {\bibfnamefont {M.}~\bibnamefont {Kamionkowski}},\ }\href {\doibase
  10.1103/PhysRevD.98.083525} {\bibfield  {journal} {\bibinfo  {journal} {Phys.
  Rev.}\ }\textbf {\bibinfo {volume} {D98}},\ \bibinfo {pages} {083525}
  (\bibinfo {year} {2018}{\natexlab{b}})},\ \Eprint
  {http://arxiv.org/abs/1806.10608} {arXiv:1806.10608 [astro-ph.CO]}
  \BibitemShut {NoStop}%
\bibitem [{\citenamefont {Turner}(1983)}]{Turner:1983he}%
  \BibitemOpen
  \bibfield  {author} {\bibinfo {author} {\bibfnamefont {M.~S.}\ \bibnamefont
  {Turner}},\ }\href {\doibase 10.1103/PhysRevD.28.1243} {\bibfield  {journal}
  {\bibinfo  {journal} {Phys. Rev.}\ }\textbf {\bibinfo {volume} {D28}},\
  \bibinfo {pages} {1243} (\bibinfo {year} {1983})}\BibitemShut {NoStop}%
\bibitem [{\citenamefont {Agrawal}\ \emph {et~al.}(2019)\citenamefont
  {Agrawal}, \citenamefont {Cyr-Racine}, \citenamefont {Pinner},\ and\
  \citenamefont {Randall}}]{Agrawal:2019lmo}%
  \BibitemOpen
  \bibfield  {author} {\bibinfo {author} {\bibfnamefont {P.}~\bibnamefont
  {Agrawal}}, \bibinfo {author} {\bibfnamefont {F.-Y.}\ \bibnamefont
  {Cyr-Racine}}, \bibinfo {author} {\bibfnamefont {D.}~\bibnamefont {Pinner}},
  \ and\ \bibinfo {author} {\bibfnamefont {L.}~\bibnamefont {Randall}},\
  }\href@noop {} {\  (\bibinfo {year} {2019})},\ \Eprint
  {http://arxiv.org/abs/1904.01016} {arXiv:1904.01016 [astro-ph.CO]}
  \BibitemShut {NoStop}%
\bibitem [{\citenamefont {Lozanov}\ and\ \citenamefont
  {Amin}(2018)}]{Lozanov:2017hjm}%
  \BibitemOpen
  \bibfield  {author} {\bibinfo {author} {\bibfnamefont {K.~D.}\ \bibnamefont
  {Lozanov}}\ and\ \bibinfo {author} {\bibfnamefont {M.~A.}\ \bibnamefont
  {Amin}},\ }\href {\doibase 10.1103/PhysRevD.97.023533} {\bibfield  {journal}
  {\bibinfo  {journal} {Phys. Rev.}\ }\textbf {\bibinfo {volume} {D97}},\
  \bibinfo {pages} {023533} (\bibinfo {year} {2018})},\ \Eprint
  {http://arxiv.org/abs/1710.06851} {arXiv:1710.06851 [astro-ph.CO]}
  \BibitemShut {NoStop}%
\bibitem [{\citenamefont {Rigault}\ \emph {et~al.}(2015)\citenamefont {Rigault}
  \emph {et~al.}}]{Rigault:2014kaa}%
  \BibitemOpen
  \bibfield  {author} {\bibinfo {author} {\bibfnamefont {M.}~\bibnamefont
  {Rigault}} \emph {et~al.},\ }\href {\doibase 10.1088/0004-637X/802/1/20}
  {\bibfield  {journal} {\bibinfo  {journal} {Astrophys. J.}\ }\textbf
  {\bibinfo {volume} {802}},\ \bibinfo {pages} {20} (\bibinfo {year} {2015})},\
  \Eprint {http://arxiv.org/abs/1412.6501} {arXiv:1412.6501 [astro-ph.CO]}
  \BibitemShut {NoStop}%
\bibitem [{\citenamefont {Rigault}\ \emph {et~al.}(2018)\citenamefont {Rigault}
  \emph {et~al.}}]{Rigault:2018ffm}%
  \BibitemOpen
  \bibfield  {author} {\bibinfo {author} {\bibfnamefont {M.}~\bibnamefont
  {Rigault}} \emph {et~al.} (\bibinfo {collaboration} {Nearby Supernova
  Factory}),\ }\href@noop {} {\  (\bibinfo {year} {2018})},\ \Eprint
  {http://arxiv.org/abs/1806.03849} {arXiv:1806.03849 [astro-ph.CO]}
  \BibitemShut {NoStop}%
\bibitem [{\citenamefont {Jones}\ \emph {et~al.}(2018)\citenamefont {Jones}
  \emph {et~al.}}]{Jones:2018vbn}%
  \BibitemOpen
  \bibfield  {author} {\bibinfo {author} {\bibfnamefont {D.~O.}\ \bibnamefont
  {Jones}} \emph {et~al.},\ }\href {\doibase 10.3847/1538-4357/aae2b9}
  {\bibfield  {journal} {\bibinfo  {journal} {Astrophys. J.}\ }\textbf
  {\bibinfo {volume} {867}},\ \bibinfo {pages} {108} (\bibinfo {year}
  {2018})},\ \Eprint {http://arxiv.org/abs/1805.05911} {arXiv:1805.05911
  [astro-ph.CO]} \BibitemShut {NoStop}%
\bibitem [{\citenamefont {Rose}\ \emph {et~al.}(2019)\citenamefont {Rose},
  \citenamefont {Garnavich},\ and\ \citenamefont {Berg}}]{Rose:2019ncv}%
  \BibitemOpen
  \bibfield  {author} {\bibinfo {author} {\bibfnamefont {B.~M.}\ \bibnamefont
  {Rose}}, \bibinfo {author} {\bibfnamefont {P.~M.}\ \bibnamefont {Garnavich}},
  \ and\ \bibinfo {author} {\bibfnamefont {M.~A.}\ \bibnamefont {Berg}},\
  }\href {\doibase 10.3847/1538-4357/ab0704} {\bibfield  {journal} {\bibinfo
  {journal} {Astrophys. J.}\ }\textbf {\bibinfo {volume} {874}},\ \bibinfo
  {pages} {32} (\bibinfo {year} {2019})},\ \Eprint
  {http://arxiv.org/abs/1902.01433} {arXiv:1902.01433 [astro-ph.CO]}
  \BibitemShut {NoStop}%
\bibitem [{\citenamefont {Yuan}\ \emph {et~al.}(2019)\citenamefont {Yuan},
  \citenamefont {Riess}, \citenamefont {Macri}, \citenamefont {Casertano},\
  and\ \citenamefont {Scolnic}}]{Yuan:2019npk}%
  \BibitemOpen
  \bibfield  {author} {\bibinfo {author} {\bibfnamefont {W.}~\bibnamefont
  {Yuan}}, \bibinfo {author} {\bibfnamefont {A.~G.}\ \bibnamefont {Riess}},
  \bibinfo {author} {\bibfnamefont {L.~M.}\ \bibnamefont {Macri}}, \bibinfo
  {author} {\bibfnamefont {S.}~\bibnamefont {Casertano}}, \ and\ \bibinfo
  {author} {\bibfnamefont {D.}~\bibnamefont {Scolnic}},\ }\href@noop {} {\
  (\bibinfo {year} {2019})},\ \Eprint {http://arxiv.org/abs/1908.00993}
  {arXiv:1908.00993 [astro-ph.GA]} \BibitemShut {NoStop}%
\bibitem [{\citenamefont {Taubenberger}\ \emph {et~al.}(2019)\citenamefont
  {Taubenberger}, \citenamefont {Suyu}, \citenamefont {Komatsu}, \citenamefont
  {Jee}, \citenamefont {Birrer}, \citenamefont {Bonvin}, \citenamefont
  {Courbin}, \citenamefont {Rusu}, \citenamefont {Shajib},\ and\ \citenamefont
  {Wong}}]{Taubenberger:2019qna}%
  \BibitemOpen
  \bibfield  {author} {\bibinfo {author} {\bibfnamefont {S.}~\bibnamefont
  {Taubenberger}}, \bibinfo {author} {\bibfnamefont {S.~H.}\ \bibnamefont
  {Suyu}}, \bibinfo {author} {\bibfnamefont {E.}~\bibnamefont {Komatsu}},
  \bibinfo {author} {\bibfnamefont {I.}~\bibnamefont {Jee}}, \bibinfo {author}
  {\bibfnamefont {S.}~\bibnamefont {Birrer}}, \bibinfo {author} {\bibfnamefont
  {V.}~\bibnamefont {Bonvin}}, \bibinfo {author} {\bibfnamefont
  {F.}~\bibnamefont {Courbin}}, \bibinfo {author} {\bibfnamefont {C.~E.}\
  \bibnamefont {Rusu}}, \bibinfo {author} {\bibfnamefont {A.~J.}\ \bibnamefont
  {Shajib}}, \ and\ \bibinfo {author} {\bibfnamefont {K.~C.}\ \bibnamefont
  {Wong}},\ }\href@noop {} {\  (\bibinfo {year} {2019})},\ \Eprint
  {http://arxiv.org/abs/1905.12496} {arXiv:1905.12496 [astro-ph.CO]}
  \BibitemShut {NoStop}%
\bibitem [{\citenamefont {Collett}\ \emph {et~al.}(2019)\citenamefont
  {Collett}, \citenamefont {Montanari},\ and\ \citenamefont
  {Rasanen}}]{Collett:2019hrr}%
  \BibitemOpen
  \bibfield  {author} {\bibinfo {author} {\bibfnamefont {T.}~\bibnamefont
  {Collett}}, \bibinfo {author} {\bibfnamefont {F.}~\bibnamefont {Montanari}},
  \ and\ \bibinfo {author} {\bibfnamefont {S.}~\bibnamefont {Rasanen}},\
  }\href@noop {} {\  (\bibinfo {year} {2019})},\ \Eprint
  {http://arxiv.org/abs/1905.09781} {arXiv:1905.09781 [astro-ph.CO]}
  \BibitemShut {NoStop}%
\bibitem [{\citenamefont {Schutz}(1986)}]{Schutz:1986gp}%
  \BibitemOpen
  \bibfield  {author} {\bibinfo {author} {\bibfnamefont {B.~F.}\ \bibnamefont
  {Schutz}},\ }\href {\doibase 10.1038/323310a0} {\bibfield  {journal}
  {\bibinfo  {journal} {Nature}\ }\textbf {\bibinfo {volume} {323}},\ \bibinfo
  {pages} {310} (\bibinfo {year} {1986})}\BibitemShut {NoStop}%
\bibitem [{\citenamefont {Holz}\ and\ \citenamefont
  {Hughes}(2005)}]{Holz:2005df}%
  \BibitemOpen
  \bibfield  {author} {\bibinfo {author} {\bibfnamefont {D.~E.}\ \bibnamefont
  {Holz}}\ and\ \bibinfo {author} {\bibfnamefont {S.~A.}\ \bibnamefont
  {Hughes}},\ }\href {\doibase 10.1086/431341} {\bibfield  {journal} {\bibinfo
  {journal} {Astrophys. J.}\ }\textbf {\bibinfo {volume} {629}},\ \bibinfo
  {pages} {15} (\bibinfo {year} {2005})},\ \Eprint
  {http://arxiv.org/abs/astro-ph/0504616} {arXiv:astro-ph/0504616 [astro-ph]}
  \BibitemShut {NoStop}%
\bibitem [{\citenamefont {Abbott}\ \emph {et~al.}(2017)\citenamefont {Abbott}
  \emph {et~al.}}]{Abbott:2017xzu}%
  \BibitemOpen
  \bibfield  {author} {\bibinfo {author} {\bibfnamefont {B.~P.}\ \bibnamefont
  {Abbott}} \emph {et~al.} (\bibinfo {collaboration} {LIGO Scientific, Virgo,
  1M2H, Dark Energy Camera GW-E, DES, DLT40, Las Cumbres Observatory, VINROUGE,
  MASTER}),\ }\href {\doibase 10.1038/nature24471} {\bibfield  {journal}
  {\bibinfo  {journal} {Nature}\ }\textbf {\bibinfo {volume} {551}},\ \bibinfo
  {pages} {85} (\bibinfo {year} {2017})},\ \Eprint
  {http://arxiv.org/abs/1710.05835} {arXiv:1710.05835 [astro-ph.CO]}
  \BibitemShut {NoStop}%
\bibitem [{\citenamefont {Mortlock}\ \emph {et~al.}(2018)\citenamefont
  {Mortlock}, \citenamefont {Feeney}, \citenamefont {Peiris}, \citenamefont
  {Williamson},\ and\ \citenamefont {Nissanke}}]{Mortlock:2018azx}%
  \BibitemOpen
  \bibfield  {author} {\bibinfo {author} {\bibfnamefont {D.~J.}\ \bibnamefont
  {Mortlock}}, \bibinfo {author} {\bibfnamefont {S.~M.}\ \bibnamefont
  {Feeney}}, \bibinfo {author} {\bibfnamefont {H.~V.}\ \bibnamefont {Peiris}},
  \bibinfo {author} {\bibfnamefont {A.~R.}\ \bibnamefont {Williamson}}, \ and\
  \bibinfo {author} {\bibfnamefont {S.~M.}\ \bibnamefont {Nissanke}},\
  }\href@noop {} {\  (\bibinfo {year} {2018})},\ \Eprint
  {http://arxiv.org/abs/1811.11723} {arXiv:1811.11723 [astro-ph.CO]}
  \BibitemShut {NoStop}%
\bibitem [{\citenamefont {Arvanitaki}\ \emph {et~al.}(2010)\citenamefont
  {Arvanitaki}, \citenamefont {Dimopoulos}, \citenamefont {Dubovsky},
  \citenamefont {Kaloper},\ and\ \citenamefont
  {March-Russell}}]{Arvanitaki:2009fg}%
  \BibitemOpen
  \bibfield  {author} {\bibinfo {author} {\bibfnamefont {A.}~\bibnamefont
  {Arvanitaki}}, \bibinfo {author} {\bibfnamefont {S.}~\bibnamefont
  {Dimopoulos}}, \bibinfo {author} {\bibfnamefont {S.}~\bibnamefont
  {Dubovsky}}, \bibinfo {author} {\bibfnamefont {N.}~\bibnamefont {Kaloper}}, \
  and\ \bibinfo {author} {\bibfnamefont {J.}~\bibnamefont {March-Russell}},\
  }\href {\doibase 10.1103/PhysRevD.81.123530} {\bibfield  {journal} {\bibinfo
  {journal} {Phys. Rev.}\ }\textbf {\bibinfo {volume} {D81}},\ \bibinfo {pages}
  {123530} (\bibinfo {year} {2010})},\ \Eprint {http://arxiv.org/abs/0905.4720}
  {arXiv:0905.4720 [hep-th]} \BibitemShut {NoStop}%
\bibitem [{\citenamefont {Marsh}(2016)}]{Marsh:2015xka}%
  \BibitemOpen
  \bibfield  {author} {\bibinfo {author} {\bibfnamefont {D.~J.~E.}\
  \bibnamefont {Marsh}},\ }\href {\doibase 10.1016/j.physrep.2016.06.005}
  {\bibfield  {journal} {\bibinfo  {journal} {Phys. Rept.}\ }\textbf {\bibinfo
  {volume} {643}},\ \bibinfo {pages} {1} (\bibinfo {year} {2016})},\ \Eprint
  {http://arxiv.org/abs/1510.07633} {arXiv:1510.07633 [astro-ph.CO]}
  \BibitemShut {NoStop}%
\bibitem [{\citenamefont {Kappl}\ \emph {et~al.}(2016)\citenamefont {Kappl},
  \citenamefont {Nilles},\ and\ \citenamefont {Winkler}}]{Kappl:2015esy}%
  \BibitemOpen
  \bibfield  {author} {\bibinfo {author} {\bibfnamefont {R.}~\bibnamefont
  {Kappl}}, \bibinfo {author} {\bibfnamefont {H.~P.}\ \bibnamefont {Nilles}}, \
  and\ \bibinfo {author} {\bibfnamefont {M.~W.}\ \bibnamefont {Winkler}},\
  }\href {\doibase 10.1016/j.physletb.2015.12.073} {\bibfield  {journal}
  {\bibinfo  {journal} {Phys. Lett.}\ }\textbf {\bibinfo {volume} {B753}},\
  \bibinfo {pages} {653} (\bibinfo {year} {2016})},\ \Eprint
  {http://arxiv.org/abs/1511.05560} {arXiv:1511.05560 [hep-th]} \BibitemShut
  {NoStop}%
\bibitem [{\citenamefont {Dong}\ \emph {et~al.}(2011)\citenamefont {Dong},
  \citenamefont {Horn}, \citenamefont {Silverstein},\ and\ \citenamefont
  {Westphal}}]{Dong:2010in}%
  \BibitemOpen
  \bibfield  {author} {\bibinfo {author} {\bibfnamefont {X.}~\bibnamefont
  {Dong}}, \bibinfo {author} {\bibfnamefont {B.}~\bibnamefont {Horn}}, \bibinfo
  {author} {\bibfnamefont {E.}~\bibnamefont {Silverstein}}, \ and\ \bibinfo
  {author} {\bibfnamefont {A.}~\bibnamefont {Westphal}},\ }\href {\doibase
  10.1103/PhysRevD.84.026011} {\bibfield  {journal} {\bibinfo  {journal} {Phys.
  Rev.}\ }\textbf {\bibinfo {volume} {D84}},\ \bibinfo {pages} {026011}
  (\bibinfo {year} {2011})},\ \Eprint {http://arxiv.org/abs/1011.4521}
  {arXiv:1011.4521 [hep-th]} \BibitemShut {NoStop}%
\bibitem [{\citenamefont {Kallosh}\ and\ \citenamefont
  {Linde}(2013)}]{Kallosh:2013hoa}%
  \BibitemOpen
  \bibfield  {author} {\bibinfo {author} {\bibfnamefont {R.}~\bibnamefont
  {Kallosh}}\ and\ \bibinfo {author} {\bibfnamefont {A.}~\bibnamefont
  {Linde}},\ }\href {\doibase 10.1088/1475-7516/2013/07/002} {\bibfield
  {journal} {\bibinfo  {journal} {JCAP}\ }\textbf {\bibinfo {volume} {1307}},\
  \bibinfo {pages} {002} (\bibinfo {year} {2013})},\ \Eprint
  {http://arxiv.org/abs/1306.5220} {arXiv:1306.5220 [hep-th]} \BibitemShut
  {NoStop}%
\bibitem [{\citenamefont {Carrasco}\ \emph {et~al.}(2015)\citenamefont
  {Carrasco}, \citenamefont {Kallosh},\ and\ \citenamefont
  {Linde}}]{Carrasco:2015pla}%
  \BibitemOpen
  \bibfield  {author} {\bibinfo {author} {\bibfnamefont {J.~J.~M.}\
  \bibnamefont {Carrasco}}, \bibinfo {author} {\bibfnamefont {R.}~\bibnamefont
  {Kallosh}}, \ and\ \bibinfo {author} {\bibfnamefont {A.}~\bibnamefont
  {Linde}},\ }\href {\doibase 10.1007/JHEP10(2015)147} {\bibfield  {journal}
  {\bibinfo  {journal} {JHEP}\ }\textbf {\bibinfo {volume} {10}},\ \bibinfo
  {pages} {147} (\bibinfo {year} {2015})},\ \Eprint
  {http://arxiv.org/abs/1506.01708} {arXiv:1506.01708 [hep-th]} \BibitemShut
  {NoStop}%
\bibitem [{\citenamefont {Griest}(2002)}]{Griest:2002cu}%
  \BibitemOpen
  \bibfield  {author} {\bibinfo {author} {\bibfnamefont {K.}~\bibnamefont
  {Griest}},\ }\href {\doibase 10.1103/PhysRevD.66.123501} {\bibfield
  {journal} {\bibinfo  {journal} {Phys. Rev.}\ }\textbf {\bibinfo {volume}
  {D66}},\ \bibinfo {pages} {123501} (\bibinfo {year} {2002})},\ \Eprint
  {http://arxiv.org/abs/astro-ph/0202052} {arXiv:astro-ph/0202052 [astro-ph]}
  \BibitemShut {NoStop}%
\bibitem [{\citenamefont {Marsh}\ and\ \citenamefont
  {Ferreira}(2010)}]{Marsh:2010wq}%
  \BibitemOpen
  \bibfield  {author} {\bibinfo {author} {\bibfnamefont {D.~J.~E.}\
  \bibnamefont {Marsh}}\ and\ \bibinfo {author} {\bibfnamefont {P.~G.}\
  \bibnamefont {Ferreira}},\ }\href {\doibase 10.1103/PhysRevD.82.103528}
  {\bibfield  {journal} {\bibinfo  {journal} {Phys. Rev.}\ }\textbf {\bibinfo
  {volume} {D82}},\ \bibinfo {pages} {103528} (\bibinfo {year} {2010})},\
  \Eprint {http://arxiv.org/abs/1009.3501} {arXiv:1009.3501 [hep-ph]}
  \BibitemShut {NoStop}%
\bibitem [{\citenamefont {Lesgourgues}(2011)}]{Lesgourgues:2011re}%
  \BibitemOpen
  \bibfield  {author} {\bibinfo {author} {\bibfnamefont {J.}~\bibnamefont
  {Lesgourgues}},\ }\href@noop {} {\  (\bibinfo {year} {2011})},\ \Eprint
  {http://arxiv.org/abs/1104.2932} {arXiv:1104.2932 [astro-ph.IM]} \BibitemShut
  {NoStop}%
\bibitem [{\citenamefont {Blas}\ \emph {et~al.}(2011)\citenamefont {Blas},
  \citenamefont {Lesgourgues},\ and\ \citenamefont {Tram}}]{Blas:2011rf}%
  \BibitemOpen
  \bibfield  {author} {\bibinfo {author} {\bibfnamefont {D.}~\bibnamefont
  {Blas}}, \bibinfo {author} {\bibfnamefont {J.}~\bibnamefont {Lesgourgues}}, \
  and\ \bibinfo {author} {\bibfnamefont {T.}~\bibnamefont {Tram}},\ }\href
  {\doibase 10.1088/1475-7516/2011/07/034} {\bibfield  {journal} {\bibinfo
  {journal} {JCAP}\ }\textbf {\bibinfo {volume} {1107}},\ \bibinfo {pages}
  {034} (\bibinfo {year} {2011})},\ \Eprint {http://arxiv.org/abs/1104.2933}
  {arXiv:1104.2933 [astro-ph.CO]} \BibitemShut {NoStop}%
\bibitem [{\citenamefont {Ratra}\ and\ \citenamefont
  {Peebles}(1988)}]{Ratra:1987rm}%
  \BibitemOpen
  \bibfield  {author} {\bibinfo {author} {\bibfnamefont {B.}~\bibnamefont
  {Ratra}}\ and\ \bibinfo {author} {\bibfnamefont {P.~J.~E.}\ \bibnamefont
  {Peebles}},\ }\href {\doibase 10.1103/PhysRevD.37.3406} {\bibfield  {journal}
  {\bibinfo  {journal} {Phys. Rev.}\ }\textbf {\bibinfo {volume} {D37}},\
  \bibinfo {pages} {3406} (\bibinfo {year} {1988})}\BibitemShut {NoStop}%
\bibitem [{\citenamefont {Liddle}\ and\ \citenamefont
  {Scherrer}(1999)}]{Liddle:1998xm}%
  \BibitemOpen
  \bibfield  {author} {\bibinfo {author} {\bibfnamefont {A.~R.}\ \bibnamefont
  {Liddle}}\ and\ \bibinfo {author} {\bibfnamefont {R.~J.}\ \bibnamefont
  {Scherrer}},\ }\href {\doibase 10.1103/PhysRevD.59.023509} {\bibfield
  {journal} {\bibinfo  {journal} {Phys. Rev.}\ }\textbf {\bibinfo {volume}
  {D59}},\ \bibinfo {pages} {023509} (\bibinfo {year} {1999})},\ \Eprint
  {http://arxiv.org/abs/astro-ph/9809272} {arXiv:astro-ph/9809272 [astro-ph]}
  \BibitemShut {NoStop}%
\bibitem [{\citenamefont {Hu}(1998)}]{Hu:1998kj}%
  \BibitemOpen
  \bibfield  {author} {\bibinfo {author} {\bibfnamefont {W.}~\bibnamefont
  {Hu}},\ }\href {\doibase 10.1086/306274} {\bibfield  {journal} {\bibinfo
  {journal} {Astrophys. J.}\ }\textbf {\bibinfo {volume} {506}},\ \bibinfo
  {pages} {485} (\bibinfo {year} {1998})},\ \Eprint
  {http://arxiv.org/abs/astro-ph/9801234} {arXiv:astro-ph/9801234 [astro-ph]}
  \BibitemShut {NoStop}%
\bibitem [{\citenamefont {Ma}\ and\ \citenamefont
  {Bertschinger}(1995)}]{Ma:1995ey}%
  \BibitemOpen
  \bibfield  {author} {\bibinfo {author} {\bibfnamefont {C.-P.}\ \bibnamefont
  {Ma}}\ and\ \bibinfo {author} {\bibfnamefont {E.}~\bibnamefont
  {Bertschinger}},\ }\href {\doibase 10.1086/176550} {\bibfield  {journal}
  {\bibinfo  {journal} {Astrophys. J.}\ }\textbf {\bibinfo {volume} {455}},\
  \bibinfo {pages} {7} (\bibinfo {year} {1995})},\ \Eprint
  {http://arxiv.org/abs/astro-ph/9506072} {arXiv:astro-ph/9506072 [astro-ph]}
  \BibitemShut {NoStop}%
\bibitem [{\citenamefont {Audren}\ \emph {et~al.}(2013)\citenamefont {Audren},
  \citenamefont {Lesgourgues}, \citenamefont {Benabed},\ and\ \citenamefont
  {Prunet}}]{Audren:2012wb}%
  \BibitemOpen
  \bibfield  {author} {\bibinfo {author} {\bibfnamefont {B.}~\bibnamefont
  {Audren}}, \bibinfo {author} {\bibfnamefont {J.}~\bibnamefont {Lesgourgues}},
  \bibinfo {author} {\bibfnamefont {K.}~\bibnamefont {Benabed}}, \ and\
  \bibinfo {author} {\bibfnamefont {S.}~\bibnamefont {Prunet}},\ }\href
  {\doibase 10.1088/1475-7516/2013/02/001} {\bibfield  {journal} {\bibinfo
  {journal} {JCAP}\ }\textbf {\bibinfo {volume} {1302}},\ \bibinfo {pages}
  {001} (\bibinfo {year} {2013})},\ \Eprint {http://arxiv.org/abs/1210.7183}
  {arXiv:1210.7183 [astro-ph.CO]} \BibitemShut {NoStop}%
\bibitem [{\citenamefont {Brinckmann}\ and\ \citenamefont
  {Lesgourgues}(2018)}]{Brinckmann:2018cvx}%
  \BibitemOpen
  \bibfield  {author} {\bibinfo {author} {\bibfnamefont {T.}~\bibnamefont
  {Brinckmann}}\ and\ \bibinfo {author} {\bibfnamefont {J.}~\bibnamefont
  {Lesgourgues}},\ }\href@noop {} {\  (\bibinfo {year} {2018})},\ \Eprint
  {http://arxiv.org/abs/1804.07261} {arXiv:1804.07261 [astro-ph.CO]}
  \BibitemShut {NoStop}%
\bibitem [{\citenamefont {Aguirre}\ \emph {et~al.}(2019)\citenamefont {Aguirre}
  \emph {et~al.}}]{Ade:2018sbj}%
  \BibitemOpen
  \bibfield  {author} {\bibinfo {author} {\bibfnamefont {J.}~\bibnamefont
  {Aguirre}} \emph {et~al.} (\bibinfo {collaboration} {Simons Observatory}),\
  }\href {\doibase 10.1088/1475-7516/2019/02/056} {\bibfield  {journal}
  {\bibinfo  {journal} {JCAP}\ }\textbf {\bibinfo {volume} {1902}},\ \bibinfo
  {pages} {056} (\bibinfo {year} {2019})},\ \Eprint
  {http://arxiv.org/abs/1808.07445} {arXiv:1808.07445 [astro-ph.CO]}
  \BibitemShut {NoStop}%
\bibitem [{\citenamefont {Aghanim}\ \emph {et~al.}(2016)\citenamefont {Aghanim}
  \emph {et~al.}}]{Aghanim:2015xee}%
  \BibitemOpen
  \bibfield  {author} {\bibinfo {author} {\bibfnamefont {N.}~\bibnamefont
  {Aghanim}} \emph {et~al.} (\bibinfo {collaboration} {Planck}),\ }\href
  {\doibase 10.1051/0004-6361/201526926} {\bibfield  {journal} {\bibinfo
  {journal} {Astron. Astrophys.}\ }\textbf {\bibinfo {volume} {594}},\ \bibinfo
  {pages} {A11} (\bibinfo {year} {2016})},\ \Eprint
  {http://arxiv.org/abs/1507.02704} {arXiv:1507.02704 [astro-ph.CO]}
  \BibitemShut {NoStop}%
\bibitem [{\citenamefont {Beutler}\ \emph {et~al.}(2011)\citenamefont
  {Beutler}, \citenamefont {Blake}, \citenamefont {Colless}, \citenamefont
  {Jones}, \citenamefont {Staveley-Smith}, \citenamefont {Campbell},
  \citenamefont {Parker}, \citenamefont {Saunders},\ and\ \citenamefont
  {Watson}}]{Beutler:2011hx}%
  \BibitemOpen
  \bibfield  {author} {\bibinfo {author} {\bibfnamefont {F.}~\bibnamefont
  {Beutler}}, \bibinfo {author} {\bibfnamefont {C.}~\bibnamefont {Blake}},
  \bibinfo {author} {\bibfnamefont {M.}~\bibnamefont {Colless}}, \bibinfo
  {author} {\bibfnamefont {D.~H.}\ \bibnamefont {Jones}}, \bibinfo {author}
  {\bibfnamefont {L.}~\bibnamefont {Staveley-Smith}}, \bibinfo {author}
  {\bibfnamefont {L.}~\bibnamefont {Campbell}}, \bibinfo {author}
  {\bibfnamefont {Q.}~\bibnamefont {Parker}}, \bibinfo {author} {\bibfnamefont
  {W.}~\bibnamefont {Saunders}}, \ and\ \bibinfo {author} {\bibfnamefont
  {F.}~\bibnamefont {Watson}},\ }\href {\doibase
  10.1111/j.1365-2966.2011.19250.x} {\bibfield  {journal} {\bibinfo  {journal}
  {Mon. Not. Roy. Astron. Soc.}\ }\textbf {\bibinfo {volume} {416}},\ \bibinfo
  {pages} {3017} (\bibinfo {year} {2011})},\ \Eprint
  {http://arxiv.org/abs/1106.3366} {arXiv:1106.3366 [astro-ph.CO]} \BibitemShut
  {NoStop}%
\bibitem [{\citenamefont {Ross}\ \emph {et~al.}(2015)\citenamefont {Ross},
  \citenamefont {Samushia}, \citenamefont {Howlett}, \citenamefont {Percival},
  \citenamefont {Burden},\ and\ \citenamefont {Manera}}]{Ross:2014qpa}%
  \BibitemOpen
  \bibfield  {author} {\bibinfo {author} {\bibfnamefont {A.~J.}\ \bibnamefont
  {Ross}}, \bibinfo {author} {\bibfnamefont {L.}~\bibnamefont {Samushia}},
  \bibinfo {author} {\bibfnamefont {C.}~\bibnamefont {Howlett}}, \bibinfo
  {author} {\bibfnamefont {W.~J.}\ \bibnamefont {Percival}}, \bibinfo {author}
  {\bibfnamefont {A.}~\bibnamefont {Burden}}, \ and\ \bibinfo {author}
  {\bibfnamefont {M.}~\bibnamefont {Manera}},\ }\href {\doibase
  10.1093/mnras/stv154} {\bibfield  {journal} {\bibinfo  {journal} {Mon. Not.
  Roy. Astron. Soc.}\ }\textbf {\bibinfo {volume} {449}},\ \bibinfo {pages}
  {835} (\bibinfo {year} {2015})},\ \Eprint {http://arxiv.org/abs/1409.3242}
  {arXiv:1409.3242 [astro-ph.CO]} \BibitemShut {NoStop}%
\bibitem [{\citenamefont {Alam}\ \emph {et~al.}(2017)\citenamefont {Alam} \emph
  {et~al.}}]{Alam:2016hwk}%
  \BibitemOpen
  \bibfield  {author} {\bibinfo {author} {\bibfnamefont {S.}~\bibnamefont
  {Alam}} \emph {et~al.} (\bibinfo {collaboration} {BOSS}),\ }\href {\doibase
  10.1093/mnras/stx721} {\bibfield  {journal} {\bibinfo  {journal} {Mon. Not.
  Roy. Astron. Soc.}\ }\textbf {\bibinfo {volume} {470}},\ \bibinfo {pages}
  {2617} (\bibinfo {year} {2017})},\ \Eprint {http://arxiv.org/abs/1607.03155}
  {arXiv:1607.03155 [astro-ph.CO]} \BibitemShut {NoStop}%
\bibitem [{\citenamefont {Scolnic}\ \emph {et~al.}(2018)\citenamefont {Scolnic}
  \emph {et~al.}}]{Scolnic:2017caz}%
  \BibitemOpen
  \bibfield  {author} {\bibinfo {author} {\bibfnamefont {D.~M.}\ \bibnamefont
  {Scolnic}} \emph {et~al.},\ }\href {\doibase 10.3847/1538-4357/aab9bb}
  {\bibfield  {journal} {\bibinfo  {journal} {Astrophys. J.}\ }\textbf
  {\bibinfo {volume} {859}},\ \bibinfo {pages} {101} (\bibinfo {year}
  {2018})},\ \Eprint {http://arxiv.org/abs/1710.00845} {arXiv:1710.00845
  [astro-ph.CO]} \BibitemShut {NoStop}%
\bibitem [{\citenamefont {Lewis}(2013)}]{Lewis:2013hha}%
  \BibitemOpen
  \bibfield  {author} {\bibinfo {author} {\bibfnamefont {A.}~\bibnamefont
  {Lewis}},\ }\href {\doibase 10.1103/PhysRevD.87.103529} {\bibfield  {journal}
  {\bibinfo  {journal} {Phys. Rev.}\ }\textbf {\bibinfo {volume} {D87}},\
  \bibinfo {pages} {103529} (\bibinfo {year} {2013})},\ \Eprint
  {http://arxiv.org/abs/1304.4473} {arXiv:1304.4473 [astro-ph.CO]} \BibitemShut
  {NoStop}%
\bibitem [{\citenamefont {Gelman}\ and\ \citenamefont
  {Rubin}(1992)}]{Gelman:1992zz}%
  \BibitemOpen
  \bibfield  {author} {\bibinfo {author} {\bibfnamefont {A.}~\bibnamefont
  {Gelman}}\ and\ \bibinfo {author} {\bibfnamefont {D.~B.}\ \bibnamefont
  {Rubin}},\ }\href {\doibase 10.1214/ss/1177011136} {\bibfield  {journal}
  {\bibinfo  {journal} {Statist. Sci.}\ }\textbf {\bibinfo {volume} {7}},\
  \bibinfo {pages} {457} (\bibinfo {year} {1992})}\BibitemShut {NoStop}%
\bibitem [{\citenamefont {Hu}\ \emph {et~al.}(2000)\citenamefont {Hu},
  \citenamefont {Barkana},\ and\ \citenamefont {Gruzinov}}]{Hu:2000ke}%
  \BibitemOpen
  \bibfield  {author} {\bibinfo {author} {\bibfnamefont {W.}~\bibnamefont
  {Hu}}, \bibinfo {author} {\bibfnamefont {R.}~\bibnamefont {Barkana}}, \ and\
  \bibinfo {author} {\bibfnamefont {A.}~\bibnamefont {Gruzinov}},\ }\href
  {\doibase 10.1103/PhysRevLett.85.1158} {\bibfield  {journal} {\bibinfo
  {journal} {Phys. Rev. Lett.}\ }\textbf {\bibinfo {volume} {85}},\ \bibinfo
  {pages} {1158} (\bibinfo {year} {2000})},\ \Eprint
  {http://arxiv.org/abs/astro-ph/0003365} {arXiv:astro-ph/0003365 [astro-ph]}
  \BibitemShut {NoStop}%
\bibitem [{\citenamefont {Hwang}\ and\ \citenamefont
  {Noh}(2009)}]{Hwang:2009js}%
  \BibitemOpen
  \bibfield  {author} {\bibinfo {author} {\bibfnamefont {J.-c.}\ \bibnamefont
  {Hwang}}\ and\ \bibinfo {author} {\bibfnamefont {H.}~\bibnamefont {Noh}},\
  }\href {\doibase 10.1016/j.physletb.2009.08.031} {\bibfield  {journal}
  {\bibinfo  {journal} {Phys. Lett.}\ }\textbf {\bibinfo {volume} {B680}},\
  \bibinfo {pages} {1} (\bibinfo {year} {2009})},\ \Eprint
  {http://arxiv.org/abs/0902.4738} {arXiv:0902.4738 [astro-ph.CO]} \BibitemShut
  {NoStop}%
\bibitem [{\citenamefont {Park}\ \emph {et~al.}(2012)\citenamefont {Park},
  \citenamefont {Hwang},\ and\ \citenamefont {Noh}}]{Park:2012ru}%
  \BibitemOpen
  \bibfield  {author} {\bibinfo {author} {\bibfnamefont {C.-G.}\ \bibnamefont
  {Park}}, \bibinfo {author} {\bibfnamefont {J.-c.}\ \bibnamefont {Hwang}}, \
  and\ \bibinfo {author} {\bibfnamefont {H.}~\bibnamefont {Noh}},\ }\href
  {\doibase 10.1103/PhysRevD.86.083535} {\bibfield  {journal} {\bibinfo
  {journal} {Phys. Rev.}\ }\textbf {\bibinfo {volume} {D86}},\ \bibinfo {pages}
  {083535} (\bibinfo {year} {2012})},\ \Eprint {http://arxiv.org/abs/1207.3124}
  {arXiv:1207.3124 [astro-ph.CO]} \BibitemShut {NoStop}%
\bibitem [{\citenamefont {Hlozek}\ \emph {et~al.}(2015)\citenamefont {Hlozek},
  \citenamefont {Grin}, \citenamefont {Marsh},\ and\ \citenamefont
  {Ferreira}}]{Hlozek:2014lca}%
  \BibitemOpen
  \bibfield  {author} {\bibinfo {author} {\bibfnamefont {R.}~\bibnamefont
  {Hlozek}}, \bibinfo {author} {\bibfnamefont {D.}~\bibnamefont {Grin}},
  \bibinfo {author} {\bibfnamefont {D.~J.~E.}\ \bibnamefont {Marsh}}, \ and\
  \bibinfo {author} {\bibfnamefont {P.~G.}\ \bibnamefont {Ferreira}},\ }\href
  {\doibase 10.1103/PhysRevD.91.103512} {\bibfield  {journal} {\bibinfo
  {journal} {Phys. Rev.}\ }\textbf {\bibinfo {volume} {D91}},\ \bibinfo {pages}
  {103512} (\bibinfo {year} {2015})},\ \Eprint {http://arxiv.org/abs/1410.2896}
  {arXiv:1410.2896 [astro-ph.CO]} \BibitemShut {NoStop}%
\bibitem [{\citenamefont {Noh}\ \emph {et~al.}(2017)\citenamefont {Noh},
  \citenamefont {Hwang},\ and\ \citenamefont {Park}}]{Noh:2017sdj}%
  \BibitemOpen
  \bibfield  {author} {\bibinfo {author} {\bibfnamefont {H.}~\bibnamefont
  {Noh}}, \bibinfo {author} {\bibfnamefont {J.-c.}\ \bibnamefont {Hwang}}, \
  and\ \bibinfo {author} {\bibfnamefont {C.-G.}\ \bibnamefont {Park}},\ }\href
  {\doibase 10.3847/1538-4357/aa8366} {\bibfield  {journal} {\bibinfo
  {journal} {Astrophys. J.}\ }\textbf {\bibinfo {volume} {846}},\ \bibinfo
  {pages} {1} (\bibinfo {year} {2017})},\ \Eprint
  {http://arxiv.org/abs/1707.08568} {arXiv:1707.08568 [gr-qc]} \BibitemShut
  {NoStop}%
\bibitem [{\citenamefont {Johnson}\ and\ \citenamefont
  {Kamionkowski}(2008)}]{Johnson:2008se}%
  \BibitemOpen
  \bibfield  {author} {\bibinfo {author} {\bibfnamefont {M.~C.}\ \bibnamefont
  {Johnson}}\ and\ \bibinfo {author} {\bibfnamefont {M.}~\bibnamefont
  {Kamionkowski}},\ }\href {\doibase 10.1103/PhysRevD.78.063010} {\bibfield
  {journal} {\bibinfo  {journal} {Phys. Rev.}\ }\textbf {\bibinfo {volume}
  {D78}},\ \bibinfo {pages} {063010} (\bibinfo {year} {2008})},\ \Eprint
  {http://arxiv.org/abs/0805.1748} {arXiv:0805.1748 [astro-ph]} \BibitemShut
  {NoStop}%
\bibitem [{\citenamefont {Aghanim}\ \emph {et~al.}(2017)\citenamefont {Aghanim}
  \emph {et~al.}}]{Aghanim:2016sns}%
  \BibitemOpen
  \bibfield  {author} {\bibinfo {author} {\bibfnamefont {N.}~\bibnamefont
  {Aghanim}} \emph {et~al.} (\bibinfo {collaboration} {Planck}),\ }\href
  {\doibase 10.1051/0004-6361/201629504} {\bibfield  {journal} {\bibinfo
  {journal} {Astron. Astrophys.}\ }\textbf {\bibinfo {volume} {607}},\ \bibinfo
  {pages} {A95} (\bibinfo {year} {2017})},\ \Eprint
  {http://arxiv.org/abs/1608.02487} {arXiv:1608.02487 [astro-ph.CO]}
  \BibitemShut {NoStop}%
\bibitem [{\citenamefont {Addison}\ \emph {et~al.}(2016)\citenamefont
  {Addison}, \citenamefont {Huang}, \citenamefont {Watts}, \citenamefont
  {Bennett}, \citenamefont {Halpern}, \citenamefont {Hinshaw},\ and\
  \citenamefont {Weiland}}]{Addison:2015wyg}%
  \BibitemOpen
  \bibfield  {author} {\bibinfo {author} {\bibfnamefont {G.~E.}\ \bibnamefont
  {Addison}}, \bibinfo {author} {\bibfnamefont {Y.}~\bibnamefont {Huang}},
  \bibinfo {author} {\bibfnamefont {D.~J.}\ \bibnamefont {Watts}}, \bibinfo
  {author} {\bibfnamefont {C.~L.}\ \bibnamefont {Bennett}}, \bibinfo {author}
  {\bibfnamefont {M.}~\bibnamefont {Halpern}}, \bibinfo {author} {\bibfnamefont
  {G.}~\bibnamefont {Hinshaw}}, \ and\ \bibinfo {author} {\bibfnamefont
  {J.~L.}\ \bibnamefont {Weiland}},\ }\href {\doibase
  10.3847/0004-637X/818/2/132} {\bibfield  {journal} {\bibinfo  {journal}
  {Astrophys. J.}\ }\textbf {\bibinfo {volume} {818}},\ \bibinfo {pages} {132}
  (\bibinfo {year} {2016})},\ \Eprint {http://arxiv.org/abs/1511.00055}
  {arXiv:1511.00055 [astro-ph.CO]} \BibitemShut {NoStop}%
\bibitem [{\citenamefont {Lyth}\ and\ \citenamefont
  {Wands}(2002)}]{Lyth:2001nq}%
  \BibitemOpen
  \bibfield  {author} {\bibinfo {author} {\bibfnamefont {D.~H.}\ \bibnamefont
  {Lyth}}\ and\ \bibinfo {author} {\bibfnamefont {D.}~\bibnamefont {Wands}},\
  }\href {\doibase 10.1016/S0370-2693(01)01366-1} {\bibfield  {journal}
  {\bibinfo  {journal} {Phys. Lett.}\ }\textbf {\bibinfo {volume} {B524}},\
  \bibinfo {pages} {5} (\bibinfo {year} {2002})},\ \Eprint
  {http://arxiv.org/abs/hep-ph/0110002} {arXiv:hep-ph/0110002 [hep-ph]}
  \BibitemShut {NoStop}%
\bibitem [{\citenamefont {Kobayashi}\ \emph {et~al.}(2013)\citenamefont
  {Kobayashi}, \citenamefont {Takahashi}, \citenamefont {Takahashi},\ and\
  \citenamefont {Yamaguchi}}]{Kobayashi:2013bna}%
  \BibitemOpen
  \bibfield  {author} {\bibinfo {author} {\bibfnamefont {T.}~\bibnamefont
  {Kobayashi}}, \bibinfo {author} {\bibfnamefont {F.}~\bibnamefont
  {Takahashi}}, \bibinfo {author} {\bibfnamefont {T.}~\bibnamefont
  {Takahashi}}, \ and\ \bibinfo {author} {\bibfnamefont {M.}~\bibnamefont
  {Yamaguchi}},\ }\href {\doibase 10.1088/1475-7516/2013/10/042} {\bibfield
  {journal} {\bibinfo  {journal} {JCAP}\ }\textbf {\bibinfo {volume} {1310}},\
  \bibinfo {pages} {042} (\bibinfo {year} {2013})},\ \Eprint
  {http://arxiv.org/abs/1303.6255} {arXiv:1303.6255 [astro-ph.CO]} \BibitemShut
  {NoStop}%
\bibitem [{\citenamefont {Hlozek}\ \emph {et~al.}(2018)\citenamefont {Hlozek},
  \citenamefont {Marsh},\ and\ \citenamefont {Grin}}]{Hlozek:2017zzf}%
  \BibitemOpen
  \bibfield  {author} {\bibinfo {author} {\bibfnamefont {R.}~\bibnamefont
  {Hlozek}}, \bibinfo {author} {\bibfnamefont {D.~J.~E.}\ \bibnamefont
  {Marsh}}, \ and\ \bibinfo {author} {\bibfnamefont {D.}~\bibnamefont {Grin}},\
  }\href {\doibase 10.1093/mnras/sty271} {\bibfield  {journal} {\bibinfo
  {journal} {Mon. Not. Roy. Astron. Soc.}\ }\textbf {\bibinfo {volume} {476}},\
  \bibinfo {pages} {3063} (\bibinfo {year} {2018})},\ \Eprint
  {http://arxiv.org/abs/1708.05681} {arXiv:1708.05681 [astro-ph.CO]}
  \BibitemShut {NoStop}%
\bibitem [{\citenamefont {Akrami}\ \emph {et~al.}(2018)\citenamefont {Akrami}
  \emph {et~al.}}]{Akrami:2018odb}%
  \BibitemOpen
  \bibfield  {author} {\bibinfo {author} {\bibfnamefont {Y.}~\bibnamefont
  {Akrami}} \emph {et~al.} (\bibinfo {collaboration} {Planck}),\ }\href@noop {}
  {\  (\bibinfo {year} {2018})},\ \Eprint {http://arxiv.org/abs/1807.06211}
  {arXiv:1807.06211 [astro-ph.CO]} \BibitemShut {NoStop}%
\bibitem [{\citenamefont {Lozanov}\ and\ \citenamefont
  {Amin}(2017)}]{Lozanov:2016hid}%
  \BibitemOpen
  \bibfield  {author} {\bibinfo {author} {\bibfnamefont {K.~D.}\ \bibnamefont
  {Lozanov}}\ and\ \bibinfo {author} {\bibfnamefont {M.~A.}\ \bibnamefont
  {Amin}},\ }\href {\doibase 10.1103/PhysRevLett.119.061301} {\bibfield
  {journal} {\bibinfo  {journal} {Phys. Rev. Lett.}\ }\textbf {\bibinfo
  {volume} {119}},\ \bibinfo {pages} {061301} (\bibinfo {year} {2017})},\
  \Eprint {http://arxiv.org/abs/1608.01213} {arXiv:1608.01213 [astro-ph.CO]}
  \BibitemShut {NoStop}%
\bibitem [{\citenamefont {Amin}\ \emph
  {et~al.}(2012{\natexlab{a}})\citenamefont {Amin}, \citenamefont {Zukin},\
  and\ \citenamefont {Bertschinger}}]{Amin:2011hu}%
  \BibitemOpen
  \bibfield  {author} {\bibinfo {author} {\bibfnamefont {M.~A.}\ \bibnamefont
  {Amin}}, \bibinfo {author} {\bibfnamefont {P.}~\bibnamefont {Zukin}}, \ and\
  \bibinfo {author} {\bibfnamefont {E.}~\bibnamefont {Bertschinger}},\ }\href
  {\doibase 10.1103/PhysRevD.85.103510} {\bibfield  {journal} {\bibinfo
  {journal} {Phys. Rev.}\ }\textbf {\bibinfo {volume} {D85}},\ \bibinfo {pages}
  {103510} (\bibinfo {year} {2012}{\natexlab{a}})},\ \Eprint
  {http://arxiv.org/abs/1108.1793} {arXiv:1108.1793 [astro-ph.CO]} \BibitemShut
  {NoStop}%
\bibitem [{\citenamefont {Amin}\ \emph {et~al.}(2014)\citenamefont {Amin},
  \citenamefont {Hertzberg}, \citenamefont {Kaiser},\ and\ \citenamefont
  {Karouby}}]{Amin:2014eta}%
  \BibitemOpen
  \bibfield  {author} {\bibinfo {author} {\bibfnamefont {M.~A.}\ \bibnamefont
  {Amin}}, \bibinfo {author} {\bibfnamefont {M.~P.}\ \bibnamefont {Hertzberg}},
  \bibinfo {author} {\bibfnamefont {D.~I.}\ \bibnamefont {Kaiser}}, \ and\
  \bibinfo {author} {\bibfnamefont {J.}~\bibnamefont {Karouby}},\ }\href
  {\doibase 10.1142/S0218271815300037} {\bibfield  {journal} {\bibinfo
  {journal} {Int. J. Mod. Phys.}\ }\textbf {\bibinfo {volume} {D24}},\ \bibinfo
  {pages} {1530003} (\bibinfo {year} {2014})},\ \Eprint
  {http://arxiv.org/abs/1410.3808} {arXiv:1410.3808 [hep-ph]} \BibitemShut
  {NoStop}%
\bibitem [{\citenamefont {Raveri}\ and\ \citenamefont
  {Hu}(2018)}]{Raveri:2018wln}%
  \BibitemOpen
  \bibfield  {author} {\bibinfo {author} {\bibfnamefont {M.}~\bibnamefont
  {Raveri}}\ and\ \bibinfo {author} {\bibfnamefont {W.}~\bibnamefont {Hu}},\
  }\href@noop {} {\  (\bibinfo {year} {2018})},\ \Eprint
  {http://arxiv.org/abs/1806.04649} {arXiv:1806.04649 [astro-ph.CO]}
  \BibitemShut {NoStop}%
\bibitem [{\citenamefont {Hildebrandt}\ \emph {et~al.}(2018)\citenamefont
  {Hildebrandt} \emph {et~al.}}]{Hildebrandt:2018yau}%
  \BibitemOpen
  \bibfield  {author} {\bibinfo {author} {\bibfnamefont {H.}~\bibnamefont
  {Hildebrandt}} \emph {et~al.},\ }\href@noop {} {\  (\bibinfo {year}
  {2018})},\ \Eprint {http://arxiv.org/abs/1812.06076} {arXiv:1812.06076
  [astro-ph.CO]} \BibitemShut {NoStop}%
\bibitem [{\citenamefont {Abbott}\ \emph
  {et~al.}(2018{\natexlab{b}})\citenamefont {Abbott} \emph
  {et~al.}}]{Abbott:2017wau}%
  \BibitemOpen
  \bibfield  {author} {\bibinfo {author} {\bibfnamefont {T.~M.~C.}\
  \bibnamefont {Abbott}} \emph {et~al.} (\bibinfo {collaboration} {DES}),\
  }\href {\doibase 10.1103/PhysRevD.98.043526} {\bibfield  {journal} {\bibinfo
  {journal} {Phys. Rev.}\ }\textbf {\bibinfo {volume} {D98}},\ \bibinfo {pages}
  {043526} (\bibinfo {year} {2018}{\natexlab{b}})},\ \Eprint
  {http://arxiv.org/abs/1708.01530} {arXiv:1708.01530 [astro-ph.CO]}
  \BibitemShut {NoStop}%
\bibitem [{\citenamefont {Khlebnikov}\ and\ \citenamefont
  {Tkachev}(1996)}]{Khlebnikov:1996mc}%
  \BibitemOpen
  \bibfield  {author} {\bibinfo {author} {\bibfnamefont {S.~{\relax Yu}.}\
  \bibnamefont {Khlebnikov}}\ and\ \bibinfo {author} {\bibfnamefont {I.~I.}\
  \bibnamefont {Tkachev}},\ }\href {\doibase 10.1103/PhysRevLett.77.219}
  {\bibfield  {journal} {\bibinfo  {journal} {Phys. Rev. Lett.}\ }\textbf
  {\bibinfo {volume} {77}},\ \bibinfo {pages} {219} (\bibinfo {year} {1996})},\
  \Eprint {http://arxiv.org/abs/hep-ph/9603378} {arXiv:hep-ph/9603378 [hep-ph]}
  \BibitemShut {NoStop}%
\bibitem [{\citenamefont {Amin}\ \emph
  {et~al.}(2012{\natexlab{b}})\citenamefont {Amin}, \citenamefont {Easther},
  \citenamefont {Finkel}, \citenamefont {Flauger},\ and\ \citenamefont
  {Hertzberg}}]{Amin:2011hj}%
  \BibitemOpen
  \bibfield  {author} {\bibinfo {author} {\bibfnamefont {M.~A.}\ \bibnamefont
  {Amin}}, \bibinfo {author} {\bibfnamefont {R.}~\bibnamefont {Easther}},
  \bibinfo {author} {\bibfnamefont {H.}~\bibnamefont {Finkel}}, \bibinfo
  {author} {\bibfnamefont {R.}~\bibnamefont {Flauger}}, \ and\ \bibinfo
  {author} {\bibfnamefont {M.~P.}\ \bibnamefont {Hertzberg}},\ }\href {\doibase
  10.1103/PhysRevLett.108.241302} {\bibfield  {journal} {\bibinfo  {journal}
  {Phys. Rev. Lett.}\ }\textbf {\bibinfo {volume} {108}},\ \bibinfo {pages}
  {241302} (\bibinfo {year} {2012}{\natexlab{b}})},\ \Eprint
  {http://arxiv.org/abs/1106.3335} {arXiv:1106.3335 [astro-ph.CO]} \BibitemShut
  {NoStop}%
\bibitem [{\citenamefont {Lozanov}\ and\ \citenamefont
  {Amin}(2019)}]{Lozanov:2019ylm}%
  \BibitemOpen
  \bibfield  {author} {\bibinfo {author} {\bibfnamefont {K.~D.}\ \bibnamefont
  {Lozanov}}\ and\ \bibinfo {author} {\bibfnamefont {M.~A.}\ \bibnamefont
  {Amin}},\ }\href {\doibase 10.1103/PhysRevD.99.123504} {\bibfield  {journal}
  {\bibinfo  {journal} {Phys. Rev.}\ }\textbf {\bibinfo {volume} {D99}},\
  \bibinfo {pages} {123504} (\bibinfo {year} {2019})},\ \Eprint
  {http://arxiv.org/abs/1902.06736} {arXiv:1902.06736 [astro-ph.CO]}
  \BibitemShut {NoStop}%
\bibitem [{\citenamefont {Khlebnikov}\ and\ \citenamefont
  {Tkachev}(1997)}]{Khlebnikov:1997di}%
  \BibitemOpen
  \bibfield  {author} {\bibinfo {author} {\bibfnamefont {S.~Y.}\ \bibnamefont
  {Khlebnikov}}\ and\ \bibinfo {author} {\bibfnamefont {I.~I.}\ \bibnamefont
  {Tkachev}},\ }\href {\doibase 10.1103/PhysRevD.56.653} {\bibfield  {journal}
  {\bibinfo  {journal} {Phys. Rev.}\ }\textbf {\bibinfo {volume} {D56}},\
  \bibinfo {pages} {653} (\bibinfo {year} {1997})},\ \Eprint
  {http://arxiv.org/abs/hep-ph/9701423} {arXiv:hep-ph/9701423 [hep-ph]}
  \BibitemShut {NoStop}%
\bibitem [{\citenamefont {Svrcek}\ and\ \citenamefont
  {Witten}(2006)}]{Svrcek:2006yi}%
  \BibitemOpen
  \bibfield  {author} {\bibinfo {author} {\bibfnamefont {P.}~\bibnamefont
  {Svrcek}}\ and\ \bibinfo {author} {\bibfnamefont {E.}~\bibnamefont
  {Witten}},\ }\href {\doibase 10.1088/1126-6708/2006/06/051} {\bibfield
  {journal} {\bibinfo  {journal} {JHEP}\ }\textbf {\bibinfo {volume} {06}},\
  \bibinfo {pages} {051} (\bibinfo {year} {2006})},\ \Eprint
  {http://arxiv.org/abs/hep-th/0605206} {arXiv:hep-th/0605206 [hep-th]}
  \BibitemShut {NoStop}%
\bibitem [{\citenamefont {Cicoli}\ \emph {et~al.}(2012)\citenamefont {Cicoli},
  \citenamefont {Goodsell},\ and\ \citenamefont {Ringwald}}]{Cicoli:2012sz}%
  \BibitemOpen
  \bibfield  {author} {\bibinfo {author} {\bibfnamefont {M.}~\bibnamefont
  {Cicoli}}, \bibinfo {author} {\bibfnamefont {M.}~\bibnamefont {Goodsell}}, \
  and\ \bibinfo {author} {\bibfnamefont {A.}~\bibnamefont {Ringwald}},\ }\href
  {\doibase 10.1007/JHEP10(2012)146} {\bibfield  {journal} {\bibinfo  {journal}
  {JHEP}\ }\textbf {\bibinfo {volume} {10}},\ \bibinfo {pages} {146} (\bibinfo
  {year} {2012})},\ \Eprint {http://arxiv.org/abs/1206.0819} {arXiv:1206.0819
  [hep-th]} \BibitemShut {NoStop}%
\bibitem [{\citenamefont {Stott}\ \emph {et~al.}(2017)\citenamefont {Stott},
  \citenamefont {Marsh}, \citenamefont {Pongkitivanichkul}, \citenamefont
  {Price},\ and\ \citenamefont {Acharya}}]{Stott:2017hvl}%
  \BibitemOpen
  \bibfield  {author} {\bibinfo {author} {\bibfnamefont {M.~J.}\ \bibnamefont
  {Stott}}, \bibinfo {author} {\bibfnamefont {D.~J.~E.}\ \bibnamefont {Marsh}},
  \bibinfo {author} {\bibfnamefont {C.}~\bibnamefont {Pongkitivanichkul}},
  \bibinfo {author} {\bibfnamefont {L.~C.}\ \bibnamefont {Price}}, \ and\
  \bibinfo {author} {\bibfnamefont {B.~S.}\ \bibnamefont {Acharya}},\ }\href
  {\doibase 10.1103/PhysRevD.96.083510} {\bibfield  {journal} {\bibinfo
  {journal} {Phys. Rev.}\ }\textbf {\bibinfo {volume} {D96}},\ \bibinfo {pages}
  {083510} (\bibinfo {year} {2017})},\ \Eprint
  {http://arxiv.org/abs/1706.03236} {arXiv:1706.03236 [astro-ph.CO]}
  \BibitemShut {NoStop}%
\bibitem [{\citenamefont {Kamionkowski}\ \emph {et~al.}(2014)\citenamefont
  {Kamionkowski}, \citenamefont {Pradler},\ and\ \citenamefont
  {Walker}}]{Kamionkowski:2014zda}%
  \BibitemOpen
  \bibfield  {author} {\bibinfo {author} {\bibfnamefont {M.}~\bibnamefont
  {Kamionkowski}}, \bibinfo {author} {\bibfnamefont {J.}~\bibnamefont
  {Pradler}}, \ and\ \bibinfo {author} {\bibfnamefont {D.~G.~E.}\ \bibnamefont
  {Walker}},\ }\href {\doibase 10.1103/PhysRevLett.113.251302} {\bibfield
  {journal} {\bibinfo  {journal} {Phys. Rev. Lett.}\ }\textbf {\bibinfo
  {volume} {113}},\ \bibinfo {pages} {251302} (\bibinfo {year} {2014})},\
  \Eprint {http://arxiv.org/abs/1409.0549} {arXiv:1409.0549 [hep-ph]}
  \BibitemShut {NoStop}%
\bibitem [{\citenamefont {Ballesteros}\ and\ \citenamefont
  {Lesgourgues}(2010)}]{Ballesteros:2010ks}%
  \BibitemOpen
  \bibfield  {author} {\bibinfo {author} {\bibfnamefont {G.}~\bibnamefont
  {Ballesteros}}\ and\ \bibinfo {author} {\bibfnamefont {J.}~\bibnamefont
  {Lesgourgues}},\ }\href {\doibase 10.1088/1475-7516/2010/10/014} {\bibfield
  {journal} {\bibinfo  {journal} {JCAP}\ }\textbf {\bibinfo {volume} {1010}},\
  \bibinfo {pages} {014} (\bibinfo {year} {2010})},\ \Eprint
  {http://arxiv.org/abs/1004.5509} {arXiv:1004.5509 [astro-ph.CO]} \BibitemShut
  {NoStop}%
\bibitem [{\citenamefont {Greene}\ \emph {et~al.}(1997)\citenamefont {Greene},
  \citenamefont {Kofman}, \citenamefont {Linde},\ and\ \citenamefont
  {Starobinsky}}]{Greene:1997fu}%
  \BibitemOpen
  \bibfield  {author} {\bibinfo {author} {\bibfnamefont {P.~B.}\ \bibnamefont
  {Greene}}, \bibinfo {author} {\bibfnamefont {L.}~\bibnamefont {Kofman}},
  \bibinfo {author} {\bibfnamefont {A.~D.}\ \bibnamefont {Linde}}, \ and\
  \bibinfo {author} {\bibfnamefont {A.~A.}\ \bibnamefont {Starobinsky}},\
  }\href {\doibase 10.1103/PhysRevD.56.6175} {\bibfield  {journal} {\bibinfo
  {journal} {Phys. Rev.}\ }\textbf {\bibinfo {volume} {D56}},\ \bibinfo {pages}
  {6175} (\bibinfo {year} {1997})},\ \Eprint
  {http://arxiv.org/abs/hep-ph/9705347} {arXiv:hep-ph/9705347 [hep-ph]}
  \BibitemShut {NoStop}%
\end{thebibliography}%
\let\addcontentsline\oldaddcontentsline

\end{document}